\documentclass[aps,prd,final,amsmath,amssymb,superscriptaddress,floatfix,showpacs,footinbib]{revtex4}
\usepackage{float,graphicx}

\newcommand{\mum}{\,\mu\mathrm{m}}
\newcommand{\Oo}{{\cal O}}

\newcommand{\dd}{{\rm d}}

\newcommand\mpl{M_{\rm Pl}}

\newcommand{\GeV}{\,\mathrm{GeV}}
\newcommand{\eV}{\,\mathrm{eV}}
\newcommand{\meV}{\,\mathrm{meV}}
\newcommand{\cm}{\,\mathrm{cm}}

\begin{document}

\title{Detecting Chameleons through Casimir Force Measurements}

\author{Philippe~Brax}
\affiliation{ Service de Physique
Th\'eorique CEA/DSM/SPhT, Unit\'e de recherche
associ\'ee au CNRS, CEA-Saclay F-91191 Gif/Yvette
cedex, France.}
\author{Carsten van de Bruck}
\affiliation{ Department of Applied Mathematics, The University of Sheffield,
 Hounsfield Road, Sheffield S3 7RH, United Kingdom}
\author{Anne-Christine Davis}
\affiliation{Department of Applied Mathematics and Theoretical Physics,  Centre for Mathematical Sciences,  Cambridge CB2 0WA, United Kingdom}
\author{David F. Mota}
\affiliation{Institut f\"ur Theoretische Physik,
Universit\"at Heidelberg, Philosophenweg 16/19,  D-69120 Heidelberg, Germany}
\author{Douglas Shaw}
\affiliation{Department of Applied Mathematics and Theoretical Physics,  Centre for Mathematical Sciences,  Cambridge CB2 0WA, United Kingdom}

\begin{abstract}
The best laboratory constraints on strongly coupled chameleon
fields come not from tests of gravity per se but from precision
measurements of the Casimir force.    The chameleonic force
between two nearby bodies is more akin to a Casimir-like force
than a gravitational one: The chameleon force behaves as an
inverse power of the distance of separation between the surfaces
of two bodies, just as the Casimir force does. Additionally,
experimental tests of gravity often employ a thin metallic sheet
to shield electrostatic forces, however this sheet mask any
detectable signal due to the presence of a strongly coupled
chameleon field.  As a result of this shielding, experiments that
are designed to specifically test the behaviour of gravity are
often unable to place any constraint on chameleon fields with a
strong coupling to matter.  Casimir force measurements do not
employ a physical electrostatic shield and as such are able to put
tighter constraints on the properties of chameleons fields with a
strong matter coupling than  tests of gravity. Motivated by this,
we perform a full investigation on the possibility of testing
chameleon model with both present and future Casimir experiments.
We find that present days measurements are not able to detect the
chameleon. However, future experiments have a strong possibility
of detecting or rule out a whole class of chameleon models.
\end{abstract}


\maketitle


\section{Introduction}
One of the most common predictions made by modern theories for
physics beyond the standard model is the existence of light scalar
fields.  It is usually the case that these fields couple to matter
and hence mediate a new (or `fifth') force between bodies.  To
date, however, no such new force has been detected, despite
numerous experimental attempts to do so \cite{cwill}. Any force
associated with light scalar fields must therefore be considerably
weaker than gravity over these scales, and under the conditions,
that have so far been probed experimentally. This imposes a strong
constraint on the properties of any new scalar fields; they must
either interact with matter much more weakly than gravity does, or
they must be sufficiently massive in the laboratory so as to have
remained undetected. If the mass, $m_{\phi}$, of the scalar field
is a constant then one must require that $\hbar c/m_{\phi}
\lesssim 0.1$~mm if the field is to couple to matter with a
strength equal to that of gravity.  The bounds on fields whose
interactions with matter have a super-gravitational strength are
even tighter \cite{cwill}.

It has recently be shown, however, that the most stringent
experimental limits on the properties of light scalar fields can
be exponentially relaxed if the scalar field theory in question
possesses a \emph{chameleon mechanism} \cite{chamKA,chamstrong}.
The chameleon mechanism provides a way to suppress the forces
mediated by the scalar fields via non-linear field
self-interactions.  A direct result of these self-interactions is
that the mass of the field is no longer fixed but depends on,
amongst other things, the ambient density of matter. The
properties of these scalar fields therefore change depending on
the environment; it is for this reason that such fields have been
dubbed \emph{chameleon fields}. Importantly, Chameleon fields
could potentially also be responsible for the observed late-time
acceleration of the Universe \cite{chamcos,chamstruc}. If this
does indeed turn out to be the case, it raises the exciting
prospect of being able to directly detect, probe and potentially
even manipulate dark energy under controlled laboratory
conditions. The properties of chameleon field theories are
constrained by experimental tests of gravity, however, as a result
of their chameleonic behaviour, theories in which the fields and
matter interact with at least gravitational strength are
\emph{not} currently ruled out \cite{chamKA,chamstrong}.  Indeed,
laboratory-based gravitational tests alone cannot even place an
upper bound strength of chameleonic interactions with matter
\cite{chamstrong}.  It was recently shown that some
strongly-coupled (i.e. compared to gravity) chameleon theories
predict alterations to the way in which light propagates through
the vacuum in the presence of a magnetic field
\cite{chamPVLAS,chamPVLASlong}; the resultant birefringence and
dichroism could be detected by laboratory searches for
axion-like-particles e.g PVLAS, Q\&%
A and BMV \cite{axion}.

In Ref. \cite{chamstrong} it was shown that the best laboratory
constraints on strongly coupled chameleon fields come not from tests
of gravity per se but from precision measurements of the Casimir
force.  In some ways this is not surprising.  As we shall see, the
chameleonic force between two nearby bodies is, in many ways, more
akin to a Casimir-like force than a gravitational one. Much like the Casimir force, the chameleonic force generally depends only very weakly on the composition and density of the test masses
and in one class of theories, the chameleon force behaves as an inverse power of the
distance of separation between the \emph{surfaces} of two bodies.
Additionally, unlike gravitational forces, the chameleonic force can
be shielded \cite{chamstrong}.

Experimental tests of gravity often employ a thin metallic sheet
to shield electrostatic forces, however this sheet was also shown
in Ref. \cite{chamstrong} to mask any detectable signal due to the
presence of strongly coupled chameleon fields.  This is because,
in such theories, the shield develops what is known as a
thin-shell.  This means that the range of the chameleon field,
$\lambda_{\phi} = \hbar c / m_{\phi}$, inside the metallic sheet
is much smaller than the thickness of the sheet, $d_{\rm shield}$.
In experimental tests of gravity, one measures the force or torque
on one test mass (`the detector') due to the movement or rotation
of another ('the attractor').  The electrostatic shield sits
between the two.  The shield is held fixed relative to the the
detector and is uniform.  As a result, residual forces due to the
shield itself do not result in any detectable effect. In chameleon
theories, the electrostatic shield attenuates the chameleonic
force (or torque) due to the attractor by a factor of
$\exp(-m_{\phi}d_{\rm shield})$.  If $m_{\phi} d_{\rm shield} \gg
1$, the electrostatic shield therefore acts as a near perfect
shield of the chameleonic force due to the attractor.   Since
$m_{\phi}$ is larger for strongly coupled fields than it is in
more weakly interacting ones, experiments that are designed to
specifically test the behaviour of gravity are often unable to
place \emph{any} constraint on chameleon fields with a strong
coupling to matter.  Casimir force measurements, on the other
hand, do not employ a physical electrostatic shield and as such
are able to put tighter constraints on the properties of
chameleons fields with a strong matter coupling than  tests of
gravity.

A preliminary analysis of the constraints on chameleon fields provided
by Casimir force measurements was made in Ref. \cite{chamstrong}.  In
this paper we refine, extend and generalize this earlier study.  Our
primary aim is to extract the bounds that measurements of the Casimir force
currently place on chameleon theories and to make predictions for what
near future Casimir experiments will be able to detect.  We shall see that there is a very real prospect that the next generation of Casimir force experiments will be able to detect or rule out most chameleonic models of dark energy.

This paper is organized as follows: In Section \ref{sec:chammodel}
we introduce the chameleon model in greater detail as well as the
concept of a \emph{thin-shell}. When dealing with gravitational
tests, it is generally the case that if the test masses have
thin-shells, then all detectable effects due to chameleon fields
are exponentially attenuated.  In Casimir force experiments,
however, the opposite is true.  This is because when very small
separations are used, the gradient of chameleon force is largest
for thin-shelled test bodies, and Casimir tests are generally most
sensitive not to the magnitude of any new forces but to their
gradients. In Section \ref{sec:thinshell} we present the
conditions that must be satisfied for a test body to have a
thin-shell.  In Section \ref{sec:force} we derive the form of the
chameleonic force between two nearby bodies such as those used to
measure the Casimir force. These results are applied in Section
\ref{sec:predict} to predict the extra force that should be
detected by Casimir force measurements if chameleon fields exist.
We also consider to what extent current experiments constrain two
of the simplest and most widely studied classes of chameleon
theories. In the penultimate section, we then consider the extent
to which planned future experiments will be able to extend the
constraints on chameleon theories, and identify two proposed tests
that have the sensitivity to detect or rule out most chameleon
theories in which the chameleon potential is associated with dark
energy. We conclude in \ref{sec:con} with a discussion of our
results.

\section{Chameleon Theories}\label{sec:chammodel}
\subsection{The Action}
As was mentioned above, chameleon theories are essentially scalar field theories with a self-interaction potential and a coupling to matter; they are specified by the action
\begin{eqnarray}
S&=&\int d^4x \sqrt{-g}\left(\frac{1}{2\kappa_4^2}R-
g^{\mu\nu}\partial_\mu\phi \partial_\nu \phi -V(\phi)\right) \\ &+& S_m( e^{\phi/M_{i}}
g_{\mu\nu},\psi_{m}) \label{action},
\end{eqnarray}
where $\phi$ is the chameleon field, $S_{m}$ is the matter action and $\psi_{m}$ are the matter fields; $V(\phi)$ is the self-interaction potential.

The strength of the interaction between $\phi$ and the matter fields is determined
by the one or more mass scales $M_{i}$.  In general, we expect different
particle species to couple with different strengths to the chameleon field
i.e. a different $M_{i}$ for each $\psi_{m}$.  Such a differential coupling
generally leads to violations of the weak equivalence principle (WEP hereafter).
Constraints on any WEP violation are very tight \cite{cwill}.
Importantly though, it has been shown that $V(\phi)$ can be chosen so that any
violations of WEP are too small to be have been detected thus far
\cite{chamKA,chamstrong}.  Even though the $M_{i}$ are generally different for
different species, if $M_{i} \neq 0$, we expect $M_{i} \sim \Oo(M)$ where
$M$ being some mass scale associated with the theory.
In this paper we are concerned with those signatures of chameleon theories
that could be detected through measurements of the Casimir force.
Since these measurements place bounds  on the magnitude (or gradient) of
close range forces rather than on any violation of WEP, and since also all that matters in this context is the coupling of the chameleon field to atoms rather than any more exotic form of matter, allowing for different $M_{i}$ is an usually an unnecessary complication. Henceforth, we assume a universal coupling $M_{i} = M$ for all $i$ and take the matter fields to be non-relativistic. The scalar field, $\phi$, then obeys:
\begin{equation}
\square \phi = V^{\prime}(\phi) + \frac{e^{\phi/M}\rho}{M}, \label{chameqn}
\end{equation}
where $\rho$ is the background density of matter.  The coupling to matter implies that particle masses in the Einstein frame depend on the value of $\phi$
\begin{equation}
m(\phi)= e^{\phi/M} m_0
\end{equation}
where $m_0 = {\rm const}$ is the bare mass. We parametrize the strength of the chameleon to matter coupling by $\beta$ where
\begin{equation}
\beta= \frac{\mpl}{M},
\end{equation}
and $\mpl = 1/\sqrt{8\pi G} \approx 2.4 \times 10^{18}\GeV$. On
microscopic scales (and over sufficiently short distances), the
chameleon force between two particles is then $2\beta^2$ times the
strength of their mutual gravitational attraction.

If the mass, $m_{\phi} \equiv \sqrt{V^{\prime \prime}(\phi)}$, of
$\phi$ is a constant then one must either require that $m_{\phi}
\gtrsim 1\meV$ or $\beta \ll 1$ for such a theory not to have been
already ruled out by experimental tests of gravity
\cite{cwill}. If, however, the mass of the scalar field grows with the
background density of matter, then a much wider range of scenarios
is possible \cite{chamKA,chamstrong,chamcos}.  In
high density regions $m_{\phi}$ can then be large enough so as to
satisfy the constraints coming from tests of gravity. At the same
time, the mass of the field can be small enough in low density regions
to produce detectable and potentially important alterations to
standard physical laws. Scalar fields that have this property are said
to be \emph{Chameleon fields}.  Assuming $\dd \ln m(\phi) / \dd \phi
\geq 0$ as it is above, a scalar field theory possesses a chameleon
mechanism if, for some range of $\phi$, the self-interaction
potential, $V(\phi)$, has the following properties:
\begin{equation}
V^{\prime}(\phi) < 0, \quad V^{\prime \prime} > 0,\quad V^{\prime \prime \prime}(\phi) < 0, \label{chamcond}
\end{equation}
where $V^{\prime} = \dd V / \dd\phi$. Whether or not the chameleon mechanism is both active and strong enough to evade current experimental constraints depends partially on the details of the theory, i.e. $V(\phi)$ and $M$, and partially on the initial conditions (see Refs. \cite{chamKA,chamstrong,chamcos} for a more detailed discussion). For exponential matter couplings  and a potential of the form:
\begin{equation}\label{poti}
V(\phi)= \Lambda^4_0\exp (\Lambda^n/\phi^n) \approx \Lambda^4_0\left(1 +
\frac{\Lambda^{n}}{\phi^n}\right)
\end{equation}
the chameleon mechanism can in principle hide the field such that
there is no conflict with current laboratory, solar system or
cosmological experiments and observations
\cite{chamKA,chamcos}. Importantly, for a large range of values of
$\Lambda$, the chameleon mechanism is strong enough in such theories
to allow even strongly coupled theories with $M \ll M_{Pl}$ to have
remained undetected \cite{chamstrong}. The first term in $V(\phi)$
corresponds to an effective cosmological constant whilst the second
term is a Ratra-Peebles inverse power law potential. If one assumes
that $\phi$ is additionally responsible for late-time acceleration of
the universe then one must require $\Lambda_0 \approx
(2.4\pm 0.1) \times 10^{-12}\GeV$.  In the simplest theories, $\Lambda \sim \Oo(\Lambda_0)$ so that there is only
energy scale in the potential; it is arguable that this represents the most natural scenario.
The smallness of $\Lambda_0$, means that, as a dark energy candidate, Chameleon theory do not solve either the naturalless
problem or the coincidence problem. However, although it would certainly be desirable
to have a model which solved both of these problems, one cannot exclude the possibility
that the acceleration of the Universe is the first sign of some new physics associated with an $\Oo(\Lambda_0)$ energy scale.
If this is truly the case then one must look for new ways in which to probe physics
at this low energy scale. As we show in this paper, the use of Casimir force experiments to search for chameleon
fields is once such probe.

Throughout the rest of this paper, it is our aim to remain as general
as possible and assume as little about the precise form of $V(\phi)$
as is necessary.  However, when we come to more detailed discussions
and make specific numerical predictions, it will be necessary to chose
a particular form for $V(\phi)$. In these situations we assume that
$V(\phi)$ has either has the following form:
$$
V(\phi) = \Lambda^4_0\left(1 + \frac{\Lambda^{n}}{\phi^n}\right).
$$
or
$$
V(\phi) = \Lambda_0^4 \exp(\Lambda^n/\phi^n).
$$
We do this not because these forms of $V$ are in any way preferred or to be expected, but merely as they have been the most widely studied in the literature and, in the case of the power-law potential, because is the simplest with which to perform analytical calculations.  The power-law form is also useful as an example as it displays, for different values of the $n$, many the features that we expect to see in more general chameleon theories.

The evolution of the chameleon field in the presence of ambient matter with density $\rho_{\rm matter}$ is determined by the effective potential:
\begin{equation}
V_{\rm eff}(\phi)=V(\phi) + \rho_{\rm matter} e^{\phi/M}
\end{equation}
Limits on any variation of the fundamental constants of Nature mean that, under the conditions that are accessible in the laboratory, we must have $\phi / M \ll 1$ \cite{chamstrong}. Henceforth we therefore take $\exp(\phi/M) \approx 1 + \phi/M$.
Even in theories where $V(\phi)$ has no minimum of its own (e.g. where it has a runaway form), the conditions given by Eq. (\ref{chamcond}) on $V(\phi)$ ensure that the effective potential has minimum at $\phi = \phi_{\rm min}(\rho_{\rm matter})$ where
\begin{equation}
V^{\prime}_{\rm eff}(\phi_{\rm min}) = 0 = V^{\prime}(\phi_{\rm min}) + \frac{\rho_{\rm matter}}{M}.
\end{equation}

\subsection{Thin-shells}
In chameleon field theories, macroscopic bodies may develop what has
been called a `thin-shell'.  Generally speaking, a body of density
$\rho_{c}$ is said to have a thin-shell if, deep inside that body,
$\phi$ is at, or lies very close to, the minimum of its effective
potential (where $\phi = \phi_c \equiv \phi_{min}(\rho_c)$ say).  We
take the density of matter outside the body to be $\rho_b$; far
outside the body $\phi \approx \phi_{b} \equiv \phi_{min}(\rho_b)$.

Thin-shelled bodies are so-called because, for such bodies, almost all
of the change in $\phi$ (from $\phi_b$ to $\phi_c$) occurs in a thin
region near the surface of the body.  The thickness of the part of
this thin region that lies inside the body is generally $\approx
\Oo(1/m_c)$ where $m_c \equiv m_{\phi}(\phi_c)$.  If the body has
thickness $R$ then, a necessary (but not sufficient) condition for a
body to have a thin-shell is $m_c R \gg 1$.

As a rule of thumb, larger bodies tend to have thin-shells whereas
smaller bodies do not.  Precisely what is meant by `larger' and
`smaller', however, depends on the details of the theory. We discuss
this further in Section \ref{sec:thinshell} below. If $m_b =
m_{\phi}(\phi_b)$ is the mass of the chameleon in the background then
the chameleon force between two non-thin-shelled bodies, separated by
a distance $r$, is $2\beta^2 e^{-m_b r}$ times as strong as their
mutual gravitational attraction. Importantly, the force between two
thin-shelled bodies is much weaker \cite{chamKA}. Moreover, it has
been shown that the chameleonic force between two thin-shelled bodies
is, to leading order, independent of the strength, $\beta$, with which the chameleon
field couples to either body \cite{chamstrong}.  In a body with a
thin-shell, it is as if the chameleon field and the resultant force
only interact with and act on the matter that is in the thin-shell
region near the body's surface.

\section{Thin-Shell in Casimir Force Experiments}\label{sec:thinshell}
Before we can consider the form or magnitude of chameleonic force we
need to know whether or not the test masses used to measure the
Casimir force are predicted to have thin-shells.  In subsection
\ref{sec:thincond} we state the thin-shell conditions for an isolated
spherical body in the context of general chameleon theory, which are
themselves derived in Appendix \ref{appA}. We then consider whether
these conditions hold for the test masses used in those Casimir force
measurements that have been conducted thus far in subsection
\ref{sec:thincas}.

\subsection{Thin-Shell Conditions}\label{sec:thincond}
In general, `larger' bodies have thin-shells whereas `smaller' ones do not.
How small is `small', however, generally depends on the details of the theory.
The test-masses used in Casimir force experiments come in a number of different shapes
and sizes.  Some experiments used relatively small test masses, with
typical length scales of $\Oo(10^{2}\mum)$, whilst others, perhaps
most notably that performed by Lamoreaux in 1997 \cite{lam97}, used
relatively large test masses with length scales of $1 - 10 \cm$.

The condition that must be satisfied for an isolated spherical body to
have a thin-shell was first derived, for $V \propto \phi^{-n}$ with
$n>0$ potentials, in Ref. \cite{chamKA}. In Ref. \cite{chamstrong},
the thin-shell conditions for such potentials were re-derived (via a
different method) and extended to theories with $n \leq -4$.  The
thin-shell conditions for theories with $n < -4$, $n = -4$ and $n > 0$
were found to be qualitatively different.

In Appendix \ref{appA}, we derive the thin-shell condition for general
$V(\phi)$. We consider an isolated spherical body with density
$\rho_c$, radius $R$ in a background with density $\rho_b$.  We define
$\phi_b$ by $V^{\prime}(\phi_b) = -\rho_b/M$ and $\phi_c$ by
$V^{\prime}(\phi_c) = -\rho_c/M$. We also define $m_b =
m_{\phi}(\phi_b)$ and $m_c = m_{\phi}(\phi_c)$. If $m_b R \gg 1$ then
the body will always have a thin-shell of some description as almost
all variation in $\phi$ will take place in a thin region (of thickness
at most $\sim 1/m_b$) near the surface of the body.  However, if $m_c
\approx m_b$ then this thin-shell would be \emph{linear}, i.e. we
would see almost the same behaviour if we considered a Yukawa theory
with mass $m_b$ (for which the field equations would be linear).  In
the cases we consider, however, $\rho_c \gg \rho_b$ and so necessarily
$m_c \gg m_b$.  If $m_c R$ is large enough then, whatever the value of
$m_b R$, a body may have a \emph{non-linear} thin-shell.  Non-linear
thin-shells are associated with the dominance, near the surface of the
body, of non-linear terms in the field equations.  Such behaviour
would \emph{not} occur in theories where $\phi$ has only a Yukawa coupling to
matter. This non-linear behaviour is key in allowing chameleon
theories to evade the stringent experimental constraints on the
coupling of a scalar field to matter that exist for Yukawa theories.
In Appendix \ref{appA} we find that a necessary and sufficient condition for a
non-linear thin-shell, in a general chameleon theory, is:
\begin{equation}
{\cal  C} = \frac{(\rho_c-\rho_b) f(m_b R) R^2}{2 M[m_b^2R^2 + m_bR+1]}\gtrsim  \phi_b - \phi_c - \frac{(\rho_{c}-\rho_b)(1-f(m_b R))}{M m_c^2}, \label{thin1}
\end{equation}
where this defines ${\cal C}$. An equivalent statement of this condition is:
\begin{equation}
\frac{m_c^2 R^2}{m_b^2R^2 + m_bR+1} \gtrsim  \frac{2M m_c^2\left(\phi_b -\phi_c\right)}{(\rho_c-\rho_b) f(m_b R)} - \frac{2(1-f(m_b R))}{f(m_b R)} \geq 2.
\end{equation}
We have defined:
$$
f(m_b R) = 2e^{-m_b R} \cosh(m_b R)\left[1+\frac{1}{m_b R} + \frac{1}{m_b^2 R^2}\right]\left(1-\frac{\tanh m_b R}{m_b R}\right).
$$
As $m_b R \rightarrow 0$, $f(m_b R) \rightarrow 2/3$ and as $m_b R \rightarrow \infty$, $f(m_b R) \rightarrow 1$.

Far from a spherical body with a non-linear thin-shell $\phi$ has the form:
$$
\phi \approx \phi_b - \frac{{\cal C}_{\rm thin} Re^{m_b(R-r)}}{r},
$$
where
$$
\frac{{\cal C}_{\rm thin}(1 + m_b R + m_b^2 R^2)}{V^{\prime}(\phi_b) - V^{\prime}(\phi_b-{\cal C}_{\rm thin})}  = \frac{R^2}{2}.
$$
We note that ${\cal C}_{\rm thin} = {\cal C}_{\rm thin}(R, \phi_b)$ and is therefore independent of $\rho_c/M$.  It follows that, in all chameleon theories, far from a body with a non-linear thin-shell $\phi$ is independent of $\rho_c/M$ i.e. it is independent of the strength with which the chameleon field couples to the matter in the body.  This, more than anything else, is what makes it so difficult for experimental tests of gravity to place a lower bound on $M$.

In many cases, it is only necessary to consider the following sufficient condition for a thin-shell:
$$
{\cal C} \gtrsim \phi_b - \phi_c,
$$
where ${\cal C}$ is given by Eq. (\ref{thin1}).

\subsection{Applying the Thin-Shell Conditions}\label{sec:thincas}
We now consider whether or not the test masses used in experimental
measurements of the Casimir force are generally predicted to have
non-linear thin-shells. The above thin-shell conditions are valid for
isolated, spherical bodies.  Generally, however, at least one of the
test masses used in Casimir force measurements is neither isolated nor
spherical.  Isolated in this case means that there is enough space
between the body in consideration and any other bodies for, in all
directions, $m_{\phi}(\phi)$ to have decreased to be about $m_b$
before any other body is encountered.

Generally speaking, $m_{\phi} \rightarrow m_{b}$ over a distance of
about $1/m_b$. We therefore take an isolated body to be one outside
which there is a region of thickness at least $\gtrsim 1/m_b$ in which
$\rho \approx \rho_b$.

For isolated non-spherical bodies, such as rectangular plates with
volume $V$ and longest dimension $2D$, the above thin-shell conditions
still apply (to a good approximation) provided one replaces $R$ by
$\sqrt{3V/4\pi D}$ (i.e. one should replace $R^2$ by the volume
divided by the longest distance from the centre of mass of the body
and its surface).  If two plates with volumes $V_1$ and $V_2$ and
longest dimensions $2D_1$ and $2D_2$ respectively are placed a
distance $d$ apart, with $d \gtrsim 1/m_b$, Eq. (\ref{thin1}) gives
the thin-condition for each plate with $R$ replaced by
$\sqrt{3V_{i}/4\pi D_{i}}$ for $i=1,\,2$. If, however, $d \ll 1/m_b$,
then plates are \emph{not} isolated and they effectively count as one
mass for the purposes of applying the thin-shell conditions.  Provided
$1/m_c$ is small compared to the smallest dimension of each plate, the
thin-shell condition for both plates is then given by
Eq. (\ref{thin1}) but with $R = \sqrt{3V_{\rm tot}/4\pi D_{\rm tot}}$ where
$V_{\rm tot}$ is the total volume of the plates (excluding the space in
between them) and $D_{\rm tot}$ is the half longest dimension of the two
plates when considered as a single object.

The overall geometry of the set-ups used for Casimir force measurement
is generally quite complicated .  Even in experiments where the test
masses are themselves relatively small and thin, the apparatus that
surrounds them is not. Furthermore, the test masses are generally not
isolated in the sense defined above. This complicates the application
of the thin-shell conditions, and generally it can only be done
thoroughly within the context of a specific chameleon theory or class.

For definiteness and as an example we consider theories
where $V(\phi)$ has a Ratra-Pebbles
form $V(\phi) = \Lambda^{4}_0(1 + (\Lambda/\phi)^{n})$, where $n
> 0$.  We take $\Lambda_0^4 = 2.4 \times 10^{-3}\,{\rm eV}$ so that the constant term in the potential
is responsible for the late time acceleration of the Universe and
specifically consider theories where $\Lambda \approx \Lambda_0$. In
Casimir force experiments the test masses are much denser than the
laboratory vacuum in which they sit and so we take $\rho_c \gg
\rho_{b}$; this implies that $\phi_{c} \ll \phi_{b}$ since $n >0$.
Given these considerations, the thin-shell condition,
Eq. (\ref{thin1}), simplifies to:
$$
{\cal C} \equiv \frac{\rho_{c}R^2 f(m_b R)}{2M(1+m_b R + m_b^2 R^2)} \gtrsim \phi_{b}.
$$
Since $\rho_c \gg \rho_{b}$ if $m_b R \gtrsim 1$ then this condition
is automatically satisfied. We therefore restrict our attention to
those cases where $m_b R \ll 1$.  For $\rho_c \gg \rho_b$ and $m_b R
\ll 1$, the thin-shell condition is:
$$
\frac{GM_{\rm body}}{R} \gtrsim \frac{\phi_{b}}{2\beta^2 M},
$$
where $M_{\rm body}$ is the mass of the body and $\beta = M_{Pl}/M$.  Note
that such a simplification of the thin-shell condition will generally
occur for all theories where ($\phi$ can be shifted so that)
$\phi(\rho) \rightarrow 0$ as $\rho \equiv -MV^{\prime}(\phi(\rho))
\rightarrow \infty$.

We take the pressure of the laboratory vacuum to be $p\times
10^{-4}\,\mathrm{torr}$, $p \sim \Oo(1)$ or greater for all
Casimir force measurements made to date. We then find that:
$$
\frac{\phi_{b}}{2 \beta^2 M} = \beta^{-\frac{n+2}{n+1}}
p^{-\frac{1}{n+1}} {\cal B}_{n},
$$
where
$$
{\cal B}_{n} = 4.9 \times 10^{-31} \left(5.1 n\times 10^{10}\right)^{\frac{1}{n+1}}.
$$
The largest value of ${\cal B}_{n}$ occurs for $n \approx 0.048$ at which
${\cal B}_{n} \approx 4.5 \times 10^{-22}$ and ${\cal B}_{n}$ decreases very quickly
to $4.9 \times 10^{-31}$ as $n \rightarrow \infty$.  Additionally, as
is discussed more fully in Refs. \cite{chamKA,chamstrong}, one must be
aware that the smallest value that $m_{b}$ can take in laboratory
vacuum which has a smallest length scale $L_{\rm vac}$ is $\Oo(1/L_{\rm vac})$.  If a typical value of $L_{\rm vac} = 1\,{\rm m}$, we therefore have
$$
\phi_{b}/2\beta^2 M \lesssim \beta^{-1}\left[ 4.9 \times 10^{-31}\left(1.5n(n+1)\times 10^{8}\right)^{\frac{1}{n+2}}\right] < 1.5 \times 10^{-27}.
$$
For bodies with $\rho_c \geq \rho_{\rm glass} \approx
3\,{\rm g\, cm}^{-3}$, we have:
$$
\frac{GM_{\rm body}}{R} \geq 6.7 \times 10^{-27} \left(\frac{V}{R\,cm^2}\right),
$$
where $V$ is the volume of the body.  If we take $V$ to be the volume of the smallest isolated system associated with either of the test masses, and $R=D_{\rm long}$ the longest distance from the surface of this system to its centre of mass, then the test bodies will certainly have thin-shells if:
$$
R_{\rm eff} \equiv \sqrt{\frac{V}{D_{\rm long}}} \gtrsim \frac{1}{2\sqrt{\beta}}\cm.
$$
For this choice of potential and $\vert n \vert \sim \Oo(1)$, the
E\"{o}t-Wash experiment \cite{EotWash} requires that if $\beta
\geq 10^{-2}$ and $\Lambda = 2.3 \times 10^{-3} \eV$ then $\beta$
must be larger than $10^{2}$ \cite{chamstrong}. For
$V=\Lambda^{4}_0(1 + \Lambda^n/\phi^n)$ with $\Lambda = \Lambda_0
= 2.4\times 10^{-3}\eV$, $\Vert n\Vert \sim \Oo(1)$, we have
checked that the thin-shell condition certainly holds for the test
masses used in the Casimir force experiments reported in Refs.
\cite{lam97,sparnaay,Bressi,Decca1,Decca2,Decca3} provided $\beta
\gtrsim 10^{3}$.  In other words, it holds for most strongly
coupled chameleon theories that are not already ruled out by tests
of gravity such as the E\"{o}t-Wash experiment.

\subsection{Discussion}
We found above for $V = \Lambda^{4}_0(1+\Lambda^n/\phi)$, $\vert n
\vert \sim \Oo(1)$ and $\Lambda = \Lambda_0 = 2.4\times
10^{3}\,eV$, the test masses used in all Casimir force experiments
conducted to date are predicted to have thin-shells in all
theories with $\beta \gtrsim 10^{3}$. In some experiments, where
particularly large test masses are used, the test masses are also
predicted to have thin-shells for $\Oo(1)$ values of $\beta$.

If $V = \Lambda^{4}_0 f((\Lambda/\phi)^{n})$, $n>0$, $\Lambda \approx
\Lambda_0 = 2.4\times 10^{-3}~eV$ where $f^{\prime \prime}>0$ and $f$ is
normalized so that $f^{\prime} = 1$ (e.g. $V = \Lambda^4_0
\exp(\Lambda^n/\phi^n)$, then the potential is always steeper than
the Ratra-Peebles form considered above and as such the thin-shell
conditions are less stringent.

In the next section we calculate the chameleonic force between two
nearby bodies under the assumption that they have thin-shells. In
the absence of thin-shells, the chameleon field behaves in the
same way as a Yukawa field, and the constraints on any Yukawa
coupling to matter derived from Casimir force measurements can be
directly applied to chameleon theories. There is, therefore,
nothing new to say about the non thin-shelled case, and we do not
consider it further.

We find below that the gradient in the chameleonic force between two
nearby thin-shelled bodies is generally much steeper than it would be
if there were no thin-shells present.  Casimir force experiments
generally measure gradients in forces (changes in forces between two
separations). So they are generally more sensitive to relatively small
quickly varying forces with a steep gradient than they are to large
but nearly constant ones. The presence of thin-shelled test masses is
therefore an aide rather than a hindrance to the detection of
chameleon fields via Casimir force measurements.  The opposite is
generally true of gravitational tests \cite{chamKA, chamstrong}.  The
stronger the matter coupling, the more likely it is that a given body
has a thin-shell.  Experiments designed along the lines of Casimir
force measurements are therefore far better suited to the search for
strongly coupled chameleon fields than tests that are specifically
designed to search for forces with gravitational (or
sub-gravitational) strength.

\section{The Chameleonic `Casimir' Force} \label{sec:force}
In this section we calculate the Casimir-like force between two nearby
bodies due to their interaction with a chameleon field.  The form of
both the Casimir force and the chameleonic force are highly dependent
on the geometry of the experiment \cite{gies,gies2,chamstrong}.  The form of these forces is most
easily calculated when the geometry is that of two parallel
plates. Making accurate measurements of forces using this set-up is,
however, notoriously difficult as it requires that the plates be both
very smooth and held parallel to a high precision.  For this reason,
most experiments conducted to date have measured the Casimir force
between a plate and a sphere rather than between two plates. Presently the highest precision measurements have been made using the sphere-plate geometry. By measuring the gradient of such a force
between a sphere and a plate it is possible to determine the force
between two parallel plates. In Section \ref{sec:force:plates} we calculate
the chameleon force between two parallel plates, and in Section
\ref{sec:force:sphere} we calculate the chameleon force for the
sphere-plate geometry.

\subsection{Parallel Plates Geometry}\label{sec:force:plates}
The parallel plate geometry is the easiest to study analytically. For simplicity we take both plates to have the same composition. We shall see that, provided the plates both have thin shells and are much denser than their environment, the chameleon force is in largely independent of composition of either plate. We take the plates to have thin-shells and to be separated by a distance $d$.

We define $x = 0$ to be the point midway between the two plates; the surfaces of the plates are then at $x=\pm d/2$.  In $-d/2 < x < d/2$ we have:
$$
\frac{\dd^2 \phi}{\dd x^2} = V^{\prime}(\phi) - V^{\prime}(\phi_b),
$$
where we have defined $V^{\prime}(\phi_b) = -\rho_b/M$ with $\rho_b$ the ambient density of matter outside the plates. Inside the plates we have:
$$
\frac{\dd^2 \phi}{\dd x^2} = V^{\prime}(\phi) - V^{\prime}(\phi_c),
$$
where $V^{\prime}(\phi_c) = -\rho_c/M$ and $\rho_c$ is the density of the plates. Deep inside either plate $\phi \rightarrow {\rm const} \approx \phi_c$ and $\dd \phi / \dd x = 0$ at $x=0$ by symmetry. We use the shorthand $\phi_0 \equiv \phi(x=0)$.  Integrating both of these equations once gives:
\begin{eqnarray}
&\left(\frac{\dd \phi}{\dd x}\right)^2 = 2\left(V(\phi)-V(\phi_0) - V^{\prime}(\phi_b)(\phi-\phi_0)\right) \qquad &-d/2 < x < d/2, \\
&\left(\frac{\dd \phi}{\dd x}\right)^2 = 2\left(V(\phi)-V(\phi_c) - V^{\prime}(\phi_c)(\phi-\phi_c)\right) \qquad &x^2 > d/4.
\end{eqnarray}
We define $\phi_{s} = \phi(x=\pm d/2)$, so that $\phi_s$ is the value of $\phi$ on the surface of the plates.  By matching the above equations are $x=\pm d/2$, we arrive at:
\begin{equation}
\phi_{s} = \frac{V(\phi_c) - V^{\prime}(\phi_c)\phi_c - V(\phi_0) + V^{\prime}(\phi_b)\phi_0}{V^{\prime}(\phi_b)-V^{\prime}(\phi_c)}. \label{phiseqn}
\end{equation}
If one of the plates were to be removed then $\phi_{s}$ on the surface of the remaining plate ($=\phi_{s0}$ say) would be given by Eq. (\ref{phiseqn}) but with $\phi_0 \rightarrow \phi_b$.  The perturbation, $\delta \phi_{s} = \phi_{s}-\phi_{s0}$, in $\phi_{s}$ due to presence of the second plate is therefore:
$$
\delta \phi_{s} = \frac{V(\phi_b)-V(\phi_0) -V^{\prime}(\phi_b)(\phi_b-\phi_0)}{V^{\prime}(\phi_b)-V^{\prime}(\phi_c)}.
$$
Deep inside either plate the perturbation, $\delta \phi$, in $\phi$ due to the presence of the second plate is exponentially attenuated. This is because the chameleon mass inside either plate $m_c \equiv m_{\phi}(\phi_c)$ is, by the thin-shell conditions, large compared to the thickness of the plate.

The attractive force per unit area, $F_{\phi}/A$, on one plate due to the other is given by:
\begin{equation}
\frac{F_{\phi}}{A} = \int^{d/2+D}_{d/2}\dd x  \frac{\rho_c}{M} \frac{\dd \delta \phi}{\dd x} \approx V^{\prime}(\phi_c)\delta\phi_{s},
\end{equation}
where $D$ is the plate thickness and we have used $\rho_c / M = -V^{\prime}(\phi_c)$. Taking $\rho_c \gg \rho_b$ we then find:
\begin{equation}
\frac{F_{\phi}}{A} = V(\phi_0) - V(\phi_b) + V^{\prime}(\phi_b)(\phi_b - \phi_0) \leq V(\phi_0)-V(\phi_b). \label{FAeqn}
\end{equation}
To leading order in $\rho_b/\rho_c$, $F_{\phi}/A$ depends only on $\phi_0$ and $\phi_b$.

To a first approximation, we calculate $\phi_0$ by linearizing the equation for $\phi$ in $-d/2 < x < d/2$ about $\phi_0$.  In $-d/2 < x < d/2$ we find:
\begin{equation}
\phi - \phi_{0} = \frac{2\left(V^{\prime}_{0} - V^{\prime}_{b}\right)}{m_{0}^2} \sinh^2\left(\frac{m_{0}x}{2}\right). \label{linapprox}
\end{equation}
where $V^{\prime}_0 = V^{\prime}(\phi_0)$ and $V^{\prime}_b=V^{\prime}(\phi_b)$. If, as is the case in theories with $V = \Lambda^4_0 f( (\Lambda/\phi)^n)$, $n>0$ and $f^{\prime} > 0$, we expect that $\phi$ on the surface of the plates is very small compared to $\phi_0$, then $m_0$ is given approximately by:
\begin{equation}
\sinh^2\left(\frac{m_{0}d}{4}\right) \approx \frac{m_0^2 \phi_0}{2\left(V^{\prime}_b - V^{\prime}_{0}\right)}.
\end{equation}
If $V = \Lambda^4_0(1 + \Lambda^n/\phi^n)$ (and $n>0$) when $m_{b}d \gg 1$ this gives:
$$
m_{0}d \approx 4\sinh^{-1}\sqrt{\frac{(n+1)}{2}}.
$$
This is a good approximation for $n \sim \Oo(1)$, but it breaks down for larger values of $n$. More generally, we must calculate $m_0 d$ by a more complicated method that takes proper account of the non-linear nature of the potential.  We define $y = \sqrt{V-V_{0} - V^{\prime}_b(\phi-\phi_{0})}$ and $1/W(y) = (V^{\prime}_b-V^{\prime}(\phi)) \geq 0$. From the chameleon field equation we then have:
\begin{equation}
\sqrt{2} \int_{0}^{y_{s}} W(y)\dd y = \frac{d}{2}, \label{inteqn}
\end{equation}
where $y_{s} = y(\phi=\phi_s)$. The above integral can then either be calculated numerically for a given $V(\phi)$ or, as is often more helpful, via an analytical approximation.  We show below that for $m_b d \gg 1$, $m_0 d \sim \Oo(1)$. When $m_{c} \gg m_0$, which therefore corresponds to $m_c d \gg 1$, we have $m_0 y_{s} W_{0} \gg 1$.  It can be checked that, $W(y)$ always decreases faster than $1/y$ as $y$ increases for $y \gg 1/m_0 W_0$. We can therefore approximate Eq. (\ref{inteqn}) by replacing $y_s$ with $\infty$ as the upper limit of the integral:
$$
\sqrt{2} \int_{0}^{\infty} W(y)\dd y = \frac{d}{2},
$$
Since $m_0 \leq m_c$, if $m_c d \lesssim 1$ then we must have $m_0 \approx m_c$.   In the $m_c d \gg 1$ case, we proceed by defining
\begin{equation}
k^2 = \frac{V^{\prime \prime \prime}_{0}(V_{0}^{\prime}-V_{b}^{\prime})}{m_{0}^4},
\end{equation}
where a subscript $0$ indicates that the quantity is evaluated for
$\phi=\phi_0$, and a subscript $b$ means that it is evaluated for
$\phi=\phi_b$.  When $m_0 \gg m_b$, which we shall see corresponds to
$m_b d \ll 1$, $k^2$ is, for many choices of potentials,
almost independent of $\phi_0$.  The value of $k^2$ dictates the dynamics of the theory, and determines how $\phi_0$
depends on $d$.  We evaluate Eq. (\ref{inteqn}) approximately in
 Appendix \ref{appB}.  We find that if, as is often the case, $1/3 \lesssim k^2
\leq 2$, then we can define
\begin{equation}
n_{\rm eff} = (2-k^2)/(k^2-1)
\end{equation}
leading to
\begin{equation}
m_{0} d \approx \sqrt{\frac{2n_{\rm eff} + 2}{n_{\rm eff}}} B\left(\frac{1}{2},\frac{1}{2}+\frac{1}{n_{\rm eff}}\right), \label{kmideqn}
\end{equation}
where $B(\cdot,\, \cdot)$ is the Beta function. $1/3 \lesssim k^2 \leq 2$ implies that $n_{eff} \geq 0$ or $n_{eff} \lesssim  -5/2$.  This approximation becomes exact when $V = \Lambda^4 + \Lambda^4(\Lambda/\phi)^n$ and $m_0 \gg m_b$. For such a potential $n_{eff} = n$, and the requirement that $m_0 \gg m_b$ implies $m_b d \ll 2$ (or all $n$).  For very steep potentials, e.g. $V = \Lambda^4 \exp( (\Lambda/\phi)^n)$ when $\phi_0 \ll \Lambda$, $k^2 \approx 1$ when $m_0 \gg m_b$. This corresponds to $n_{eff}^2 \rightarrow \infty$.
It is clear that $m_0 d \sim \Oo(1)$.

If $k^2 \geq 2$ then we find in Appendix \ref{appB} that:
\begin{equation} m_{0} d \approx
\frac{\pi^{3/2}}{2\sqrt{2}(k^2-2)^(1/2)}\left[J_{-1/4}^2\left(\frac{1}{2\sqrt{k^2-2}}\right)
+ Y_{-1/4}^2\left(\frac{1}{2\sqrt{k^2-2}}\right)\right],
\label{klargeeqn}
\end{equation} where $J_{-1/4}(\cdot)$ and $Y_{-1/4}(\cdot)$ are
Bessel functions.  For small $4(k^2-2)$ this gives:
$$
m_{0} d \approx \sqrt{2\pi}\left(1-3(k^2-2)/8\right),
$$
and if $k^2 \gg 2$ we have:
$$
m_{0}d \approx \frac{B\left(\frac{1}{4},\frac{1}{4}\right)}{\sqrt{2
k}} \approx \frac{5.24}{\sqrt{k}}.
$$
The $1/3 < k^2 \leq 2$ and $k^2 \geq 2$ approximations for $m_0 d$ are
continuous at $k^2 = 2$.

In Appendix \ref{appB}, we show that when $k^2 \lesssim 1/3$ the
analytical approximation used to evaluate $\phi_0(d)$ for $k^2 \gtrsim
1/3$ breaks down.  When $k^2$ is small it is either because at least
one of $k_0^2 =V^{(3)}_0V^{\prime}_0/m_0^4$ or
$1-V^{\prime}_b/V^{\prime}_0$ is small. Theories with small $k_0^2$
only exhibit very weak non-linear behaviour near $\phi_0$ and so we do
not consider them further.

It is important to know how $F_{\phi}/A$ behaves as $k^2 \rightarrow
0$ because as $d \rightarrow \infty$ we have $\phi_0 \rightarrow
\phi_b$.  In Appendix \ref{appB} we find that for $m_b d \gg 5$ we
have:
\begin{equation}
\frac{m_b d}{2} \approx \ln(12) - \ln(k^2),\label{mnsmall}
\end{equation}
and so
\begin{equation}
\frac{F_{\phi}}{A} \sim \frac{72 m_b^6 e^{-m_b
d}}{V^{\prime \prime \prime\,2}_b}. \label{Fmdlarge}
\end{equation}

We have now derived the expressions for $F_{\phi}(d)/A$ for different
classes of theory and for different ranges of $d$.  These expression
generally consist of an exact expression for $F_{\phi}/A$ as a
non-linear function of $\phi_0$ and an approximate implicit equation
for $\phi_0$ as a non-linear function of $d$. In these cases an
explicit equation for $F_{\phi}/A$ as a function of $d$ can only be
found once $V(\phi)$ is specified. In the limit $d \rightarrow
\infty$, it was possible to find an explicit expression for
$F_{\phi}/A$ as a function of $d$.  We note that when $d \gg
m_c^{-1}$, the leading chameleonic force is independent of $m_c$, and
hence also of the strength with which the chameleon field couples to
the body.
\begin{figure}[tbh]
\begin{center}
\includegraphics[width=8.8cm]{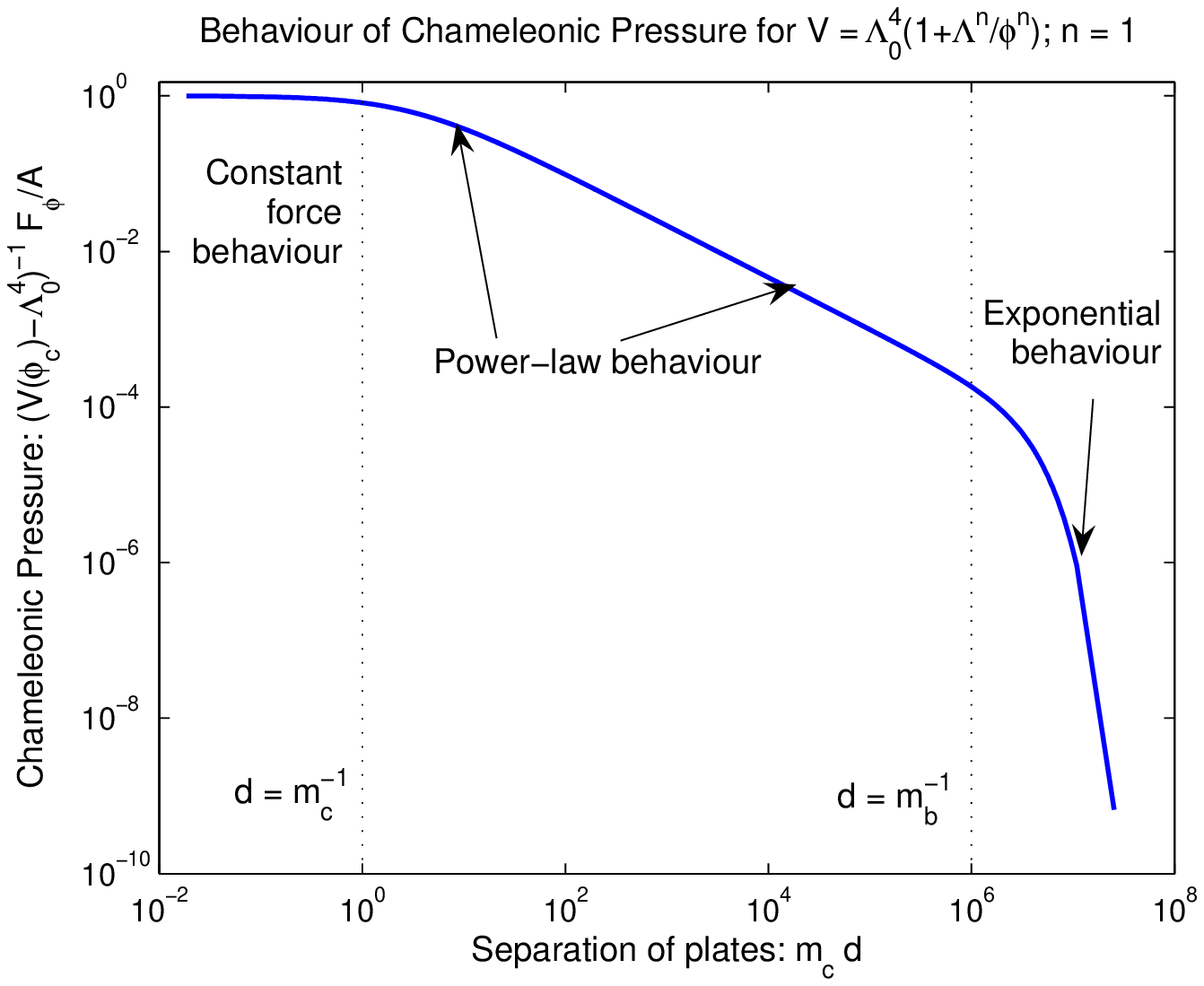}
\includegraphics [width=8.8cm]{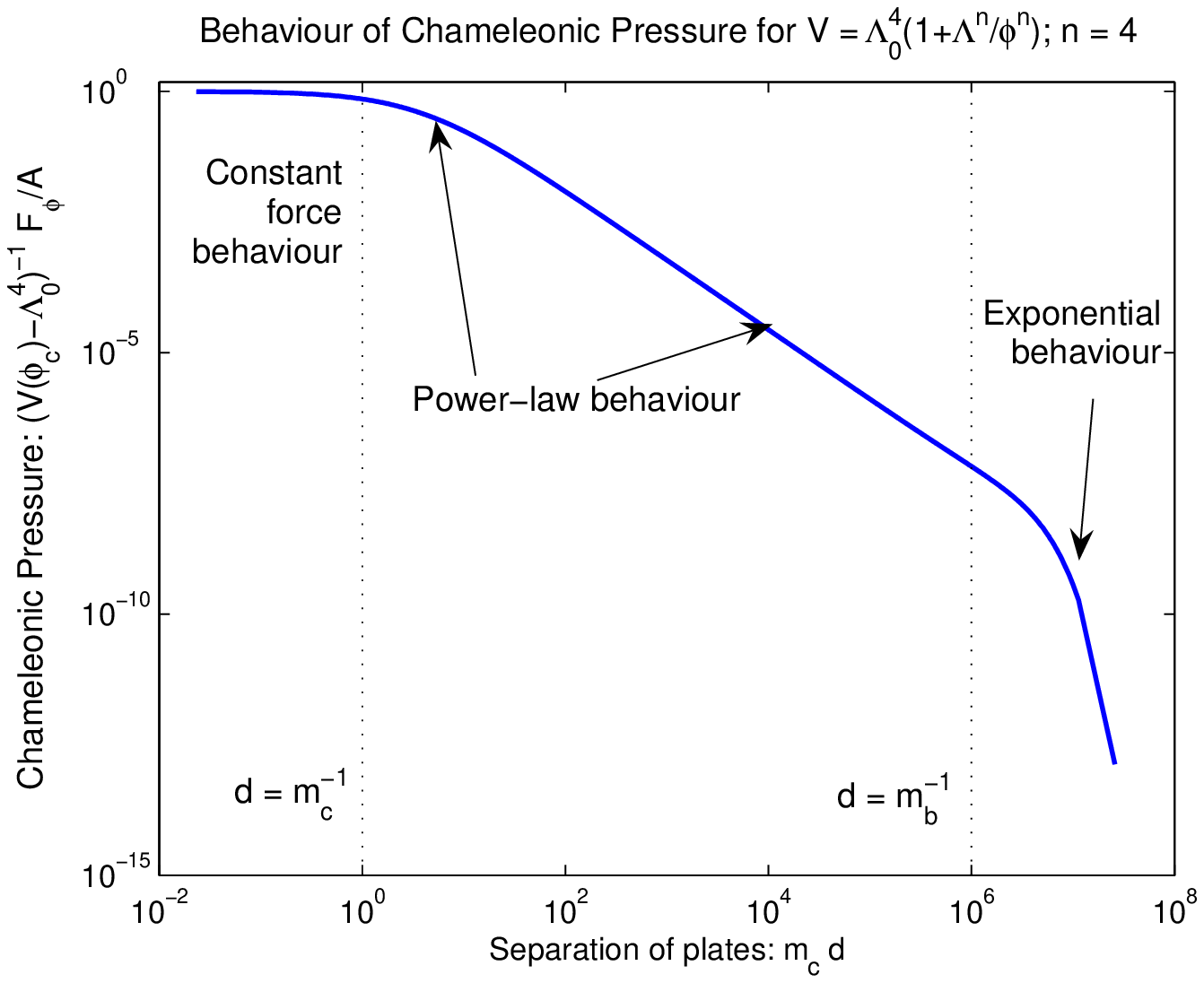}
\includegraphics [width=8.8cm]{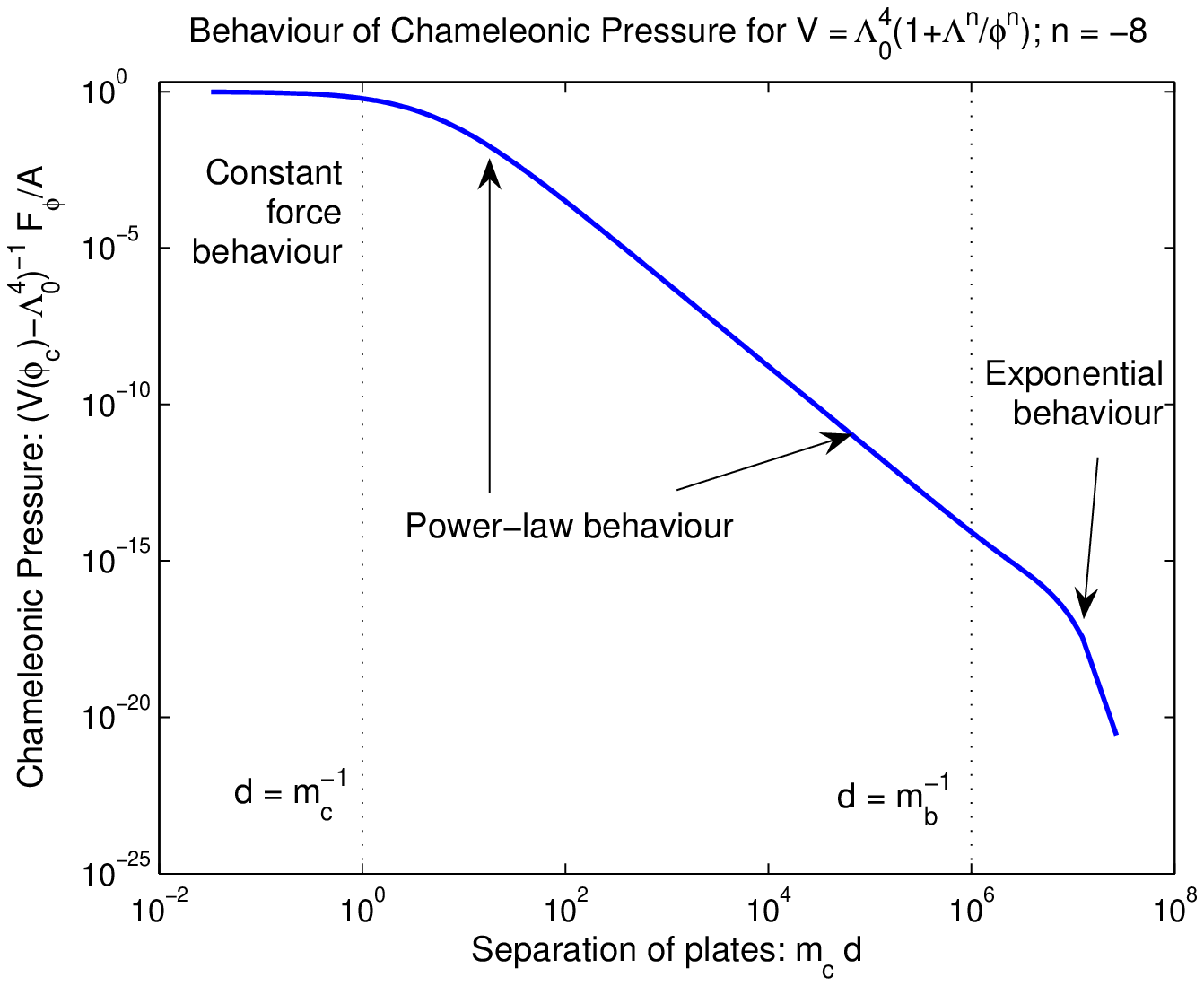}
\caption{The dependence of the chameleonic pressure, $F_{\phi}/A$, between two parallel plates on separation, $d$. We have taken $V(\phi)=\Lambda_{0}^4(1+\Lambda^{n}/\phi^n)$ and fixed $m_c/m_b= 10^{6}$.  The three plots show the behaviour of $F_{\phi}/A$ for a theory with $n = 1$, $n = 4$ and $n  = -8$.  Each of these are respectively representative of theories with $0 < n \leq 2$, $n > 2$ and $n \leq -4$.  Three types of behaviour are clearly visible in these plots.  For $d \lesssim m_{c}^{-1}$, $F_{\phi}/A \approx V_{c}-V_{b}$ which is independent of $d$: this is the `constant force behaviour'. For $m_{c}^{-1} \ll d \ll m_{b}^{-1}$, $F_{\phi}/A \propto 1/d^{p}$ for some $p$.  Theories with $0 < n \leq 2$ have $0 < p \leq 1$. If $n > 2$ then $1 < p < 2$ and if $n \leq -4$ we have $2 < p \leq -4$.  This is the `power-law behaviour'.  Finally when $d \ll m_{b}^{-1}$, $F_{\phi}/A \propto \exp(-m_{b}d)$, i.e. we have `exponential behaviour'.  Note that in a standard Yukawa scalar field theory (where $m_{\phi} = {\rm const}$) one would have $F_{\phi}/A \approx {\rm const}$ for $d \ll m_{\phi}^{-1}$ and an exponential drop-off for $d \gtrsim m_{\phi}^{-1}$; however there would be no region of power-law behaviour.}
\label{forceplotfig}
\end{center}
\end{figure}
In general, the approximate expressions for $\phi_0(d)$ derived above
should be seen as providing a good order of magnitude estimate for
$F_{\phi}/A$ rather than an accurate numerical prediction.  This said,
in some cases the expressions found above are actually exact.
Specifically, if $V(\phi) = \Lambda^4_0(1 + \Lambda^n/\phi^n)$ then
Eq. (\ref{kmideqn}) is exact (and $n_{eff}=n$) in the limit
$m_{b}^{-1} \ll d \ll m_c^{-1}$.  Additionally, if $V(\phi) =
\Lambda^4_0 \exp(g(\phi/\Lambda))$ and $\phi_{0}$ is such that
$g^{\prime \prime}(\phi_0\Lambda)/g^{\prime\,2}(\phi_0\Lambda) \ll 1$
then Eq. (\ref{kmideqn}) with $k^2=1$ provides an excellent
approximation when $m_{b}^{-1} \ll d \ll m_c^{-1}$.  When one only
wishes to consider a specific form of $V(\phi)$, numerically accurate
predictions for $F_{\phi}/A$ can be made by performing the integral in
Eq. (\ref{inteqn}) numerically.

We now consider two specific examples. Firstly, if $V(\phi)=
\Lambda^4_0(1 + \Lambda^n/\phi^n)$ and $m_{c}^{-1} \ll d \ll m_b^{-1}$
then:
\begin{equation}
\frac{F_{\phi}(d)}{A} = \Lambda^{4}_0 K_{n}
\left(\Lambda_d d\right)^{-\frac{2n}{n+2}},
\label{Vneqn}
\end{equation}
where $\Lambda_d =\Lambda_0^2/\Lambda$ and
\begin{equation}
K_n = \left(\sqrt{\frac{2}{n^2}}
B\left(\frac{1}{2},\frac{1}{2}+\frac{1}{n}\right)\right)^{\frac{2n}{n+2}}. \label{Kneqn}
\end{equation}
In all chameleon theories with power-law potentials, $F_{\phi}/A$ drops off as $1/d^{p}$ for some $p$ in this regime. In theories with $0 < n \leq 2$, $0 < p \leq 1$; if $n > 2$ $1 < p < 2$ and if $n \leq - 4$ we have $2 < p \leq 4$.

If $d \lesssim m_c^{-1}$ then $m_{0} \approx m_{c}$ and so:
$$
\frac{F_{\phi}(d)}{A} \approx V_{c} - V_{b} - V_{b}^{\prime}(\phi_c -\phi_b) \approx V_{c}-V_{b}.
$$
where the last approximation holds if $\rho_{c} \gg \rho_b$.  In this regime, the chameleonic force is independent of $d$ at leading order.
Finally if $d \gg m_{b}^{-1}$ we have from Eq. (\ref{Fmdlarge}):
$$
\frac{F_{\phi}}{A} \sim \frac{72n(n+1)V_{b} e^{-m_b
d}}{(n+2)^2}.
$$
In FIG. \ref{forceplotfig} we show the behaviour of the chameleonic pressure, $F_{\phi}/A$, in all three regimes ($d \lesssim m_{c}^{-1}$, $m_{c}^{-1}\ll d \ll m_{b}^{-1}$ and $d \gg m_{b}^{-1}$) for chameleon theories with $n = -8$, $n = 1$ and $n = 4$.  In all cases we have fixed $m_c/m_b = 10^{6}$.  The $n = -8$ plot is representative of theories with $n \leq -4$ and the $n =4$ is representative of theories with $n > 2$.  The $n =1$ graph show an example for a theory with $0 < n \leq 2$.  The three types of behaviour: constant force for $d \lesssim m_{c}^{-1}$, power-law for $m_{c}^{-1} \ll d \ll m_b^{-1}$ and exponential drop-off for $d \ll m_{b}^{-1}$ are clearly visible in these plots.  The main difference in the behaviour of the force for the different values of $n$ is the slope of $F_{\phi}/A$ in the power-law drop-off region.

If $V(\phi) = \Lambda^4_0 \exp(\Lambda^n/\phi^n)$ and  again $m_{b}^{-1} \ll d \ll m_c^{-1}$, then:
$$
n_{\rm eff} = \frac{n^2/(n+1) + 2(\phi_0/\Lambda)^n +
n(\phi_0/\Lambda)^{2n}}{(\phi_0/\Lambda)^{2n}}.
$$
For small $(\Lambda/\phi)^n$, we have $n_{\rm eff} = n$ and hence
$F_{\phi}/A$ is given by Eq. (\ref{Vneqn}); this limit corresponds to
$m_0 \ll \Lambda_d$ so $d \gg \Lambda d^{-1}$.  In
the opposite limit when $m_0 \gg \Lambda_d$ i.e. $d \ll
\Lambda_d^{-1}$, we have instead $n_{\rm eff}\approx n^2
(\Lambda/\phi)^{2n} /(n+1)$ and so:
$$
m_{0}^2 d^2 \approx 2\pi^2 \left[1+ \frac{1-4\ln(2)}{n_{\rm eff}}\right].
$$
It follows that:
\begin{equation}
\frac{F_{\phi}(d)}{A} \approx \frac{\Lambda^2 2\pi^2}{n^2 h(\Lambda_d d)^{\frac{2n +2}{n}} d^2} \left[1-\frac{n+1}{n h(\Lambda_d d)}\right]-\Lambda_0^4,
\end{equation}
where $h(\Lambda_d d)$ is a slowly varying function of $\Lambda_d d$ defined by:
\begin{equation}
h^{2+2/n} e^{h} = \frac{2\pi^2}{n^2 (\Lambda_0^2 d/\Lambda)^2}. \label{heqn}
\end{equation}
The above expression for $F_{\phi}/A$ is valid provided that
$(n+1)/n h(\Lambda_d d) \ll 1$. Note that, in all cases,
$F_{\phi}(d)/A \sim \Oo(\Lambda_0^4)$ when $d \approx
\Lambda/\Lambda_0^2$.  When $d \lesssim m_c^{-1}$, we have in all
cases that:
$$
\frac{F_{\phi}}{A} \approx V_{c}-V_b + V_b^{\prime}(\phi_b-\phi_c).
$$
When $d \ll m_{b}^{-1}$, Eq. (\ref{Fmdlarge}) gives the behaviour of $F_{\phi}/A$.

Note, for comparison,that the Casimir force per unit area between two parallel plates with separation $d$ at zero temperature is:
\begin{equation}
\frac{F_{\rm cas}(d)}{A} = \frac{\pi^2}{240 d^4}.
\end{equation}

\subsection{Sphere-Plate Geometry}\label{sec:force:sphere}
Calculating the chameleonic and Casimir forces is simplest for the
parallel plate geometry.  However, the most accurate measurements of
the Casimir force have been made using a sphere and a plate.  In this
geometry, the Casimir force depends only on the radius of curvature,
$R$, of the curved body, and the distance, $d \ll R$, between the
surfaces of the two bodies at the point of least separation.

We now calculate the chameleonic force between a sphere with radius
$R$ and a circular plate with total surface area $A$.  $d$ is defined to be
smallest separation between these bodies.  The $z$
direction is defined to be perpendicular to the plate, and we take $r$ to be a
radial coordinate which measures the distance from the point of least
separation in the plane of the plate.  Both the plate and the sphere
are assumed to have thin-shells.

In the region between the two bodies the chameleon field satisfies:
\begin{equation}
\frac{\dd \phi^2}{\dd z^2} + \frac{\dd^2 \phi}{\dd r^2} + \frac{1}{r}\frac{\dd \phi}{\dd r} = V^{\prime}(\phi) - V^{\prime}(\phi_b).
\end{equation}
At $r$, the separation between the sphere and the plate in the $z$-direction is
$$
s(r) \equiv  d + R\left(1-\left(1-\frac{r^2}{R^2}\right)^{\frac{1}{2}}\right).
$$
We define $\phi_{\rm PP}(z,s)$ to be the value of the chameleon field
in the parallel plate set-up for plates with separation $s$.  When $s
\ll R$, we can approximate $\phi(z,r)$ in the sphere-plate geometry by
$\phi_{\rm PP}(z,s(r))$.  This approximation is valid so long as:
$$
\left \vert \frac{\frac{\dd^2 \phi}{\dd r^2} + \frac{1}{r}\frac{\dd
\phi}{\dd r}}{\frac{\dd^2 \phi}{\dd z^2}}\right\vert \ll 1.
$$
Now in the parallel plate set-up we found that:
$$
\sqrt{2}\int_{0}^{y(\phi(z); \phi_0)} W(y^{\prime}; \phi_0)\dd y^{\prime}
= \frac{s}{2} - \frac{z}{2},
$$
where $z=0$ is the surface of the plate.  We then have:
$$
\frac{\dd \phi}{\dd s} = \left[\frac{W(y)}{W(0)} -
\frac{y}{W(0)}\int_0^{y} \frac{1}{y^{\prime}}\frac{\partial
W}{\partial y^{\prime}} \dd y^{\prime}\right]\frac{\dd \phi_0}{\dd s}
- \frac{y}{\sqrt{2}}.
$$
Therefore if $V^{\prime}(\phi_{s})/V^{\prime}_0 \gg 1$:
\begin{equation} \frac{\dd \phi_0}{d s} \approx
-W(0)\left[\sqrt{2}\int_{0}^{\infty} \frac{\partial W}{\partial
y^{\prime}} \frac{1}{y^{\prime}} \dd y^{\prime}\right]^{-1}.
\end{equation}
Inserting the equation for $s(r)$ we arrive at:
$$
\frac{\dd^2 \phi}{\dd z^2} = -1/W(y), \qquad \frac{\dd \phi}{\dd r} =
\frac{\sqrt{R^2-(R+d-s)^2}}{R+d-s(r)}\frac{\dd \phi}{\dd s}.
$$
It is clear then that the approximation $\phi \approx \phi_{\rm
PP}(r,s(r))$ is good provided that $d \leq s \ll R$ i.e. $r,\, d \ll
R$.  All sphere-plate Casimir measurements have $d \ll R$.

Whenever $\phi(z,r) \approx \phi_{\rm PP}(z,s(r))$ we have:
$$
\frac{\dd F_{\phi}}{\dd A} \approx V(\phi_0(s(r)))-V(\phi_{b}) - V^{\prime}(\phi_b)(\phi_0(s(r))-\phi_b),
$$
and for $r \ll R$, $\dd A = 2\pi r \,\dd r \approx 2\pi R \,\dd s$. The contribution to total force between the plate and the sphere from the points with $r\ll R$ is:
$$
F_{\phi}(r) \approx 2\pi R\int_{0}^{s(r)}  \dd s^{\prime} \left[V(\phi_0(s(r^{\prime})))-V(\phi_{b}) - V^{\prime}(\phi_b)(\phi_0(s(r^{\prime}))-\phi_b)\right].
$$
We define $r_{\rm max} = \min(\sqrt{A/\pi},R)$ i.e. $r_{\rm max}$ is the
smaller of $R$ and the radius of the circular plate.  We also define
$s_{\rm max} = s(r_{\rm max})$; $s_{\rm max}$ is then the largest separation
between the surfaces of the plate and the sphere. If $\dd F_{\phi}/\dd
A$ drops off faster than $1/s$ for all $s \gtrsim s^{\ast}$ for some
$s^{\ast} \ll R$, then the dominant contribution to the total force
between the sphere and the plate comes from the region where $s\ll
R$. To a very good approximation we therefore have:
\begin{equation}
F_{\phi}^{\rm tot} \approx 2\pi R\int_{d}^{s_{\rm max}} \dd
s^{\prime} \left[V(\phi_0(s(r^{\prime})))-V(\phi_{b}) -
V^{\prime}(\phi_b)(\phi_0(s(r^{\prime}))-\phi_b)\right]. \label{Fphitot}
\end{equation}
In some chameleon theories, however, $\dd F_{\phi}/\dd
A$ drops off more slowly than $1/s$ for all $s \lesssim \Oo(R) \ll
m_b^{-1}$.  In these cases, Eq. (\ref{Fphitot}) is no longer accurate.

In all theories $\dd F_{\phi}^{\rm tot}(d) / \dd d \approx 2\pi R \dd
F_{\phi}(d) / \dd A$ provided $d \ll R$.  By measuring the gradient of
$F_{\phi}^{\rm tot}(d)$, it is therefore possible to extract the form of
$\dd F_{\phi}(d)/\dd A$.  $\dd F_{\phi}(d)/\dd A$ is equal to the
force per unit area between two parallel plates with distance of
separation $d$.

It is clear that the dependence of $\dd F_{\phi}/\dd A$ on $s$ plays
an important role. In particular, theories where it drops off more
slower than $1/s$ behave differently from those where the drop off is
faster.  When $m_0 \gg m_b$ we have:
$$
\frac{\dd F_{\phi}}{\dd A} \approx V(\phi_0) - V(\phi_b).
$$
We define $Q(\phi_0; V) = \dd \ln (V(\phi_0)-V(\phi_b)) / \dd \ln
(1/s)$.  If $Q(\phi_0(s);V) > 1$ for all $s > s^{\ast}$ (and $< 1$
otherwise), where $s^{\ast} \ll \min(s_{\rm max}, 1/m_{b})$, then the
dominant contribution to the total force comes from points with
separations $\approx s^{\ast}$. In these cases Eq. (\ref{Fphitot})
provides a good approximation to $F_{\phi}^{\rm tot}$. If no such
$s^{\ast}$ exists but $m_b s_{\rm max} \gg 1$ then the dominant
contribution to $F_{\phi}^{\rm tot}$ comes from points with separations
$\approx 1/m_b$; in these cases we may also use Eq. (\ref{Fphitot}) to
calculate $F_{\phi}^{\rm tot}$.  If neither of these conditions hold, then
dominant contribution to $F_{\phi}^{\rm tot}$ comes from separations $\sim
\Oo(s_{\rm max})$. The assumption that $\phi \approx \phi_{\rm PP}(z,s(r))$
fails for $s \sim \Oo(s_{\rm max})$ and it is particularly bad if $s_{max}
\sim \Oo(R)$.  In these cases the chameleon field equations are too
complicated to solve analytically. However, Eq. (\ref{Fphitot}) is still expected to provide an order of magnitude
estimate for $F_{\phi}^{\rm tot}$. This is because the assumption that
$\phi \approx \phi_{\rm PP}(z,s(r))$ only breaks down for $s \sim
\Oo(s_{\rm max})$ but holds for all smaller values of $s$.  We therefore,
do not expect $\phi(z,s_{\rm max})$ to be very different from
$\phi_{\rm PP}(z,s_{\rm max})$ or $F_{\phi}^{\rm tot}$ to be very different from the
form given by Eq. (\ref{Fphitot}).

If $V = \Lambda^4_0(1 + \Lambda^n/\phi^n)$ then $\dd
F_{\phi}(s)/\dd A$ drops off more slowly than $1/s$ when $m_b s \ll 1$
for $-2 < n < 2$. Theories with $-2 \leq n \leq 0$ are not valid
chameleon theories.  Theories where $0 < n < 2$ therefore make
qualitatively different predictions for $F_{\phi}^{tot}$ than do those
where $n > 2$ or $n \leq -4$.

In the sphere-plate geometry (with $m_{c}^{-1} \ll d \ll R,\,m_b^{-1}$ and $m_b d \ll 1$), the total chameleonic force for theories with $n > 2$ or $n \leq -4$ is given to a very good approximation by:
\begin{equation}
F_{\phi}^{\rm tot}(d) \approx 2 \pi \Lambda_0^2 \Lambda R \left(\frac{n+2}{n-2}\right) K_{n} \left(\Lambda_d d\right)^{-\frac{n-2}{n+2}}, \label{Ftotnl}
\end{equation}
where $K_n$ is given by Eq. (\ref{Kneqn}) and, as above, $\Lambda_d = \Lambda_0^2/\Lambda$.

In theories with $0 < n \leq 2$, however, we have:
\begin{equation}
F_{\phi}^{\rm tot}(d) \approx F_{0}(s_{\rm max},m_{b}) - 2\pi \Lambda_0^2 \Lambda R \left(\frac{n+2}{2-n}\right) \left(\Lambda_d d\right)^{\frac{2-n}{n+2}}. \label{Ftotns}
\end{equation}
where $F_0$ is independent of $d$ and is calculated in Appendix \ref{appC}. If $m_b s_{\rm max} \ll 1$ then we are only able to find the order of magnitude of $F_0$:
\begin{equation}
F_0 \sim 2\pi \Lambda_0^2 \Lambda R\left(\frac{n+2}{2-n}\right) K_{n} \left(\Lambda_d s_{\rm max} \right)^{\frac{2-n}{n+2}}, \quad m_b s_{\rm max} \ll 1. \label{F01}
\end{equation}
If $m_b s_{\rm max} \gg 1$, however, we are able to calculate $F_0$:
\begin{equation}
F_0 = 2\pi \Lambda_0^2 \Lambda R\left(\frac{n+2}{2-n}\right) K_n D_n \left(\frac{a_n \Lambda_d}{m_b}\right)^{\frac{2-n}{n+2}}, \quad m_b s_{\rm max} \gg 1. \label{F02}
\end{equation}
where
\begin{eqnarray}
a_{n} &=& \sqrt{\frac{2(n+1)}{n}} B\left(\frac{1}{2},\frac{1}{2}+\frac{1}{n}\right), \\
D_{n} &=& \frac{4n(n+1)}{(n+4)(n+2)}\left(1+\frac{2-n}{3(n+2)}\beta_n\right),
\end{eqnarray}
and
$$
\beta_n = \frac{n+2}{2n^2}\left[2(n+1)\left(\Psi\left(\frac{1}{n}\right) - \Psi\left(\frac{1}{2}+\frac{1}{n}\right)+n\right)-n\right].
$$
$\Psi\left(\,\cdot\,\right)$ is the Digamma function.

If $V = \Lambda^4_0 \exp(\Lambda^n/\phi^n)$ then for $\Lambda_d d \ll 1$
so that $(n+1)h(\Lambda_d d)/n \ll 1$ where $h$ is given by Eq. (\ref{heqn}), we found that:
$$
\frac{\dd F_{\phi}(d)}{\dd A} = \frac{\Lambda^2 2\pi^2}{n^2 h(\Lambda_d d)^{\frac{2n +2}{n}} d^2} \left[1-\frac{n+1}{n h(\Lambda_d d)} + \Oo(1/f^2)\right],
$$
Since $2(n+1)/(n h) < 1$, $\dd F_{\phi}/\dd A$ drops off faster than $1/d$ for all $n$.  For $(n+1)/(n h(\Lambda_d d)) \ll 1$and $m_c^{-1}\ll d \ll m_b^{-1}$, the total force between a sphere and a plate is:
\begin{equation}
F_{\phi}^{\rm tot} = F_{1}(s_{\rm max}, m_b) + \frac{4\pi^3 \Lambda^2 R}{n^2 h(\Lambda_d d)^{\frac{2n +2}{n}} d}\left[1+3\frac{n+1}{n h(\Lambda_d d)} +\Oo(1/h^2)\right].
\end{equation}
If $\Lambda_d s_{\rm max} \ll 1$ or $m_b/\Lambda_d \gg 1$ then the
$F_{1}$ term is negligible relative to the $d$-dependant term.
If however $\Lambda_d s_{\rm max} \gg 1$ and $m_b/\Lambda_d \ll 1$,
$F_1 = F_0(s_{\rm max},m_b)$ as given by Eqs. (\ref{F01} and \ref{F02}).

For comparison, note that, the total Casimir force between a sphere and a plate is
\begin{equation} F_{\rm tot} = \frac{\pi^3 R}{360 d^3}.
\end{equation}

\section{Predictions and Constraints}\label{sec:predict}
We now use the results derived above to make specific predictions and
derive constraints on theories with either $V(\phi) = \Lambda^4_0(1 +
\Lambda^n/\phi^n)$ or $V(\phi) = \Lambda_0^4\exp(\Lambda^n/\phi^n)$;
$\Lambda_0 = 2.4\times 10^{-3}\eV$.  The simplest and most natural
scenario is $\Lambda \approx \Lambda_0$. With these potentials, the
energy density of the
chameleon field can be identified with dark energy cosmologically. If $\Lambda =
2.4\times 10^{-3}\eV$ then
$$\Lambda d = d / \left(82.2\mum\right), \quad \Lambda^4 = 6.92 \times 10^{-3}\, \mu{\rm dyne}\,\cm^{-2} = 6.92 \times 10^{-7}\,{\rm mPa}.$$

For either of the potentials given above one must require $n > 0$ or $n \leq -4$
for a valid chameleon theory to emerge \cite{chamstrong}.
When $m_c^{-1} \ll d \ll m_b^{-1}$, Eq. (\ref{kmideqn}) is actually exact
for theories with power-law potentials.  In these power-law theories, the force per unit
area between two parallel plates when $m_c^{-1} \ll d \ll m_b^{-1}$ is given by
Eq. (\ref{Vneqn}).  When $m_b d \gtrsim 5$, $\dd F_{\phi}/\dd A$ is
given by Eq. (\ref{mnsmall}).  The total chameleon force in the
sphere-plate geometry is (when $m_b d \ll 1$) given by Eq.
(\ref{Ftotnl}) if $n > 2$ or $n \leq -4$ or by Eq. (\ref{Ftotns}) if
$0 < n \leq 2$.

We begin by considering how Casimir force measurements presently
constrain chameleon theories, and then discuss the prospects for the
detection of chameleon fields by the next generation of such tests
in the next section.

The first attempt to measure the Casimir force between two parallel
plates was made in 1958 by Sparnaay \cite{sparnaay}. The data he found
contained large systematic errors, due mostly to the determination of
$d$, and so was only said to "not contradict Casimir's theoretical
prediction".  At the largest separations probed ($\sim 2\mum$) this
experiment was sensitive to pressures between the two plates of
$0.1\,{\rm mPa}$ (=$1\,{\rm mdyne}\,{\rm cm}^{-2}$), however the
inaccuracy in determining $d$ was generally $\pm 0.12 \mum$. This
measurement was conducted in a vacuum with pressure $10^{-2}\,{\rm torr}$.

The Casimir force between two parallel plates was successfully
measured by Bressi et. al. \cite{Bressi}.  They measured the Casimir
force between $0.5-3.0\mum$ to an average precision of $15\%$.  This
corresponds to a sensitivity of approximately $1\,{\rm mPa}$.  A
vacuum pressure with $10^{-3}\,{\rm torr}$ was used in this
experiment.

\begin{figure}[tbh]
\begin{center}
\includegraphics[width=8.8cm]{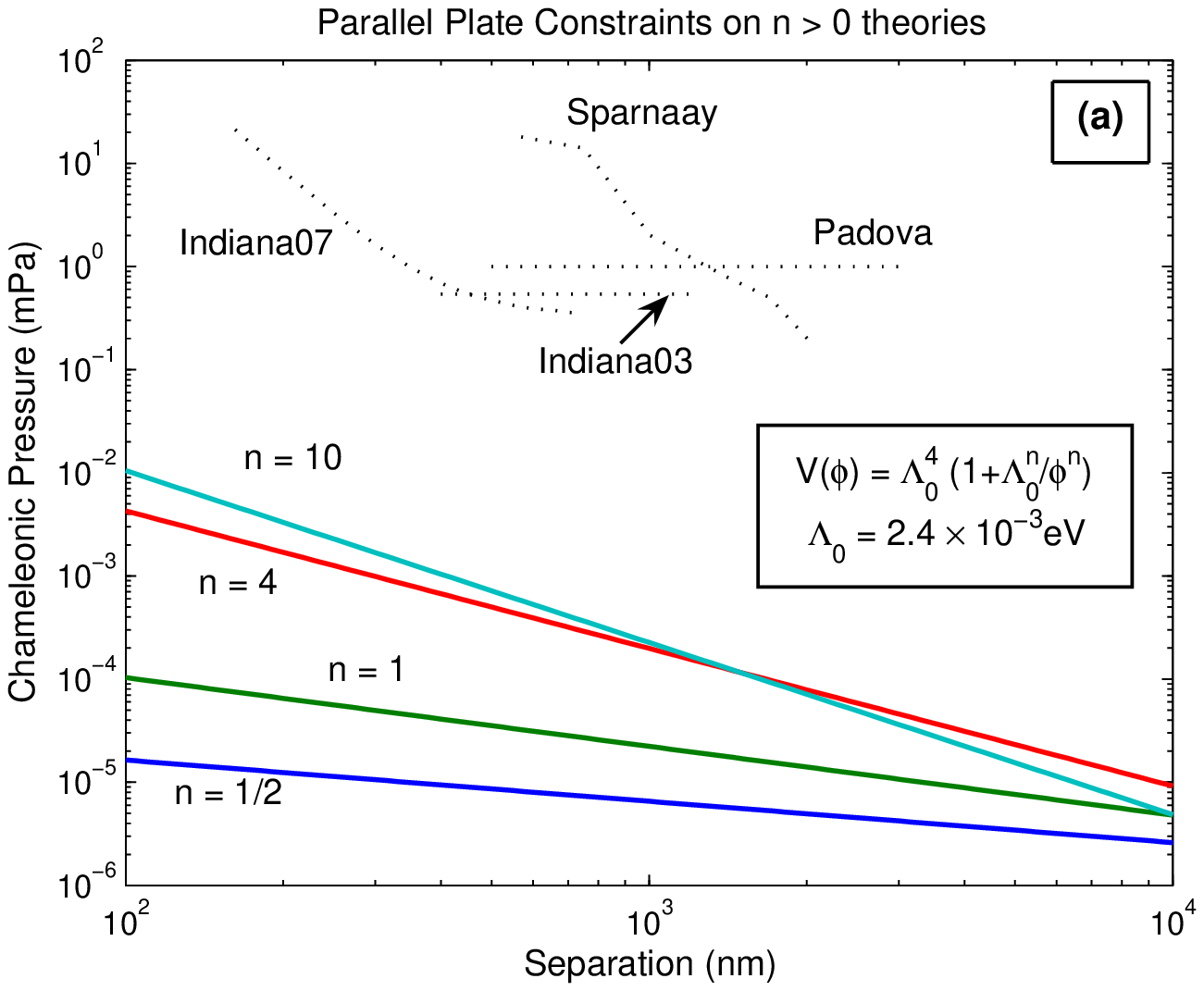}
\includegraphics [width=8.8cm]{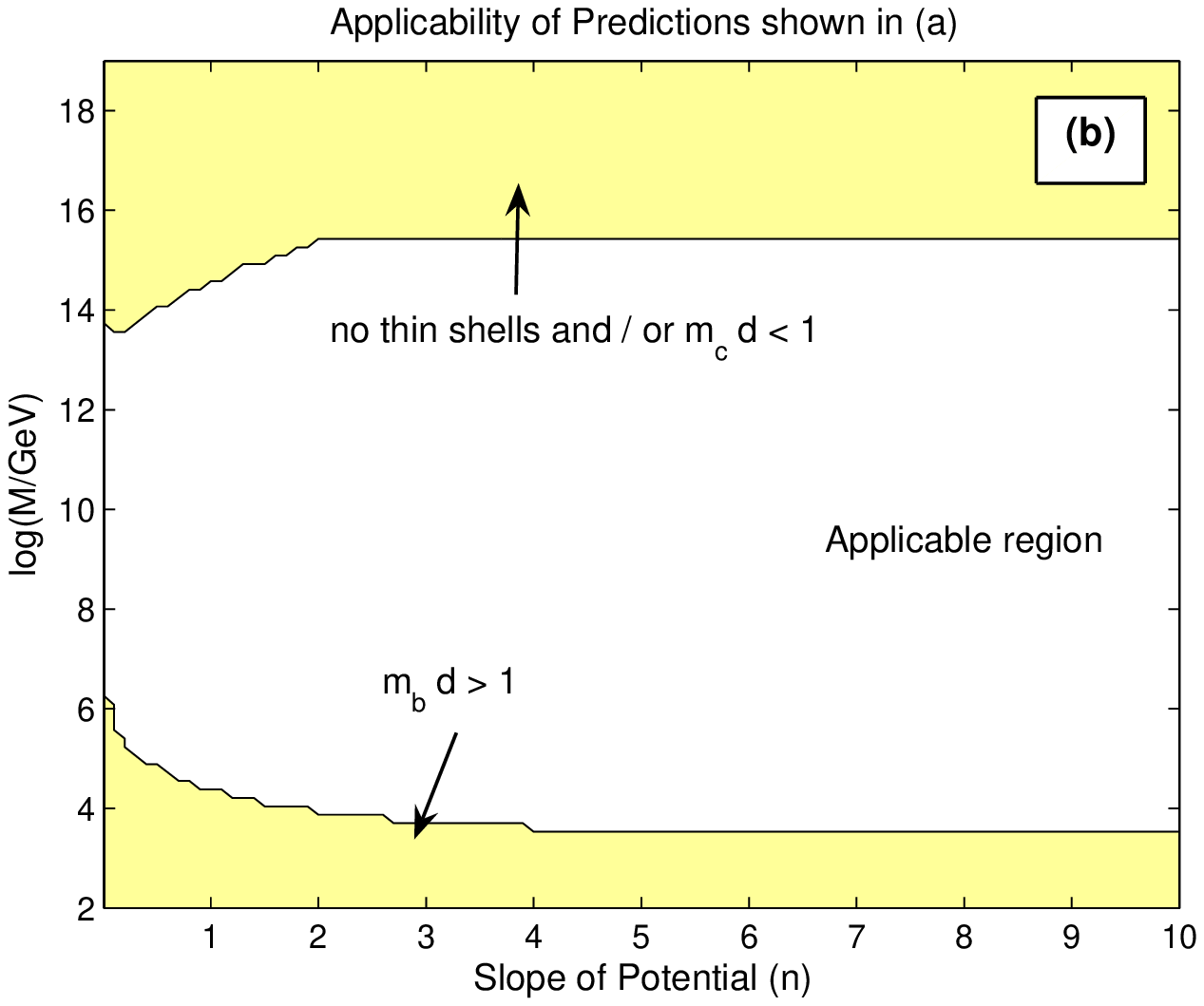}
\caption{The solid lines in Figure (a) show the predicted chameleonic pressure between two parallel plates for $V = \Lambda^4_0\left(1 + \Lambda^{n}/\phi^n\right)$, $n>0$ and $\Lambda = \Lambda_0 =  2.4 \times 10^{-3} \eV$.  The dotted lines show the current experimental constraints on any such pressure. Sparnaay, Padova, Indiana03 and Indiana07 refer to Refs. \cite{sparnaay}, \cite{Bressi}, \cite{Decca1} and \cite{Decca3} respectively. The predictions shown in Figure (a) only apply when the test masses have thin-shells and for $m_c^{-1}\ll d \ll m_b^{-1}$; $m_c$ is the chameleon mass inside the test masses and $m_b$ is the chameleon mass in the background. The white region in Figure (b) shows the values of the chameleon to matter coupling, $M$, for which the predictions shown in Figure (a) are applicable to the most recent experiment conducted by Decca \emph{et al.}, labeled Indiana07 in Figure (a).  }
\label{fig1}
\end{center}
\end{figure}

The most accurate measurements of the Casimir force between two
parallel plates have, however, been made by measuring the gradient of
the force between a sphere and a plate.  Dynamical measurements of the
force between a sphere and a plate would detect not
$F_{\phi}^{\rm tot}(d)$ but $\dd F_{\phi}^{\rm tot}/\dd d$ and hence, by
Eq. (\ref{Fphitot}), $\dd F_{\phi}/ \dd A$.

We define $P_{\phi} = \dd F_{\phi}/\dd A$ to be the chameleonic
pressure between two parallel plates.  The Casimir pressure between
two such plates is similarly defined to be $P_{c} = \pi^2 / 240 d^4$.
Thermal corrections to the Casimir force \cite{thermal} are
sub-leading order at the separations that have been probed thus far,
and so we do not consider them at this point.  To date, the most
accurate measurements of $P_c$ over separations $d \sim
0.16\mum-1.2\mum$ have been made by Decca \emph{et al.} in a series of
three experiments taking place between 2003 and 2007
\cite{Decca1,Decca2,Decca3}. We define $\bar{P}$ to be the total
measured pressure between two parallel plates. Using their most recent
experiment, described in Ref. \cite{Decca3}, Decca \emph{et al.} found
the following $95\%$ confidence intervals on $\Delta P = \bar{P}-P_c$:
at $d = 162{\rm nm}$, $\vert \Delta P \vert < 21.2{\rm mPa}$, at $d =
400{\rm nm}$, $\vert \Delta P \vert < 0.69{\rm mPa}$ and at $d =
746{\rm nm}$, $\vert \Delta P \vert < 0.35{\rm mPa}$.  In the first
experiment \cite{Decca1}, measurements were also made for larger
separations. For $450 {\rm nm} \leq z < 1200{\rm nm}$, they found
$\vert \Delta P \vert < 0.54{\rm mPa}$. At $d=162{\rm nm}$, $400{\rm
nm}$ and $746{\rm nm}$ the results of Decca {\emph et al.} represent a
detection of the Casimir force to an accuracy of $0.19\%$, $0.9\%$ and
$9.0\%$ respectively.  A vacuum pressure of $10^{-4}\,{\rm torr}$ was
used in making all of these measurements.

\begin{figure}[tbh]
\begin{center}
\includegraphics[width=8.8cm]{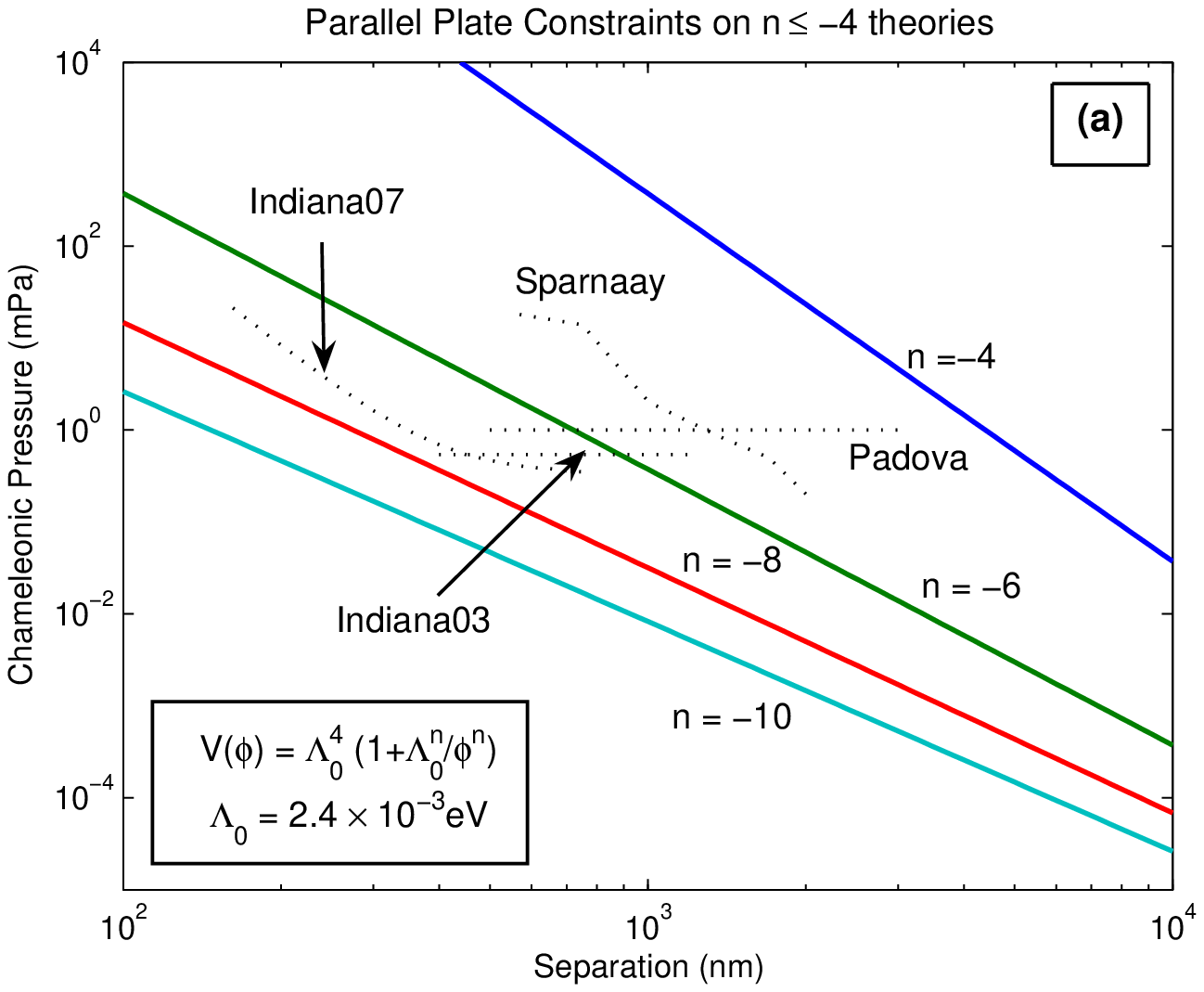}
\includegraphics [width=8.8cm]{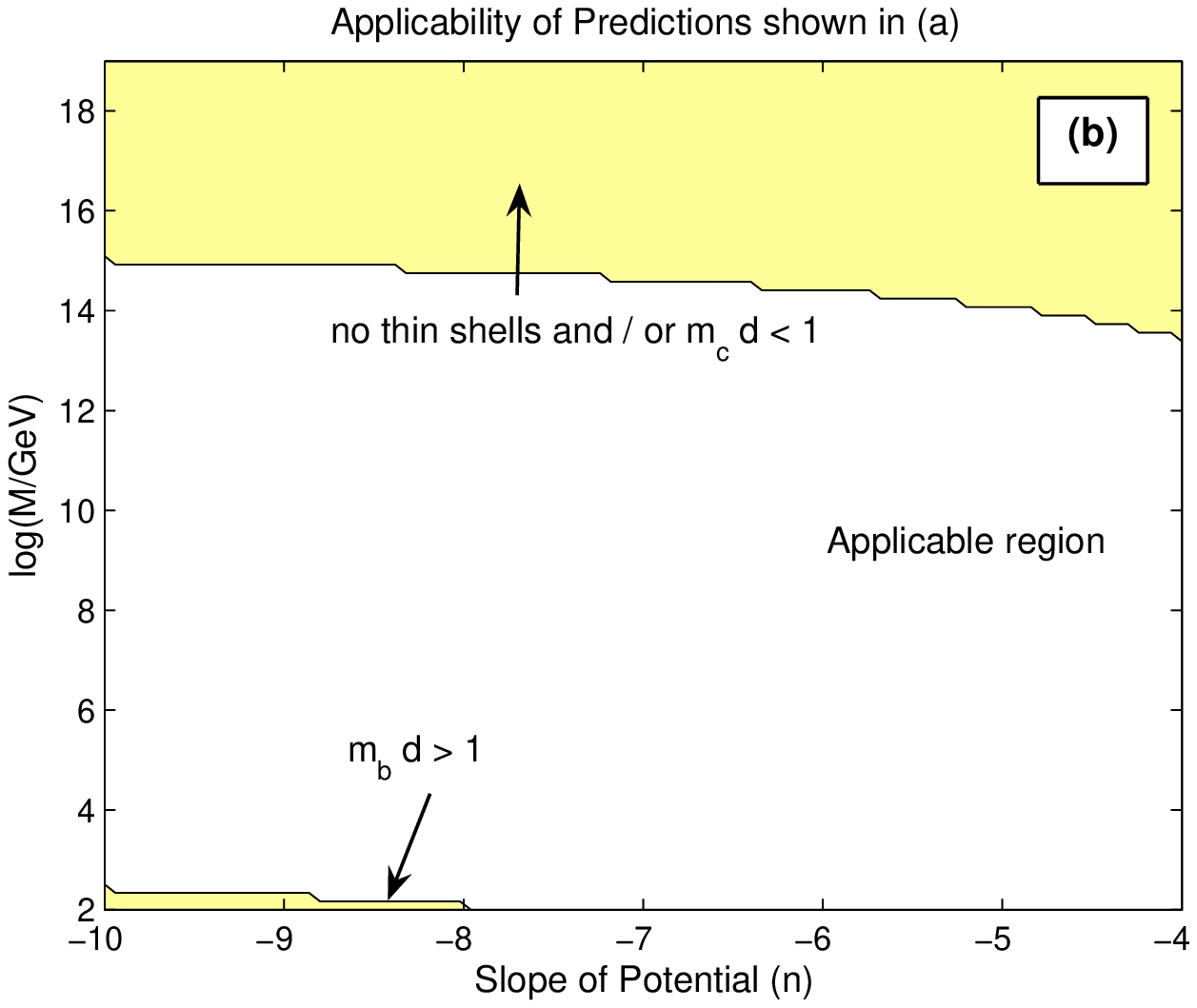}
\caption{The solid lines in Figure (a) show the predicted chameleonic pressure between two parallel plates for $V = \Lambda^4 + \Lambda^{4+n}/\phi^n$, $n\leq -4$ and $\Lambda = \Lambda_0 = 2.4 \times 10^{-3} \eV$.  The dotted lines show the current experimental constraints on any such pressure. Sparnaay, Padova, Indiana03 and Indiana07 refer to Refs. \cite{sparnaay}, \cite{Bressi}, \cite{Decca1} and \cite{Decca3} respectively. The predictions shown in Figure (a) only apply when the test masses have thin-shells and for $m_c^{-1}\ll d \ll m_b^{-1}$; $m_c$ is the chameleon mass inside the test masses and $m_b$ is the chameleon mass in the background. The white region in Figure (b) shows the values of the chameleon to matter coupling, $M$, for which the predictions shown in Figure (a) are applicable to the most recent experiment conducted by Decca \emph{et al.}, labeled Indiana07 in Figure (a).  }
\label{fig2}
\end{center}
\end{figure}

In FIGs \ref{fig1}a and \ref{fig2}a we plot $P_{\phi}$ vs. $d$ for
$m_c^{-1} \ll d \ll m_b^{-1}$ as solid lines for representative values
of $n$: $n=1/2, 1, 4, 10$ in the former plot, and $n= -4, -6, -8, -10$
in the latter for $V(\phi) = \Lambda_0^4(1+\Lambda^n/\phi^n)$. In all
these plots we have taken $\Lambda = \Lambda_0 = 2.4 \times
10^{-3}\eV$.  The dotted lines show the experimental limits on $\vert
\Delta P(z) \vert$ and the labels Sparnaay, Padova, Indiana03 and
Indiana07 refer to Refs. \cite{sparnaay}, \cite{Bressi}, \cite{Decca1}
and \cite{Decca3} respectively.  For Padova we have taken the upper
bound on $\vert \Delta P \vert$ to be $1\,{\rm mPa}$.

It is very clear from FIG. \ref{fig1}a that the magnitude of the
chameleonic pressure, $P_{\phi}$, predicted by theories with $n > 0$ and
$\Lambda = \Lambda_0 = 2.4 \times 10^{-3}\eV$ currently lies well below the
experimental limits. The predicted $P_c(d)$ is everywhere at least $2$
orders of magnitude smaller than the current experiment bounds.  The
story is very different for $n \leq -4$ theories with the same value of $\Lambda$. FIG. \ref{fig2}a clearly shows that the $n
=-4$ and $-6$ theories are strongly ruled out by the latest
$95\%$ confidence limits found by Decca et al. \cite{Decca3} (labeled
Indiana07 on the plot).  Indeed the $n=-4$ theories is even
ruled out by the 1958 measurements made by Sparnaay \cite{sparnaay}.
The value of $P_{\phi}$ predicted by the $n= -8$ theory is close to edge of what is currently allowed. For $n > 0$,
the larger $n$ is, the steeper the drop-off in $P_{\phi}(d)$ with $d$ and,
as a result, the larger $P_{\phi}(d)$ is at separations $<\Lambda^{-1} \approx 82\mum$. Very shallow
potentials ($0 < n < 2$) predict the smallest $P_{\phi}(d)$.  This is
disappointing, as the shallower the potential is, the larger the
difference ($\Delta \phi$) between the value of $\phi$ here on Earth
and in the cosmological background.  Both variations in the traditional `constants' of Nature
and the magnitude of the chameleon force between distant thin-shelled
bodies grow with $\Delta \phi$ \cite{chamstrong}. The larger $\Delta \phi$ is, then, the
more scope there is for the presence of a chameleon field to produce
non-negligible and potentially detectable alterations to the standard
cosmological model.

The values of $P_{\phi}$ plotted in FIGs. \ref{fig1}a and
\ref{fig2}a are accurate provided $m_c^{-1} \ll d \ll m_b^{-1}$
and the test masses have thin-shells. It is clear that the
strongest constraint comes from the 2007 experiment of Decca
\emph{et al.}.  For these constraints to actually be comparable
with the plotted values of $P_c$, it must therefore be the case
that $m_c^{-1} \ll d \ll m_b^{-1}$ for $d \sim 0.2-0.8\mum$. We
must also require that the test bodies used in this experiment
have thin-shells. If $M$ is too small ($\beta$ too large) then
$m_b d \gg 1$ for the above range of separations.  The chameleonic
force between the plates would then be exponentially suppressed
(by a factor $\approx \exp(-m_b d)$) and as such would be
negligible.  If $M$ is too large ($\beta$ too small) then the test
bodies will either lose their thin-shells or $m_c d \gtrsim
\Oo(1)$. In the absence of a thin-shell, the chameleon force has a
Yukawa form with mass $m_b$ and we would certainly have $m_b d \ll
1$. Casimir force experiments such as those conducted by Decca
\emph{et al.} are, however, only sensitive to Yukawa forces for
which $m_b d \sim \Oo(1)$.  If the test-masses do not have
thin-shells then, Casimir force experiments cannot be used to
constrain chameleon theories. If $m_c d \gtrsim \Oo(1)$, then
$P_{\phi} \approx {\rm const}$ and so, once again, Casimir force
experiments would be unable to constrain it. FIGs. \ref{fig1}b and
\ref{fig2}b shows the values of $M = M_{\rm Pl}/\beta$ for which
the values of $P_{\phi}$ shown in FIGs. \ref{fig1}a and
\ref{fig2}a can be compared with the Indiana07 constraints.

\begin{figure}[tbh]
\begin{center}
\includegraphics[width=8.8cm]{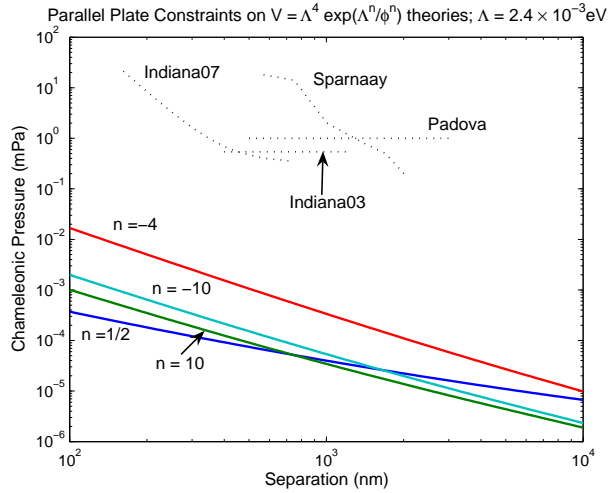}
\caption{The solid lines show the predicted chameleonic pressure between two parallel plates for $V = \Lambda^4_0 \exp \Lambda^n/\phi^n$ and $\Lambda = \Lambda_0 = 2.4 \times 10^{-3} \eV$.  The dotted lines show the current experimental constraints on any such pressure. Sparnaay, Padova, Indiana03 and Indiana07 refer to Refs. \cite{sparnaay}, \cite{Bressi}, \cite{Decca1} and \cite{Decca3} respectively. The predictions shown above only apply when the test masses have thin-shells, $m_c^{-1}\ll d \ll m_b^{-1}$; $m_c$ is the chameleon mass inside the test masses and $m_b$ is the chameleon mass in the background.}
\label{figexp}
\end{center}
\end{figure}

The predictions and constraints found above apply to theories with $V(\phi) \approx \Lambda_0^4(1+ \Lambda^{n}/\phi^n)$.  If $V
= \Lambda^4_0 g(\Lambda^n/\phi^n)$, where $g(y)$ has a Taylor expansion
about $y=0$ and $g^{\prime}(0) = 1$, then the above predictions would
apply for $(\Lambda/\phi)^n \ll 1$ i.e. $m_{\phi} \ll \Lambda$.  We
expect $\Lambda = \Lambda_0 \approx 2.4\times 10^{-3}\eV$ so that the chameleon
field is responsible for the late time acceleration of the
universe. The above predictions would then only apply if $d \gg
\Lambda^{-1} \approx 82\mum$, which is not the case.  Generally
speaking, if the potential is very steep, $P_{\phi} = F_{\phi}/A
\propto 1/d^2$ which is a stronger $d$ dependence than that exhibited
by theories with $V = \Lambda_0^4(1+ \Lambda^{n}/\phi^n)$ and $n > 0$, but
a weaker $d$-dependence than that predicted by power-law theories with $n
\leq -4$.

For concreteness, we consider the parallel plate predictions
and constraints for a theory with $V = \Lambda^4_0 \exp
\Lambda^n/\phi^n$ and again $\Lambda = \Lambda_0 = 2.4 \times 10^{-3}\eV$.  For $d \ll
82\mum$ we find if $n > 0$ the chameleonic pressure,
$P_{\phi}(d)$, predicted by such theories is larger than that
predicted by theories with $V = \Lambda_0^4(1+ \Lambda^{n}/\phi^n)$; if
$n \leq -4$, the opposite is true.  FIG. \ref{figexp} shows the
predictions for $P_{\phi}$ made by a theory with $V = \Lambda^4_0 \exp
\Lambda^n /\phi^n$.  It is clear from these plots that currently no
such models are ruled out, although the predicted chameleonic
pressure for $n =1/2$ is an order of magnitude larger than it would if
$V(\phi)$ were exactly $\Lambda_0^4(1+ \Lambda^{n}/\phi^n)$.

The experiments performed by Decca \emph{et al.} employed the
sphere-plate geometry but measured the Casimir force dynamically.
Dynamical experiments such as these directly measure not forces but
the rate of change of forces with separation i.e. $\dd F / \dd d$.
Since $\dd F / \dd d \propto \dd F / \dd A$, what is actually measured
is equivalent to the force per unit area between two parallel plates.
Other experiments that use the sphere-plate geometry have made static,
rather than dynamical, measurements of the force between the two
bodies.  Static measurements allow one to place limits on the force
itself rather than its gradient.  It should be noted that when such
measurements are performed, it is necessary to calibrate the
experiment so as to eliminate any electro-static forces. This has the
effect that at some large separation, $d_{\rm cal}$ say, the force between
the sphere and the plate is defined to be zero. If one has an
expression for the force $F(d)$ then, what would actually be measured by these
experiments is $\Delta F(d)= F(d)-F(d_{\rm cal})$. These experiments are
therefore insensitive to forces that are virtually constant for
$d < d_{cal}$.

In 1997, Lamoreaux measured the force between a spherical lens with
radius $(12.5 \pm 0.3)\cm$ and a $2.54\cm$ diameter, $0.5\cm$ thick
optical flat \cite{lam97}.  Measurements of the Casimir force where
made for separations in the $0.6$ to $6\mum$ range and the Casimir
force was measured to an overall accuracy of $15\%$.  The calibration
of the system was performed at a separation of about $10\mum$.  At the
largest separations ($d \gtrsim 1\mum$), the results of this
experiment place an upper bound on the magnitude of any residual
force, which includes any thermal corrections to the Casimir force, of
about $30$~pN.  The experiment was conducted in a vacuum with pressure
$10^{-4}\,{\rm torr}$.

\begin{figure}[tbh]
\begin{center}
\includegraphics[width=8.8cm]{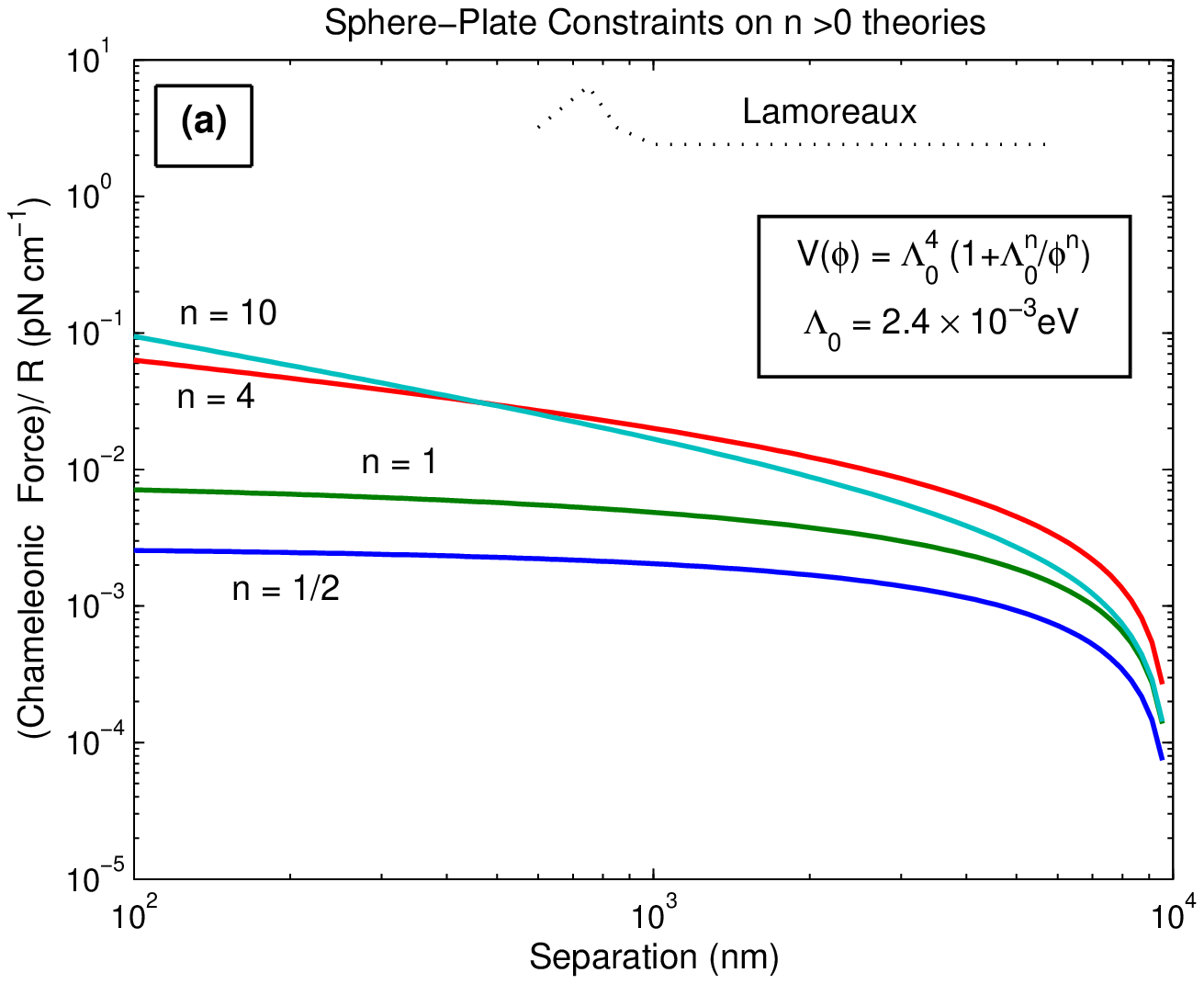}
\includegraphics [width=8.8cm]{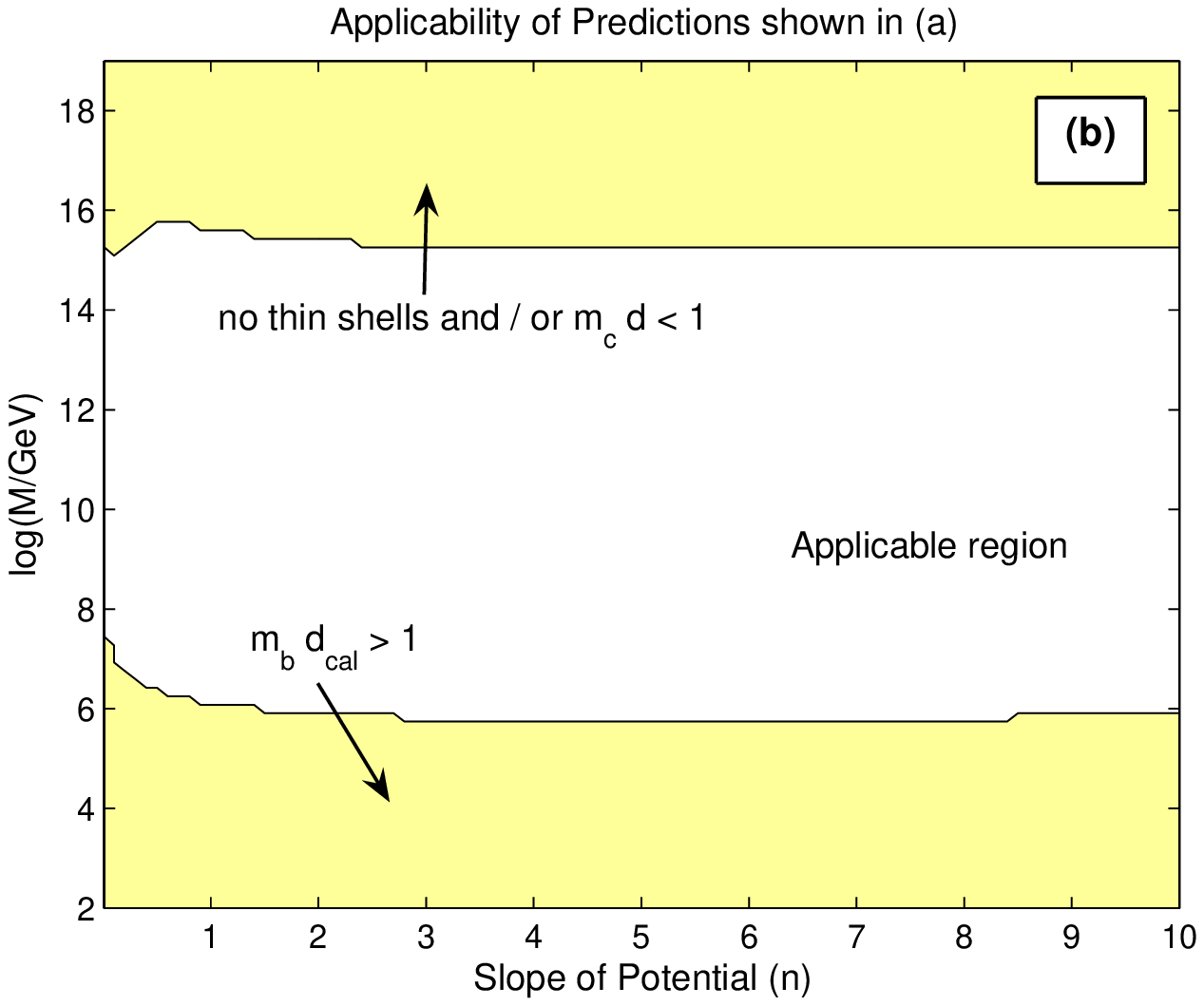}
\caption{The solid lines in Figure (a) shows $(F_{\phi}^{\rm tot}(d) -F_{\phi}^{\rm tot}(d_{cal}))/R$ where $F_{\phi}^{\rm tot}$ is the predicted chameleonic force between a sphere and a plate, for $V = \Lambda^4_0(1 + \Lambda^{n}/\phi^n)$, $n > 0$ and $\Lambda = \Lambda_0 = 2.4 \times 10^{-3} \eV$.  $R$ is the radius of the sphere and we have taken $d_{cal} = 10\mum$. The dotted line shows the current best experimental constraint on any such force pressure, which comes from Ref. \cite{lam97}. The predictions shown in Figure (a) only apply when the test masses have thin-shells, $m_c d \gg 1$ and $m_b d_{cal}\ll 1$; $m_c$ is the chameleon mass inside the test masses and $m_b$ is the chameleon mass in the background.  The white region in Figure (b) shows the values of the chameleon to matter coupling, $M$, for which the predictions shown in Figure (a) are applicable to the 1997 Casimir force measurement performed by Lamoreaux \cite{lam97}. }
\label{fig3}
\end{center}
\end{figure}

A similar measurement was made in 1998 by Mohideen \& Roy
\cite{MohRoy}. In this experiment a relatively small polystyrene
sphere was used with diameter $196\mum$, and the system was calibrated
at a separation of about $900$~nm. Mohideen \& Roy were able to
measure the Casimir force to precision of $1\%$ at the smallest
separation of about $100$~nm. They found an RMS derivation between
experiment and theory of $1.4$~pN \cite{MohRoy}. The error bars on
the measurements at individual separations were however larger, being
about $\pm 7$~pN.  This experiment was performed in a vacuum with
pressure $50{\rm mTorr}$.

In additional to dynamical force measurements, in their 2003
experiment Decca \emph{et al.} also made a static measurement of the
Casimir force \cite{Decca1}.  The experiment was calibrated at a
separation of $3\mum$.  The direct force measurements limit $\vert
\Delta F \vert \lesssim 0.5$~pN for $400\,{\rm nm} \lesssim d \lesssim
1200\,{\rm nm}$. In this experiment the sphere had radius $(296 \pm
2)\mum$ and the vacuum pressure was $10^{-4}\,{\rm torr}$.

We found in Section \ref{sec:force:sphere} that when the sphere and
the plate have thin-shells and $m_c^{-1} d \ll m_b^{-1}$ then if
$d_{\rm cal} \ll m_b^{-1}$:
$$
\Delta F_{\phi}^{\rm tot} \approx 2\pi R \Lambda^2_0 \Lambda K_{n}
\left(\frac{n+2}{n-2}\right) \left[\left(\Lambda_d
d\right)^{-\frac{n-2}{n+2}}-\left(\Lambda_d d_{\rm
cal}\right)^{-\frac{n-2}{n+2}}\right],
$$
where $\Lambda_d = \Lambda_0^2/\Lambda$. If, however, $d_{\rm cal}
\gtrsim m_b^{-1}$ then
$$
\Delta F_{\phi}^{\rm tot} \approx 2\pi R \Lambda^2_0 \Lambda K_{n}
\left(\frac{n+2}{n-2}\right) \left[\left(\Lambda_d
d\right)^{-\frac{n-2}{n+2}}-D_n\left(\frac{a_n
\Lambda_d}{m_b}\right)^{-\frac{n-2}{n+2}}\right].
$$
It should be noted the terms that for $m_{b}^{-1},\,d_{\rm cal} \gg d$, the terms that depend on $m_b$ and $d_{\rm cal}$ are only important for $0 < n < 2$ theories.  For all $n$,
the larger $R$ is (for fixed $d$), the larger $\Delta
F_{\phi}^{\rm tot}(d)$.  $R$ is largest (by several orders of magnitude)
in the 1997 Lamoreaux experiment \cite{lam97}.  The relatively large
dimensions ($\sim \Oo(cm)$) of the test masses used in this experiment (see
Ref. \cite{lam97}) ensure that they have thin-shells for a larger
range of $M$ than do the test masses used in Refs. \cite{MohRoy} and
\cite{Decca1}.  Currently then, the best constraints on chameleon
theories from static measurements of the Casimir force using the
sphere-plate geometry are provided by the Lamoreaux experiment.

\begin{figure}[tbh]
\begin{center}
\includegraphics[width=8.8cm]{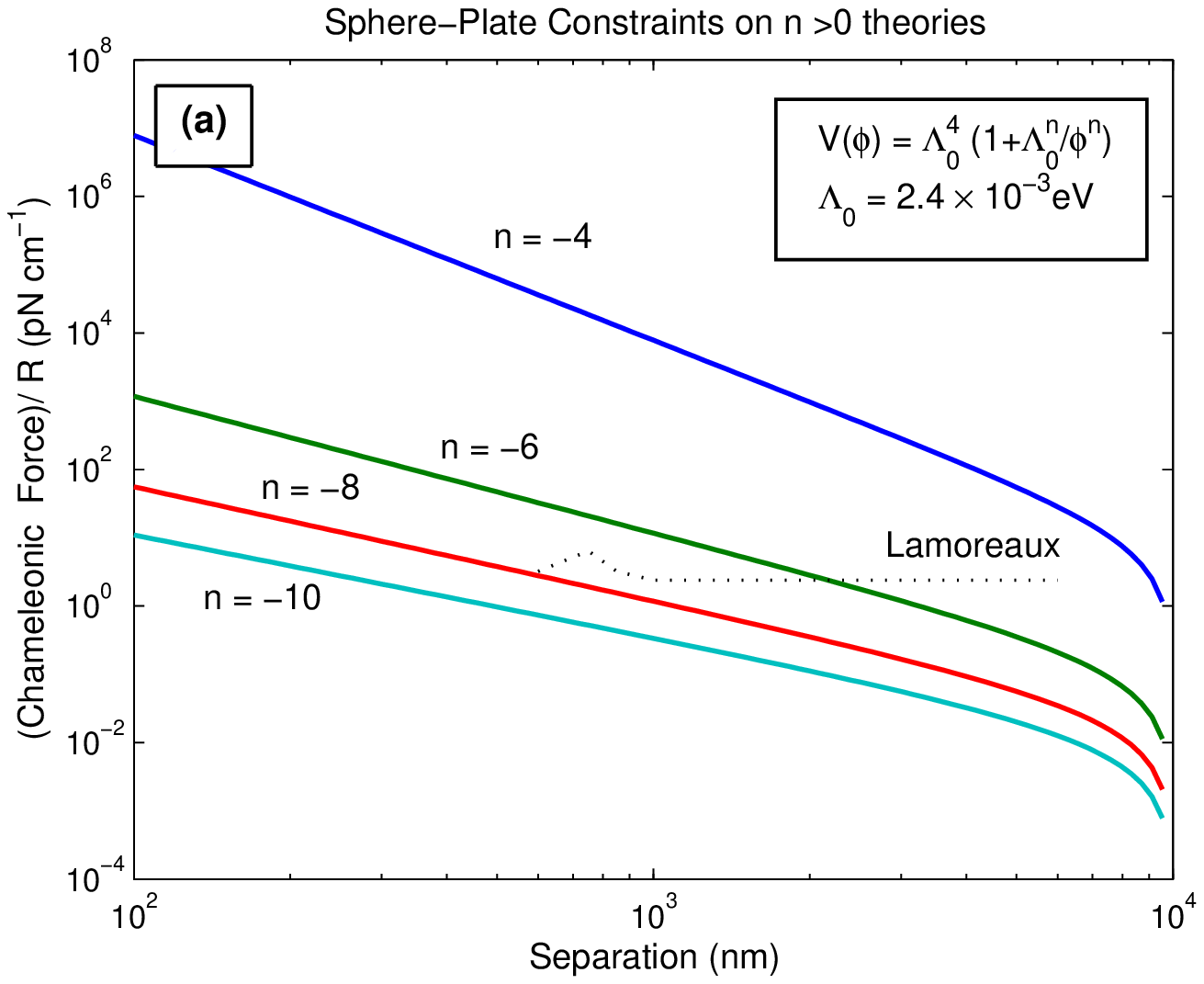}
\includegraphics [width=8.8cm]{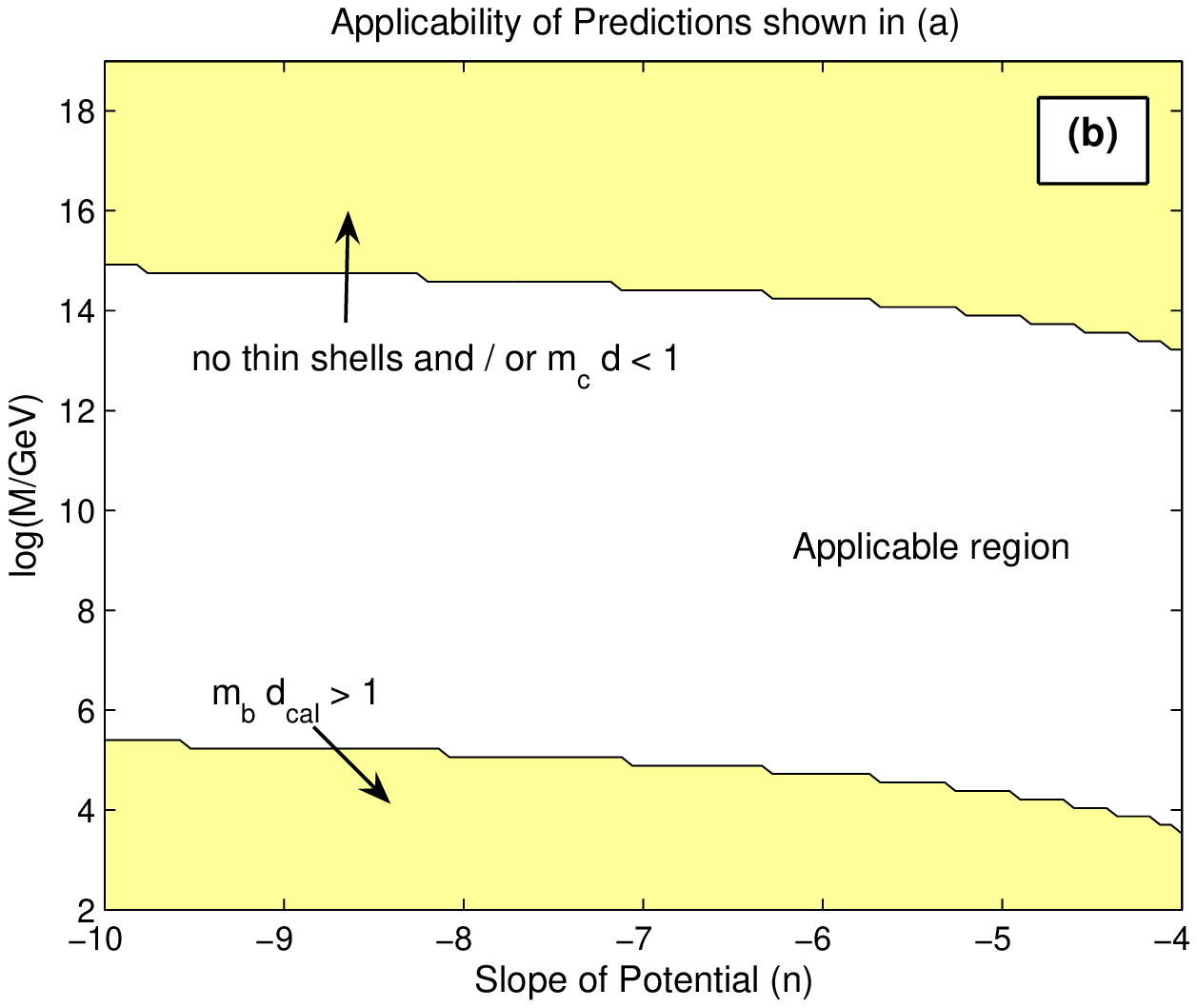}
\caption{The solid lines in Figure (a) shows $(F_{\phi}^{\rm tot}(d) -F_{\phi}^{\rm tot}(d_{\rm cal}))/R$ where $F_{\phi}^{\rm tot}$ is the predicted chameleonic force between a sphere and a plate, for $V = \Lambda^4_0(1 + \Lambda^{n}/\phi^n)$, $n \leq -4$ and $\Lambda = \Lambda_0 = 2.4 \times 10^{-3} \eV$.  $R$ is the radius of the sphere and we have taken $d_{cal} = 10\mum$. The dotted line shows the current best experimental constraint on any such force pressure, which comes from Ref. \cite{lam97}. The predictions shown in Figure (a) only apply when the test masses have thin-shells, $m_c d \gg 1$ and $m_b d_{cal} \ll 1$; $m_c$ is the chameleon mass inside the test masses and $m_b$ is the chameleon mass in the background.  The white region in Figure (b) shows the values of the chameleon to matter coupling, $M$, for which the predictions shown in Figure (a) are applicable to the 1997 Casimir force measurement performed by Lamoreaux \cite{lam97}. }
\label{fig4}
\end{center}
\end{figure}

In FIGs. \ref{fig3}a and \ref{fig4}a we plot, as solid lines, the
predicted values of $\Delta F_{\phi}^{\rm tot}(d)/R$ with $d_{cal} =
10\mum$, such as it is in Lamoreaux experiment \cite{lam97}. In making
these predictions, we have taken
$V(\phi)=\Lambda_0^4(1+\Lambda^n/\phi^n)$ and $\Lambda = \Lambda_0 =
2.4 \times 10^{-3}\eV$. These predictions are accurate provided
$m_c^{-1} < d < d_{cal} \ll m_b^{-1}$ and the test masses have
thin-shells. If either of these conditions did not hold, then $\Delta
F_{\phi}^{\rm tot}$ would be much smaller. The dotted line in each
plot is the upper bound placed on $\Delta F_{\phi}^{tot}/R$ by the
1997 experiment of Lamoreaux \cite{lam97}. FIG. \ref{fig3}a shows the
predictions for $n > 0$.  We see that if $n>0$, the predicted values
of $\Delta F_{\phi}^{tot}/R$ are several orders of magnitude smaller
than the current experiment upper bound.  Predictions for $n \leq -4$
are shown in FIG. \ref{fig4}a. It is clear to see that theories with
this potential and $\Lambda = 2.4 \times 10^{-3}\eV$ and $n = -4$ and
$-6$ are strongly ruled out by the Casimir force measurements made by
Lamoreaux \cite{lam97}.  This picture is remarkably similar to that
found by comparing measurements of the force between two parallel
plates with the predictions of chameleon theories; there too theories
with $n=-4$ and $-6$ were ruled out.

\begin{figure}[tbh]
\begin{center}
\includegraphics[width=10cm]{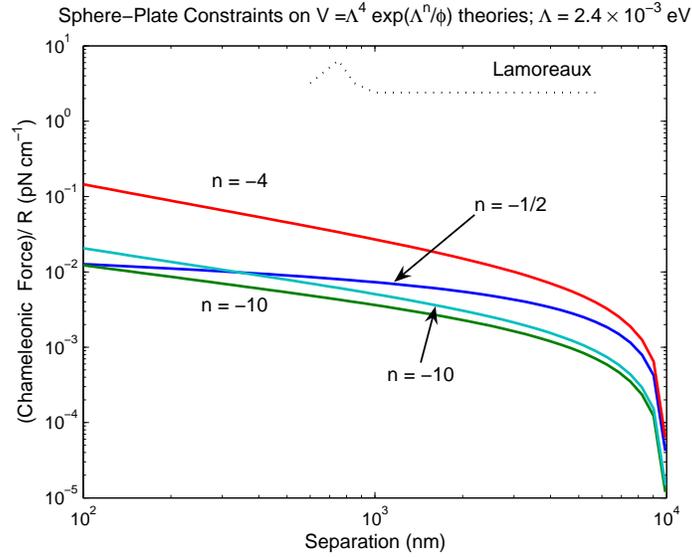}
\end{center}
\caption{The solid lines show $(F_{\phi}^{\rm tot}(d) -F_{\phi}^{\rm tot}(d_{\rm cal}))/R$ where $F_{\phi}^{\rm tot}$ is the predicted chameleonic force between a sphere and a plate, for $V = \Lambda^4_0 \exp \Lambda^n/\phi^n$ and $\Lambda = \Lambda_0 = 2.4 \times 10^{-3} \eV$.  $R$ is the radius of the sphere and we have taken $d_{cal} = 10\mum$. The dotted line shows the current best experimental constraint on any such force pressure, which comes from Ref. \cite{lam97}.  The predictions are valid for $m_c^{-1} \ll d < d_{\rm cal}\ll m_b^{-1}$; $m_c$ is the chameleon mass inside the test masses and $m_b$ is the chameleon mass in the background.}
\label{figsexp}
\end{figure}

For the predictions shown in FIGs. \ref{fig3}a and \ref{fig4}a to
apply to the 1997 experiment performed by Lamoreaux, we must require
that the test masses have thin-shells, $m_c d \gg 1$ and that $m_b
d_{\rm cal} \ll 1$.  If either of these conditions fail then the magnitude
of the force due to the presence of the chameleon field would be much
smaller than the predictions shown.  The white region in
FIGs. \ref{fig3}b and \ref{fig4}b indicates the values of the
chameleon to matter coupling, $M$, where these conditions are
predicted to hold for an experiment such as Lamoreaux's \cite{lam97}.
\begin{figure}[tbh]
\begin{center}
\includegraphics[width=8.8cm]{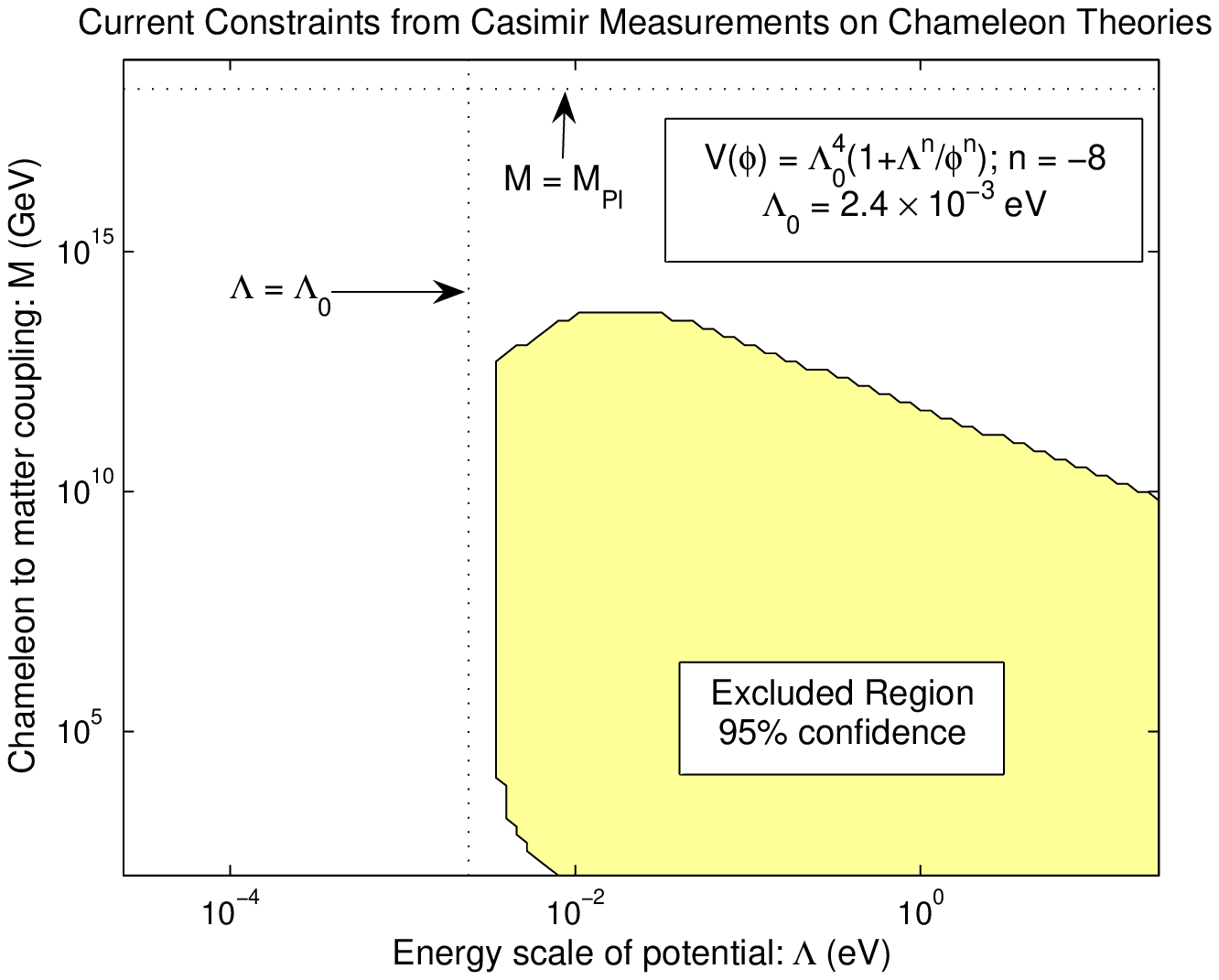}
\includegraphics [width=8.8cm]{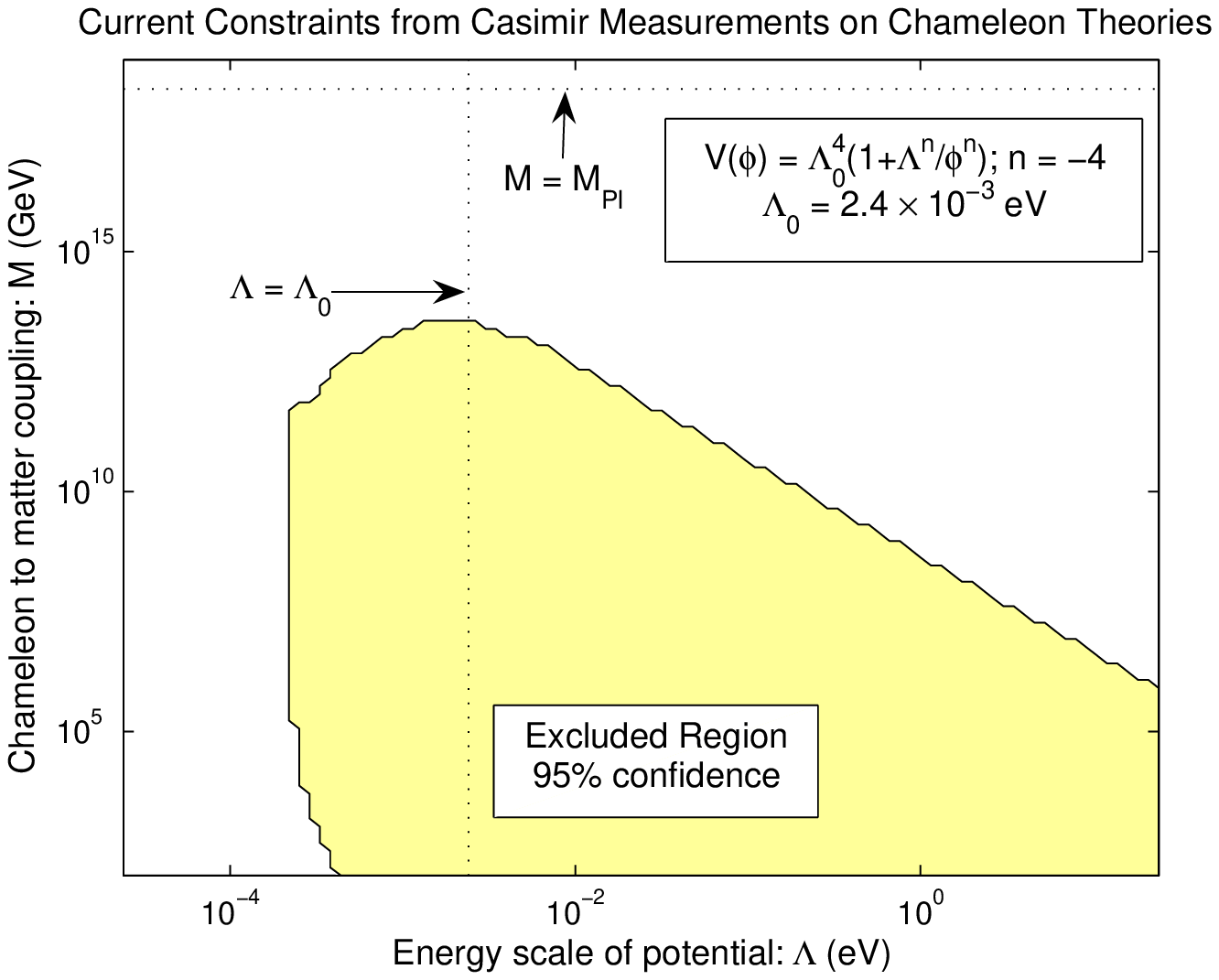}
\includegraphics [width=8.8cm]{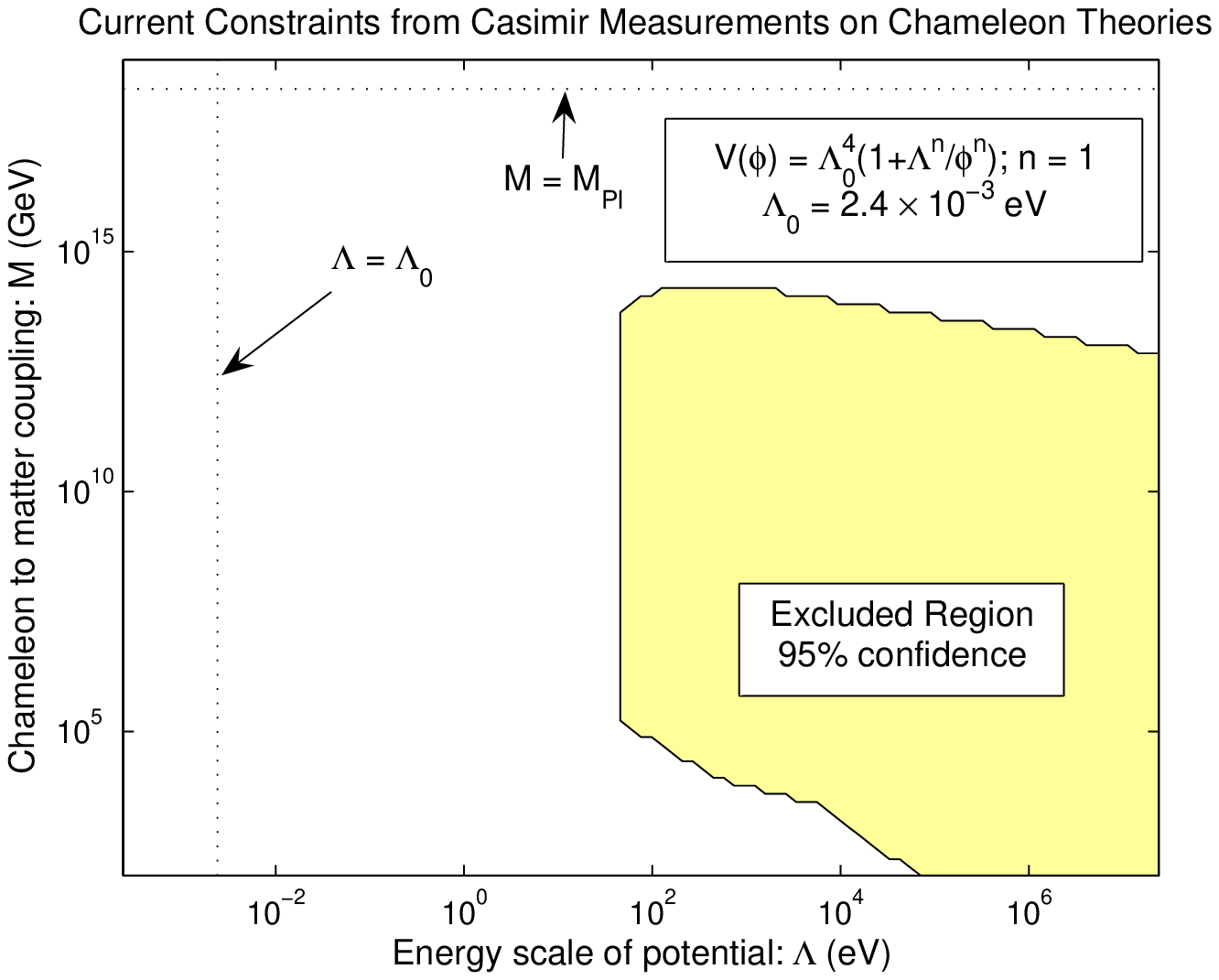}
\includegraphics [width=8.8cm]{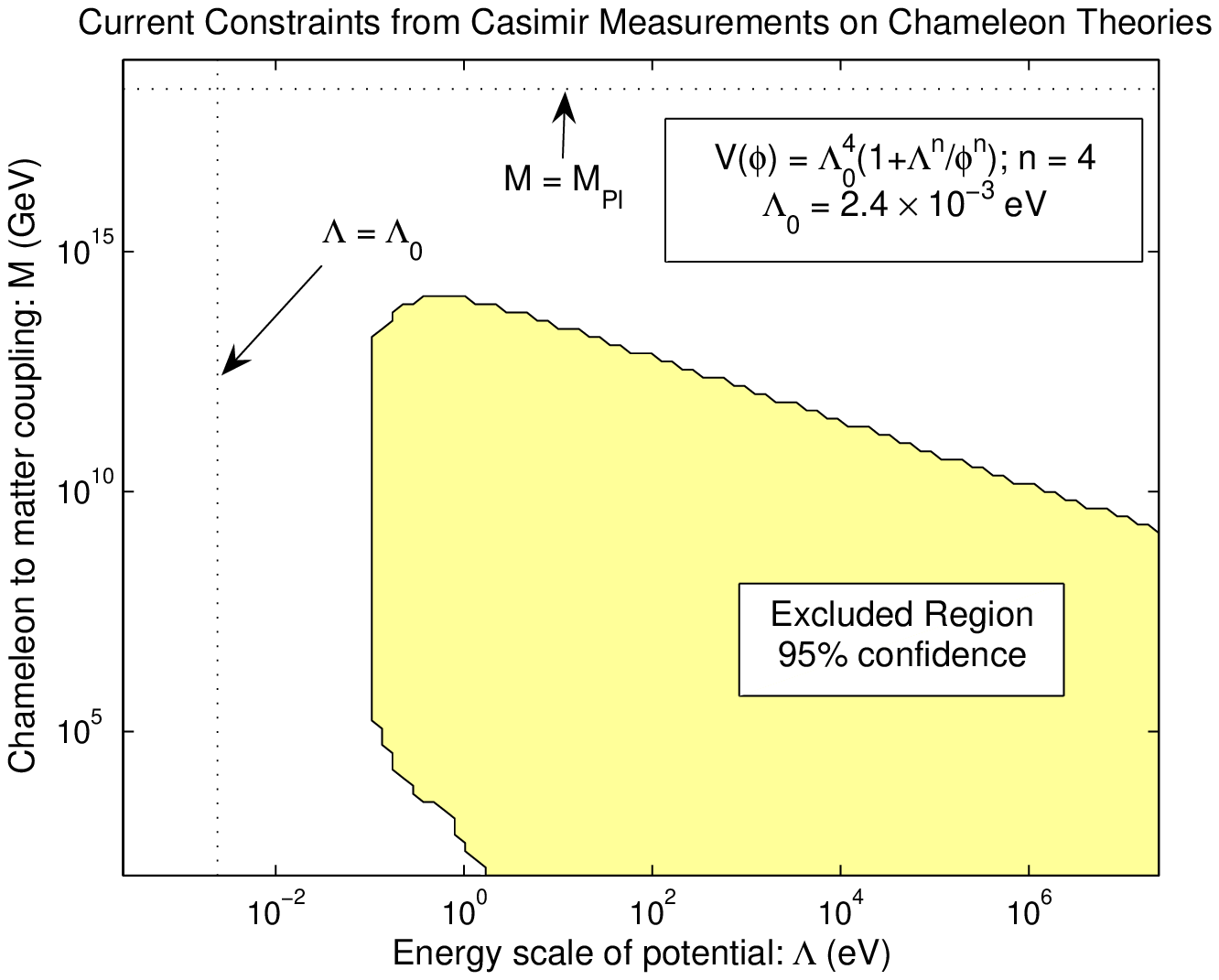}
\caption{Current constraints from Casimir force measurements on chameleon theories with $V(\phi) = \Lambda_0^4(1+\Lambda^n/\phi^n)$; $\Lambda_0 = 2.4 \times 10^{-3}\eV$.  The cases $n = -8$, $n = -4$, $n = 1$ and $n = 4$ are shown above.  Presently $\Lambda \approx \Lambda_0$ is only ruled out for theories with $n = -4$ and $n = -6$ (for which a plot is not shown).  See text for further discussion.}
\label{curcons}
\end{center}
\end{figure}
FIG. \ref{figsexp} shows how the predictions for $\Delta F_{\phi}^{\rm tot}
=(F_{\phi}^{\rm tot}(d) -F_{\phi}^{\rm tot}(d_{\rm cal}))$
with $V = \Lambda^4_0 \exp \Lambda^n/\phi^n$ and
$\Lambda = \Lambda_0 = 2.4 \times 10^{-3} \eV$ compared to the
experimental data.  It is clear that currently no such theories
are ruled out.

We conclude this section by presenting the current combined
constraints from Casimir force experiments on Chameleon theories
with $V(\phi) = \Lambda_0^4(1+\Lambda^n/\phi^n)$ and matter
coupling $M$; $\Lambda_0 = 2.4 \times 10^{-3}\eV$.   The current
combined constraints on $M$ and $\Lambda$ for theories with $n =
-8$, $n = -4$, $n = 1$ and $n = 4$ are show in FIG.
(\ref{curcons}).  As we noted above chameleon theories with $n =
-4$, $\Lambda \approx \Lambda_0$ and $M \lesssim 5\times
10^{13}\,{\rm GeV}$ are currently ruled out. The $n = -4$
corresponds to a theory where $V(\phi) = \Lambda_0^4 + \lambda
\phi^4$ where $\lambda = (\Lambda_0/\Lambda)^4$. This potential
first considered in the context of chameleon theories in Ref.
\cite{nelson}. The strongest constraint on $\lambda$ from Casimir
force measurements is $\Lambda < 2 \times 10^{-4}\eV$ i.e.
$\lambda > 10^{4}$ for $10^{4}\,{\rm GeV} \lesssim M \lesssim
10^{12}\,{\rm GeV}$.  In the $n = -8$, $n = 1$ and $n =4$ plots we
can see that $\Lambda \approx \Lambda_0$ is not currently excluded
(indeed this the case for all theories with $n \leq -8$ or $n >
0$).  In general, the steeper the drop-off of $F_{\phi}/A$ with
$d$ is, the stronger the constraints on $\Lambda$ are.  For $n >
0$, the steepness of $F_{\phi}/A$ increases as $n \rightarrow
\infty$; for $n \leq -4$ the steepness decreases as $n \rightarrow
-\infty$.  Currently the strongest constraints on $\Lambda$ are:
for $n = 1$, $\Lambda < 40\eV$ for $2 \times 10^{5}\,{\rm GeV} < M
< 3\times 10^{13}\,{\rm GeV}$; for $n = 4$, $\Lambda < 0.1\eV$ for
$2 \times 10^{5}\,{\rm GeV} <  M < 2 \times 10^{13}\,{\rm GeV}$
and for $n = -8$, $\Lambda< 3 \times 10^{-3}\eV$ for $2 \times
10^{5}\,{\rm GeV} < M < 7 \times 10^{12}\,{\rm GeV}$.  For
potentials like $V = \Lambda_0^4 \exp (\Lambda/\phi)^n$, which are
steeper than any power-law, we generally have $\Lambda \lesssim 10
- 100\Lambda_0$ for $10^{5} \lesssim M \lesssim 10^{13}\,{\rm
GeV}$. Although Casimir force measurements are generally unable to
see chameleon theories with $\Lambda \approx \Lambda_0$ at
present, they do generally require $\Lambda$ be no more than a few
orders of magnitude larger than $\Lambda_0$.  Unless we are
prepared to allow for two small but unrelated energy scales in the
theory, it therefore, if chameleon fields do exist, seems natural
that we should expect $\Lambda \approx \Lambda_0$.

\section{Prospects for Future Experiments}

Measurements of the Casimir force have so far been
unable to rule out or detect chameleon theories where the energy scale
of the potential is approximately the dark energy scale of $2.4\times
10^{-3}\,eV$. The exception  are the $n=-4$ and $n=-6$ chameleon theories with
Ratra-Peebles potentials.  In this section we consider the prospects for future
Casimir measurements detecting chameleon field theories, in particular
those with $\Lambda = 2.4 \times 10^{-3}\,eV$.  We shall identify two
proposed experiments that could potentially make such a detection.

We define $\varepsilon(d)$ to be the ratio of the chameleonic,
$F_{\phi}$, and Casimir, $F_{\rm cas}$ forces between two parallel
plates:
$$
\varepsilon(d) = \frac{F_{\phi}}{F_{\rm cas}},
$$
If the chameleonic force is to be detected or ruled out, one would
have to measure the Casimir force at a separation $d$ to an accuracy
of $100 \varepsilon(d) \%$.  If $V(\phi)=\Lambda^4_0
G((\Lambda/\phi)^n)$, for some $G$ such that $G(1)=1$ and
$G^{\prime}(1), G^{\prime \prime}(1)/2 \sim \Oo(1)$, then we would
find $F_{\phi}/A \approx A_{\phi} \Lambda^4_0 H(\Lambda_d d)$ where
$\Lambda_d = \Lambda_0^2/\Lambda$ and $H$ is some function related to
$G$; $H(1) = 1$ and $A_{\phi} \sim \Oo(1)$. It follows that
$$
\varepsilon(d) \approx \frac{240A_{\phi} \Lambda^4}{\pi^2 \Lambda_0^4}
(\Lambda_d d)^4 H(\Lambda_d d).
$$
For $A_{\phi} \sim \Oo(1)$, $240A_{\phi}/\pi^2 \sim \Oo(20)$.  Thus in
the simplest and most natural situation where there is only one energy
scale associated with the potential i.e. $\Lambda \approx \Lambda_0$,
we expect $\varepsilon(\Lambda^{-1}_0) \sim \Oo(20)$.  If $G(\Lambda_d
d) \propto d^{-2}$ and $\Lambda \approx \Lambda_0$, then $\varepsilon(d)
\sim \Oo(1)$ when $d \sim \Oo(\Lambda^{-1}_0/3)$.  If we take $\Lambda
\approx \Lambda_0 = 2.4 \times 10^{-3}\eV$ then $\Lambda^{-1}_0
\approx 82\,\mu{\rm m}$. We would then expect $\varepsilon(d) \sim
\Oo(1)$ when $d \approx 30\,\mu{\rm m}$.  At $d \approx 10\,\mu{\rm
m}$, the ratio of the chameleon to the Casimir force would then be a
few percent. This provides us with a rough estimate for the
sensitivity required to detect chameleon theories with $\Lambda
\approx \Lambda_0 \approx 2.4 \times 10^{-3}\,\eV$. A more precise
requirement can, of course, be given when $V(\phi)$ is specified.

\begin{figure}[tbh]
\begin{center}
\includegraphics[width=8.8cm]{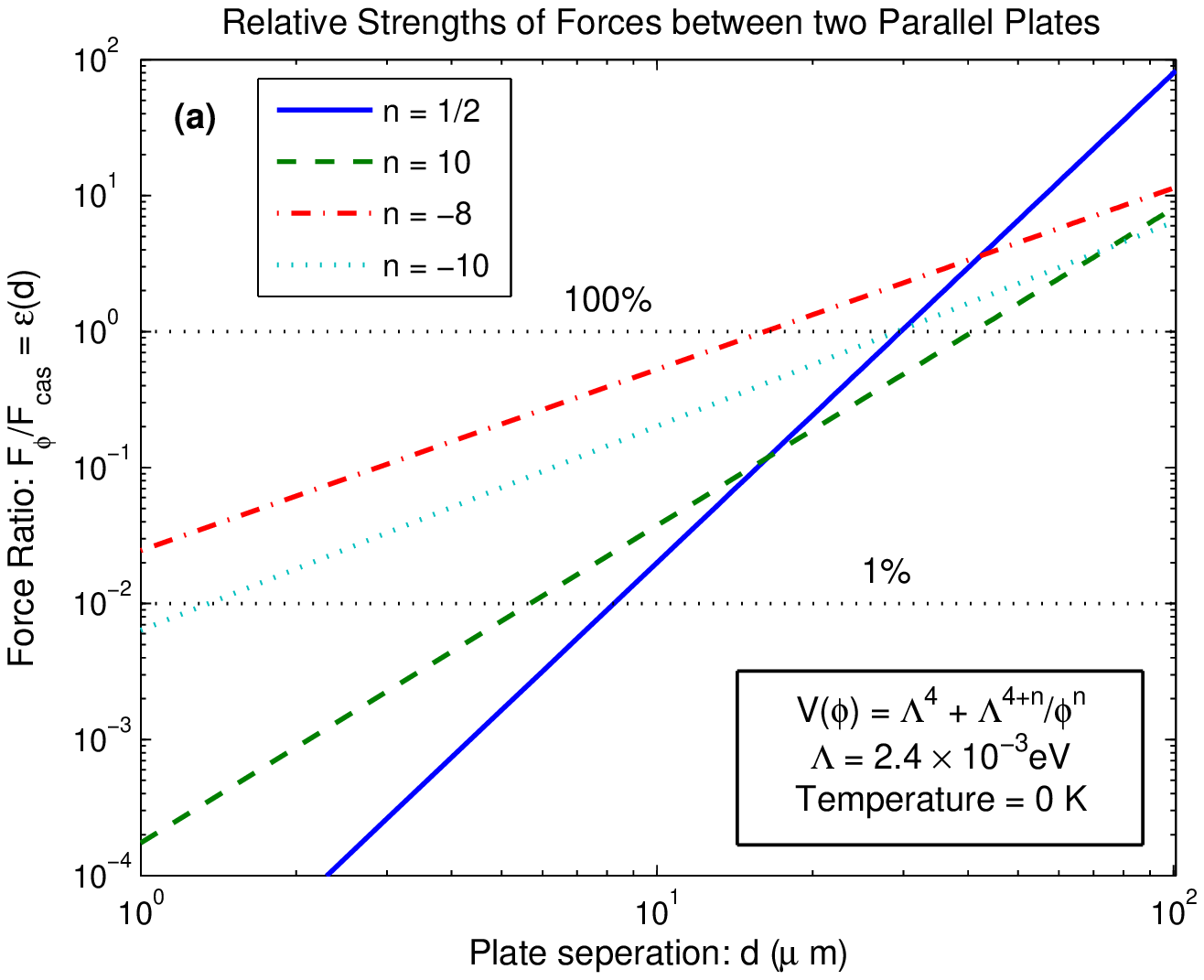}
\includegraphics [width=8.8cm]{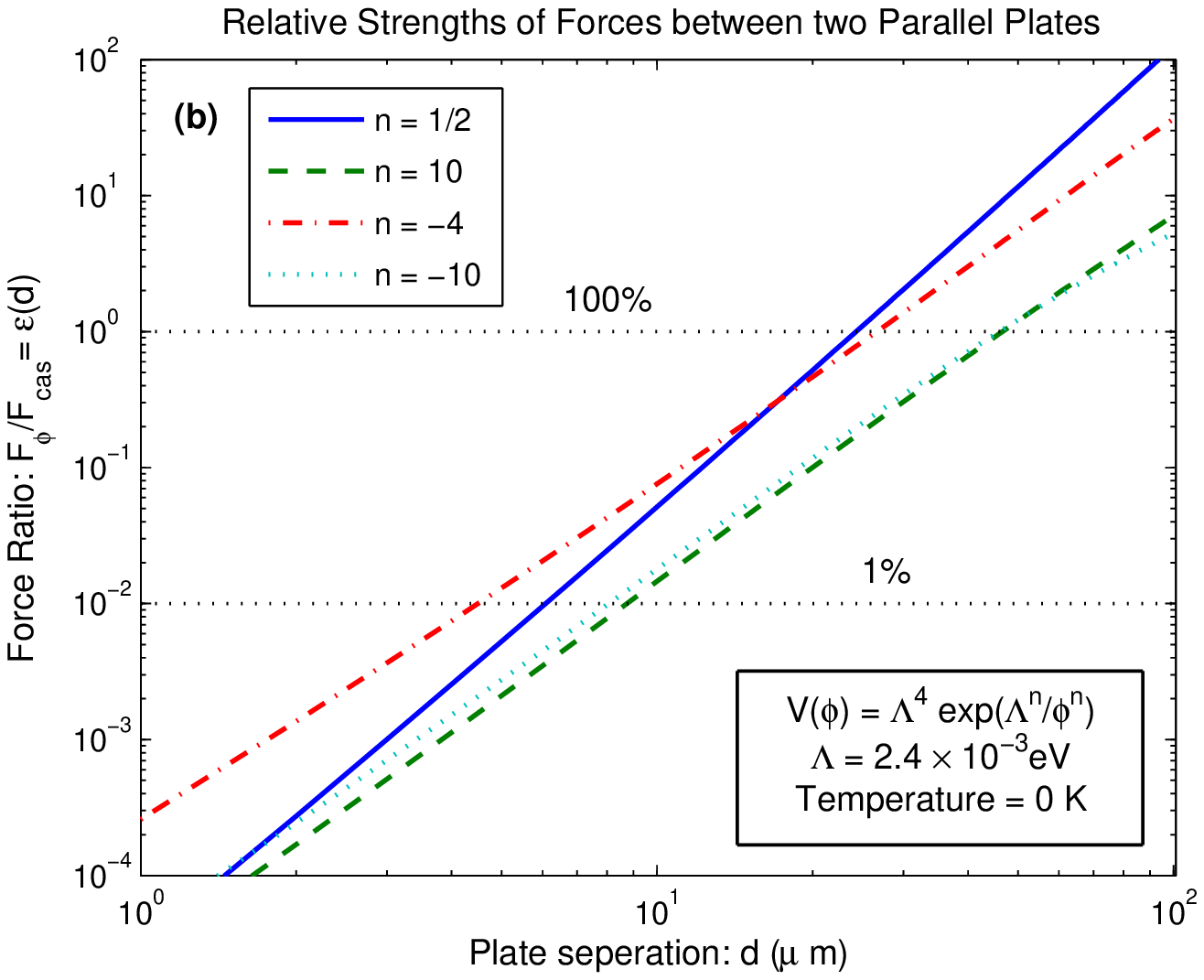}
\caption{Relative strengths of the chameleonic and Casimir forces between two parallel plates at zero temperature.  We have taken $\Lambda = 2.4 \times 10^{-3}\eV$.  Figure (a) is for $V=\Lambda^4 + \Lambda^{4+n}/\phi^n$ and figure (b) for $V = \Lambda^4 \exp(\Lambda^n/\phi^n)$ with $\Lambda = 2.4 \times 10^{-3}\eV$. For most values of $n$ we see that $\varepsilon(d)=1$ for $d \approx 30-40\,\mu{\rm m}$. At $d \approx 10\, \mu{\rm m}$, $\varepsilon \approx 0.01 - 0.1$ in most cases.}
\label{fig:req1}
\end{center}
\end{figure}

If $V = \Lambda^4_0\left(1 + \Lambda^{n}/\phi^n\right)$ then for $m_b d \ll 1$:
\begin{equation}
\varepsilon(d) = \varepsilon_{\rm pow}(d;n) \equiv \frac{240K_n \Lambda^4}{\pi^2 \Lambda_0^4}\left(\Lambda_d d\right)^{\frac{2(n+4)}{n+2}}.
\end{equation}
If $V = \Lambda^4_0 \exp \Lambda^n / \phi^n$ then for $m_b d \ll 1$ and $\Lambda_0 d \ll 1$ so that $(n+1)/nh(\Lambda_d) \ll 1$:
\begin{equation}
\varepsilon(d) \approx \varepsilon_{\rm exp}(d;n) = \frac{480 \Lambda^4}{n^2 \Lambda_0^4 h(\Lambda_d d)^{\frac{2n+2}{2}}}\left(\Lambda_d d\right)^{\frac{2(n+4)}{n+2}}\left[1-\frac{n+1}{n h(\Lambda_d d)}\right].
\end{equation}
If $V = \Lambda^4_0 \exp \Lambda^n / \phi^n$ and $d \gg
\Lambda^{-1}_d$ we have $\varepsilon = \varepsilon_{\rm pow}$. For
both of these examples $F_{\phi}/A$ grows more slowly than the Casimir
force as $d \rightarrow 0$. We plot $\varepsilon(d)$ against $d$ for
both of these potentials in FIG. \ref{fig:req1}.  In both figures we
have taken $\Lambda = \Lambda_0 = 2.4\times 10^{-3}\eV$.  For most
values of $n$ we see that $\varepsilon(d) =100\%$ for $d \approx
30-40\,\mu{\rm m}$, and $\varepsilon(d) \approx 1\% - 10\%$ when $d
\approx 10\,\mu{\rm}$.  In principle at least, all of these chameleon
theories could be detected if one were able to unambiguously measures
forces that were $1\%$ the size of zero-point Casimir force at $d
\approx 10\,{\mu m}$.  This corresponds to a sensitivity to pressure,
$P=F/A$, between the two plates of $0.13 {\rm pN}\,{\rm cm}^{-2} = 1.3
{\rm nPa}$.  As we discuss further below, such a precision is well
within the reach of the next generation of experiments.  However, in
order to actually detect the chameleonic force one must not only be
able to reach this sensitivity, but also be able to calculate and
control all background non-chameleonic forces to the same precision.
Whilst it is possible to do this for the zero-point (i.e. zero
temperature) Casimir force and any electrostatic forces, it becomes a problem at room
temperature and at separations of about $10\,{\mu m}$. Under these conditions
the thermal contribution to the Casimir force is expected to dominate over the
zero-point force.  If one were able to calculate the thermal
contribution to the Casimir force accurately then this would not be a
major problem. However, there are currently two main approaches to
calculating the thermal Casimir force and there is a great amount of
debate and disagreement as to which is correct \cite{thermal}.
The two models are often refereed to as the \emph{Drude model}
\cite{drude} and the \emph{plasma model} \cite{plasma}. For an excellent
review of the current status of this controversy see
Ref. \cite{thermal}.

\begin{figure}[tbh]
\begin{center}
\includegraphics[width=8.8cm]{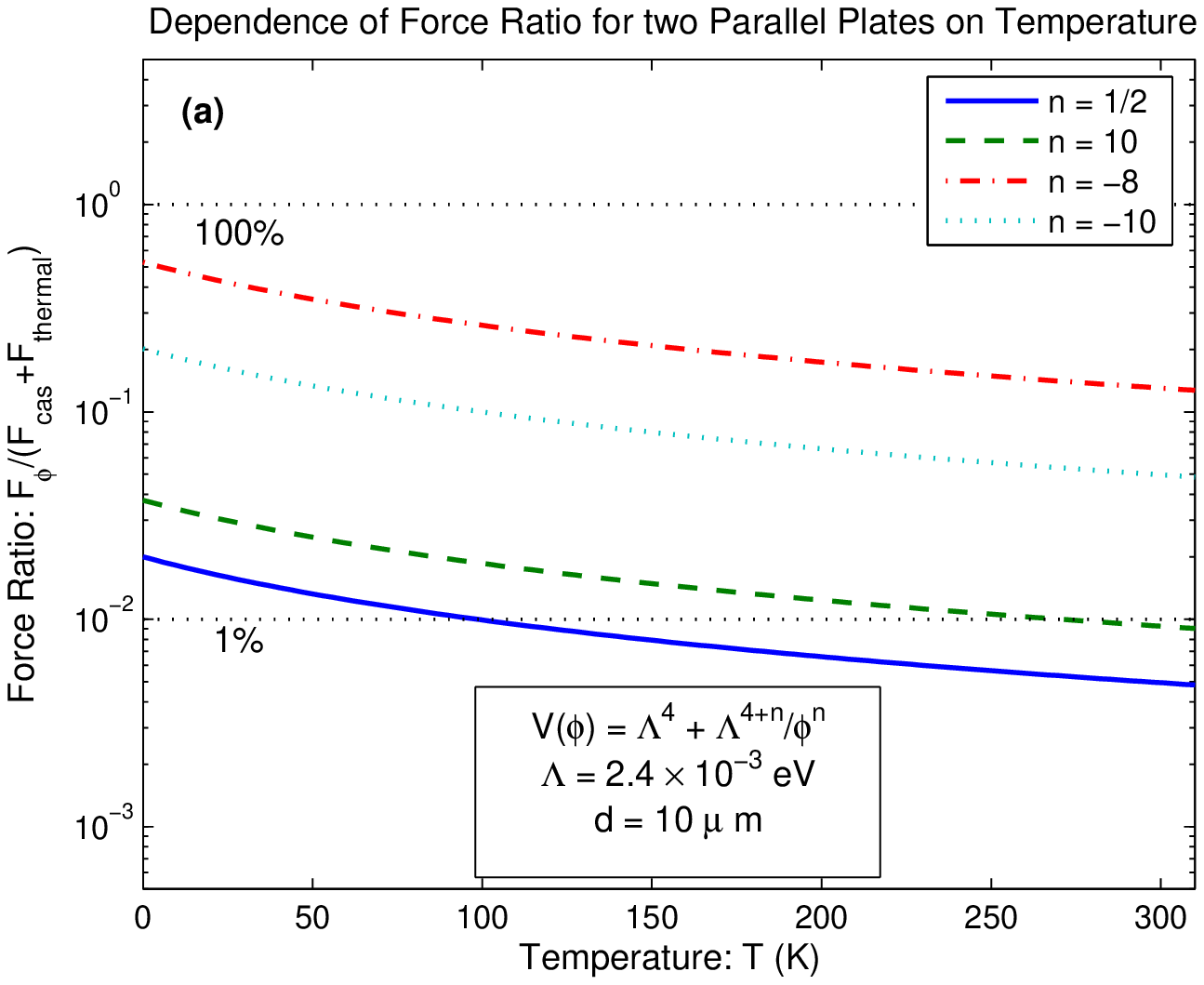}
\includegraphics [width=8.8cm]{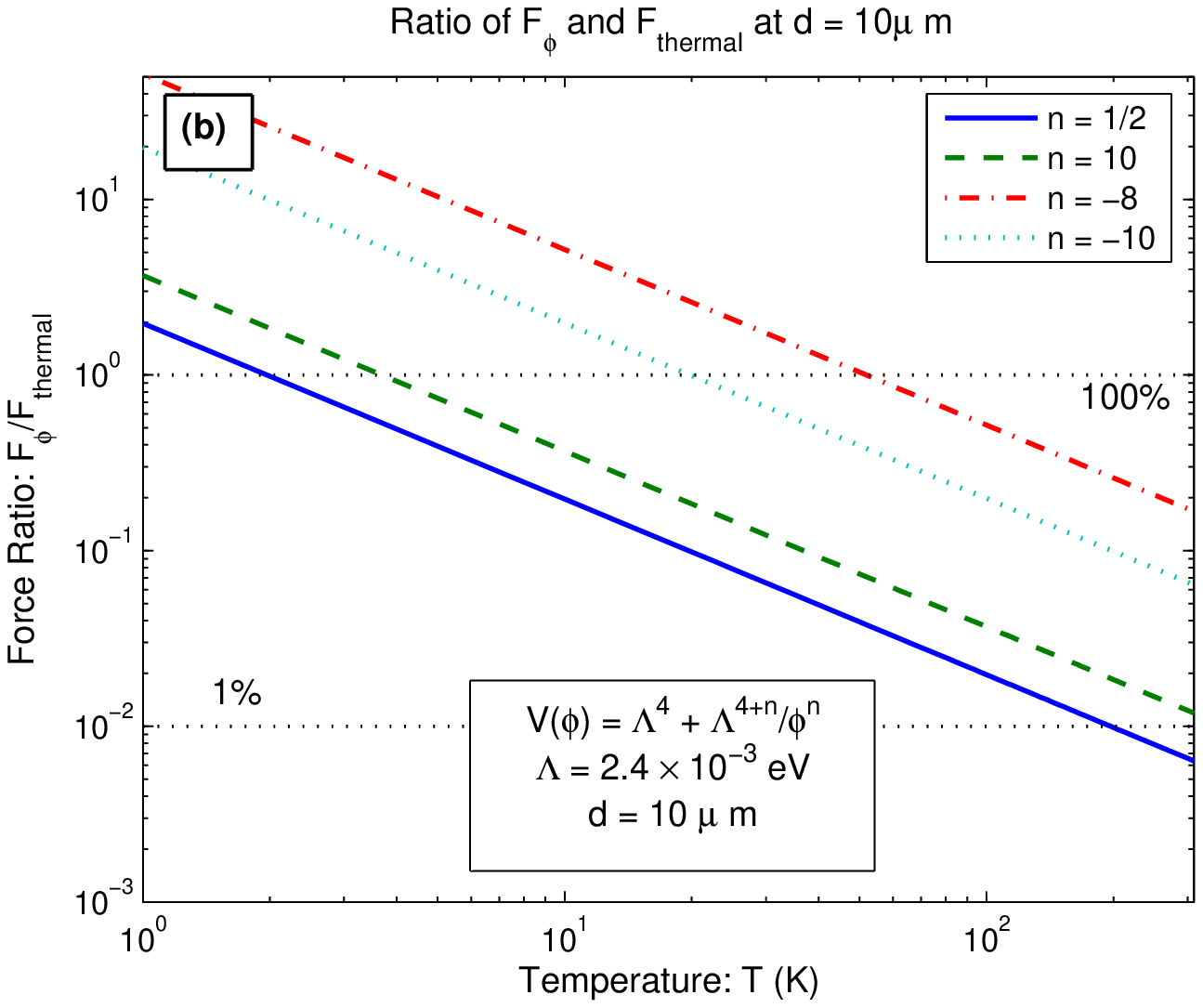}
\includegraphics [width=8.8cm]{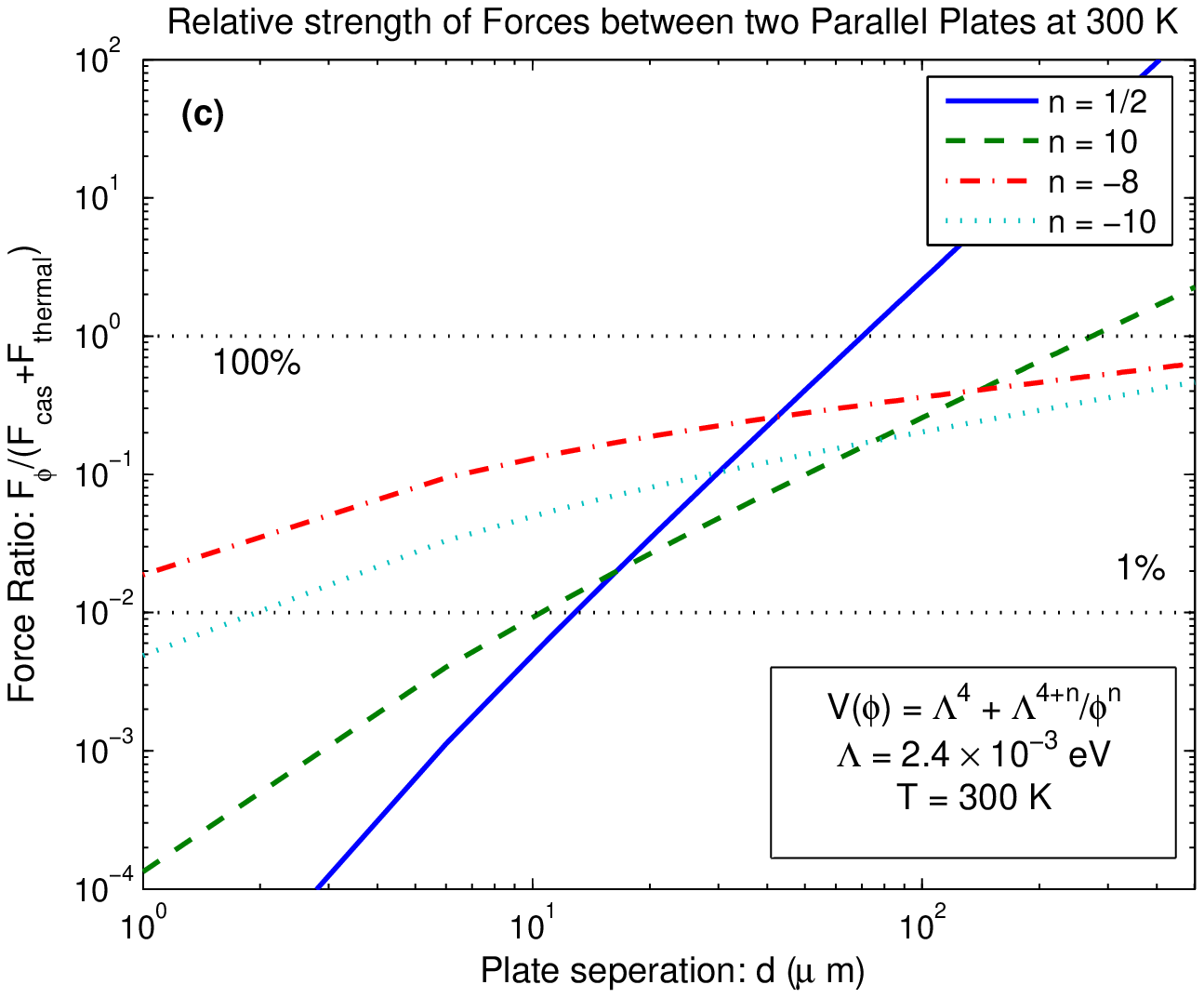}
\caption{Relative strengths of the chameleonic and the Casimir forces between two parallel plates at different temperatures.   We have taken $V=\Lambda^4_0(1 + \Lambda^{n})/\phi^n$ and $\Lambda = \Lambda_0 =  2.4 \times 10^{-3}\eV$. Figure (a) shows how $F_{\phi}/(F_{\rm cas}+\bar{F}_{\rm thermal})$ at $d=10\,\mu{\rm m}$ depends on $T$, and figure (b) shows the temperature dependence of $F_{\phi}/\bar{F}_{\rm thermal}$.  If $n=1/2$ then we see that the chameleonic force only dominates over the thermal contribution to the Casimir force for $T \lesssim 2\,{\rm K}$.  Figure (c) shows how $F_{\phi}/(F_{\rm cas} +F_{\rm thermal})$ at $300\,{\rm K}$ depends on separation, $d$.    We see that at $300$~K and with $d \approx 10\, \mu{\rm m}$, $\varepsilon \approx 0.005 - 0.1$.}
\label{fig:req2}
\end{center}
\end{figure}

For separations larger than $5{\mu m}$, the thermal Casimir force between two perfectly reflecting mirrors is:
$$
\frac{\bar{F}_{\rm thermal}}{A} \equiv \frac{\zeta(3)k_{B}T}{4\pi d^3},
$$
where $k_B$ is Boltzmann's constant. The competing more realistic
models predict $\bar{F}_{\rm thermal}/2 \lesssim F_{\rm thermal}
\lesssim \bar{F}_{\rm thermal}$. At $T=300{\rm K}$ and at $d =
10\,\mu{\rm m}$ we then have $1.5 \lesssim F_{\rm thermal}/F_{\rm cas}
\lesssim 3.0$.  At $10\,{\mu m}$ , the magnitude of the chameleonic
force for $\Lambda = \Lambda_0 \approx 2.4 \times 10^{-3}\eV$ is
generally about $1\%$ of the total Casimir force (including thermal
correction).

\begin{figure}[tbh]
\begin{center}
\includegraphics[width=8.8cm]{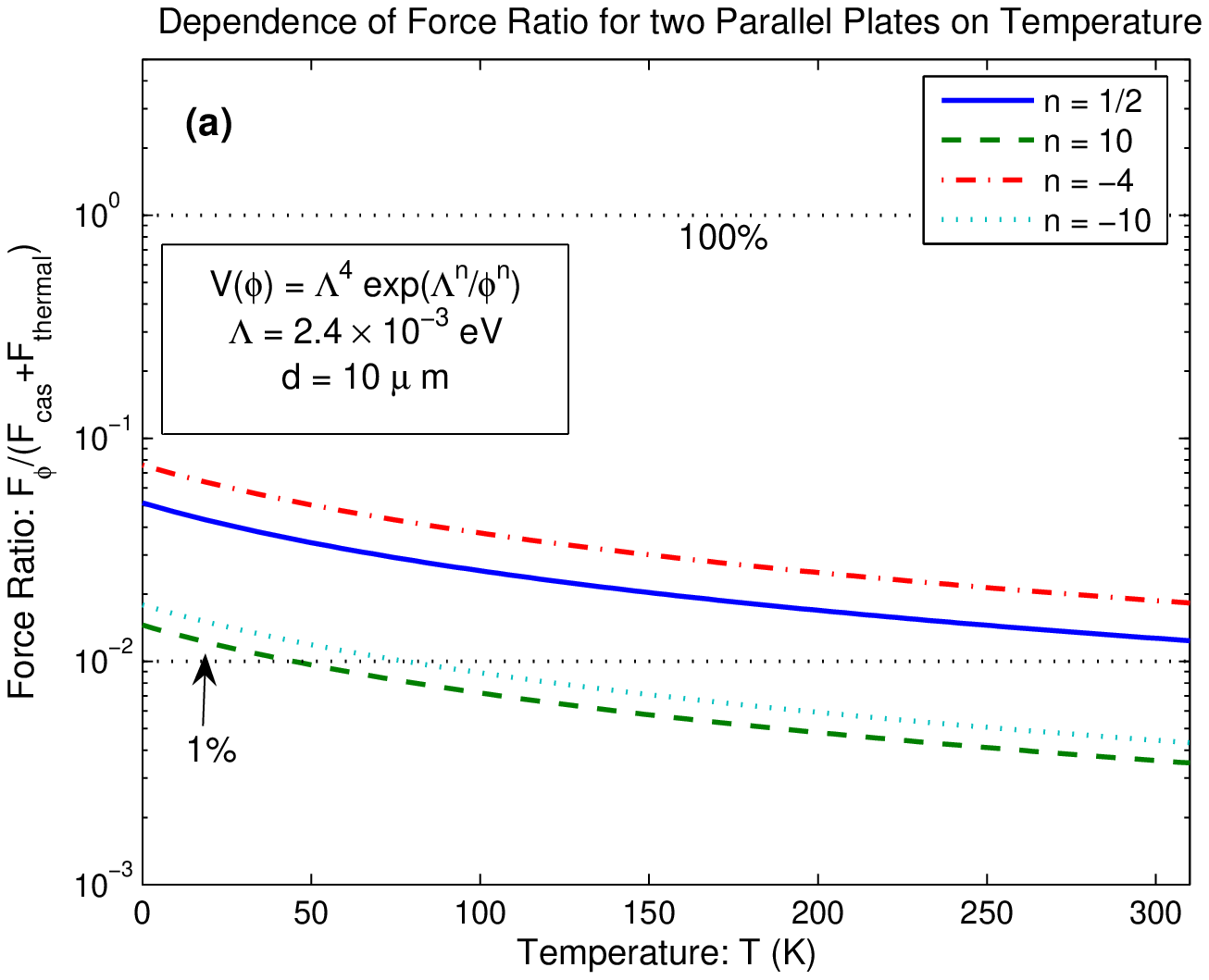}
\includegraphics [width=8.8cm]{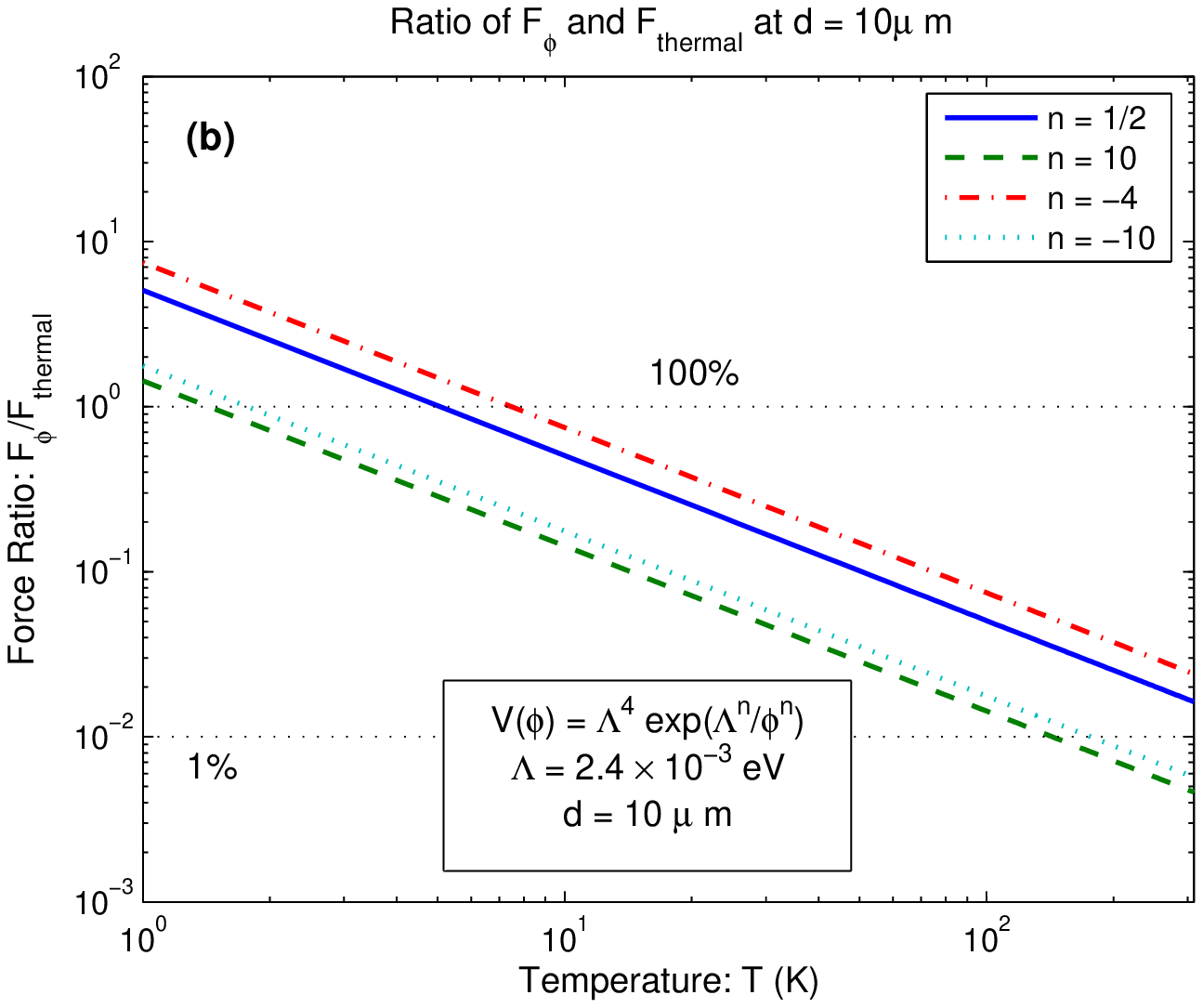}
\includegraphics [width=8.8cm]{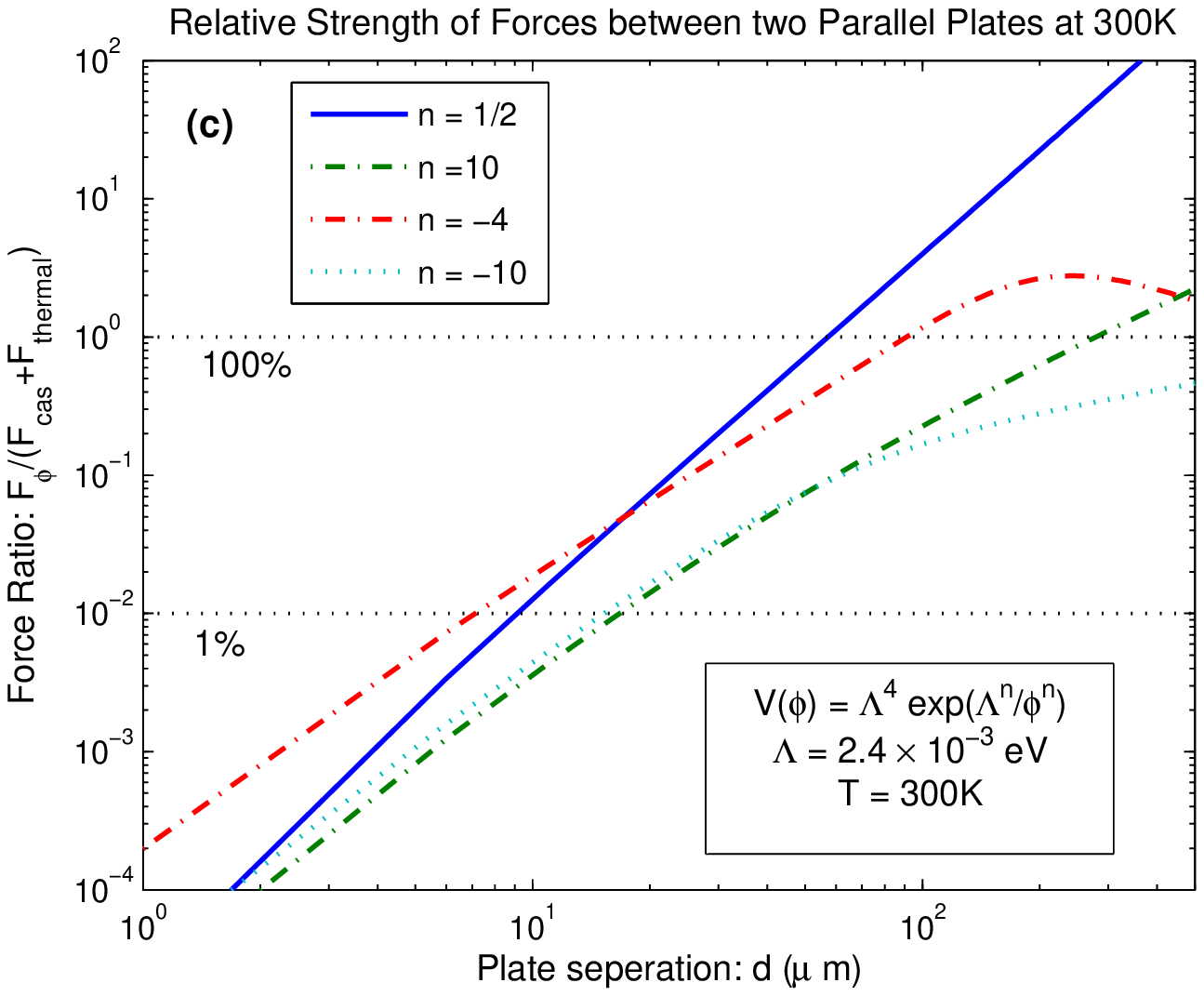}
\end{center}
\caption{Relative strengths of the chameleonic and the Casimir forces between two parallel plates at different temperatures.   We have taken $V=\Lambda^4_0\exp(\Lambda^n/\phi^)$ and $\Lambda = \Lambda_0 =  2.4 \times 10^{-3}\eV$. Figure (a) shows how $F_{\phi}/(F_{\rm cas}+\bar{F}_{\rm thermal})$ at $d=10\,\mu{\rm m}$ depends on $T$, and figure (b) shows the temperature dependence of $F_{\phi}/\bar{F}_{\rm thermal}$.  If $n=1/2$, the chameleonic force dominates over the thermal contribution to the Casimir force for $T \lesssim 5\,{\rm K}$.  Figure (c) shows how $F_{\phi}/(F_{\rm cas} +F_{\rm thermal})$ at $300\,{\rm K}$ depends on separation, $d$.    At $300$~K and with $d \approx 10\, \mu{\rm m}$, $\varepsilon \approx 0.002 - 0.02$. }
\label{fig:req3}
\end{figure}

In FIGs. \ref{fig:req2} and \ref{fig:req3} we indicate
how $F_{\phi}/(F_{cas} +\bar{F}_{\rm thermal})$ and
$F_{\phi}/\bar{F}_{\rm thermal}$ depend on $T$ and $d$.

 To detect, or rule out, the presence of chameleon fields with
$\Lambda \approx \Lambda_0 \approx 2.4\times 10^{-3}\eV$ at a
separation of $\Oo(10)\,{\mu m}$, one first has to calculate both the
zero-point Casimir force to an accuracy of better than $1\%$ and then
either to be able to do the same for the thermal Casimir force. Otherwise perform
the experiment at sufficiently low temperature so that
$\bar{F}_{thermal} \ll \bar{F}_{\phi}$. From FIGs. \ref{fig:req2} and
\ref{fig:req3} we see that to realize this latter option for most
theories considered here, the experiment would have to be run at $T \sim
\Oo(1)K$ or cooler.

The Gravitation group at the University of Birmingham \cite{gravB}
have constructed a super-conducting torsion balance that they intend
to use to measure, amongst other things, the Casimir force at $4.2\,{\rm
K}$ \cite{gravB}. Precise details of the separations they will probe
and the precision which they expect to achieve have not yet been
announced. However, even at $d \approx 10\,\mu{\rm m}$, it is clear
from Figures \ref{fig:req2}b and \ref{fig:req3}b that even with
$T=4.2\,{\rm K}$ the chameleonic force between two parallel plates is
still generally only $0.3 - 3$ times the size of the thermal
contribution to the Casimir force.  Even at this low temperature,
one would still have to be able to calculate the thermal Casimir force to
greater accuracy than is currently possible due to the controversy
over the different models.

The thermal Casimir force drops off as $1/d^3$ as $d\rightarrow
\infty$.  Of the chameleon theories considered here only those
with $n=-4$ and $n=-6$ exhibit a faster drop off for $d \ll
m_b^{-1}$ and with a Ratra-Peebles power-law potential, these
theories are already ruled out for $\Lambda \approx \Lambda_0 =
2.4 \times 10^{-3}\eV$.  If one wished to avoid having accurately
model the thermal Casimir force, one could take advantage of the
slow drop-off of the chameleonic force, and run a Casimir force
experiment at separations where the chameleonic force is predicted
to dominate over the total Casimir force. For $\Lambda \approx
\Lambda_0 \approx 2.4 \times 10^{-3}\eV$, this would generally
involve running the experiment at separations larger than $30-100
\mum$.  Although this range of separations has already been probed
by tests of gravity such as the E\"{o}t-Wash experiment
\cite{EotWash}, the physical electrostatic shield that is employed
in such tests would block any signal due to strongly coupled
chameleon fields \cite{chamstrong}.

Importantly, the way in which electrostatic forces are controlled
in Casimir force measurements, i.e. without a physical shield,
does \emph{not} shield the chameleonic force \cite{chamstrong}.
Hence any such experiment could probe relatively large separations
and could use large test masses (i.e. with lengths scales of a few
centimeters or more) so as to magnify any new forces.  Given the
size of the test masses, and the very high accuracy force
measurements that would be required, it would seem sensible for
any such experiment to make use of a torsion balance as was used
by Lamoreaux in Ref. \cite{lam97}.  Lamoreaux has discussed a
number of improvements to his 1997 experiment \cite{lampriv} which
would allow it to detect chameleon fields with $\Lambda \approx
\Lambda_0 \approx 2.4 \times 10^{-3}\eV$.  We discuss this further
below in section \ref{Lamsec}.

In the longer term, there are a number of planned experiments (not least the
new one proposed by Lamoreaux) that aim to detect the thermal Casimir
force to high accuracy.  If experiments can decide which is the best
model, then one of these new experiments, proposed in
Ref. \cite{IILnew} and already under construction at the Institute Laue-Langevin (ILL) in
Grenoble, would ultimately be able to detect or rule out the presence
of chameleonic forces at $d \approx 10\mum$.  We discuss this
experiment further below.

\subsection{Proposed Grenoble Experiment}

In Ref. \cite{IILnew},
Lambrecht \emph{et al.} proposed a new experiment to measure the
Casimir force between two parallel plates and search for thermal
corrections in the $1-10\,\mu{\rm m}$ separation range using a high
sensitivity torsion balance.  Previous attempts to make such
measurements \cite{sparnaay,Bressi} using this geometry have been
limited by the extent to which the plates can be kept parallel. The
solution that Lambrecht \emph{et al.} proposed in Ref. \cite{IILnew}
was to take advantage of the inclinometer developed for a neutron
experiment performed at the Institute Laue-Langevin (ILL) in Grenoble
\cite{neutronIIL}. This would allow them to limit deviations from
parallelism at the $10^{-6}{\rm radians}$ level. For comparison the
Padova experiment of Bressi \emph{et al.} \cite{Bressi} had estimated
deviations from parallelism of $3 \times 10^{-5}{\rm radians}$.  The
surface area of the plates that will be used in the new experiment
would be of the order of $120\,{\rm cm}^{2}$. The force measurements
made using the torsion balance would have a resolution of about $1{\rm
pN}$ for $5\,\mu{\rm m} \lesssim d \lesssim 10\,\mu{\rm m}$
\cite{IILnew}.  The aim of this proposed experiment is two-fold.
Firstly, it should contribute to the settling of the controversy that
surrounds the correct approach to calculating the thermal Casimir
force. Secondly, with a theoretical model for the thermal Casimir
force established, it would be able to probe for the presence of new
forces.

\begin{figure}[tbh]
\begin{center}
\includegraphics[width=8.8cm]{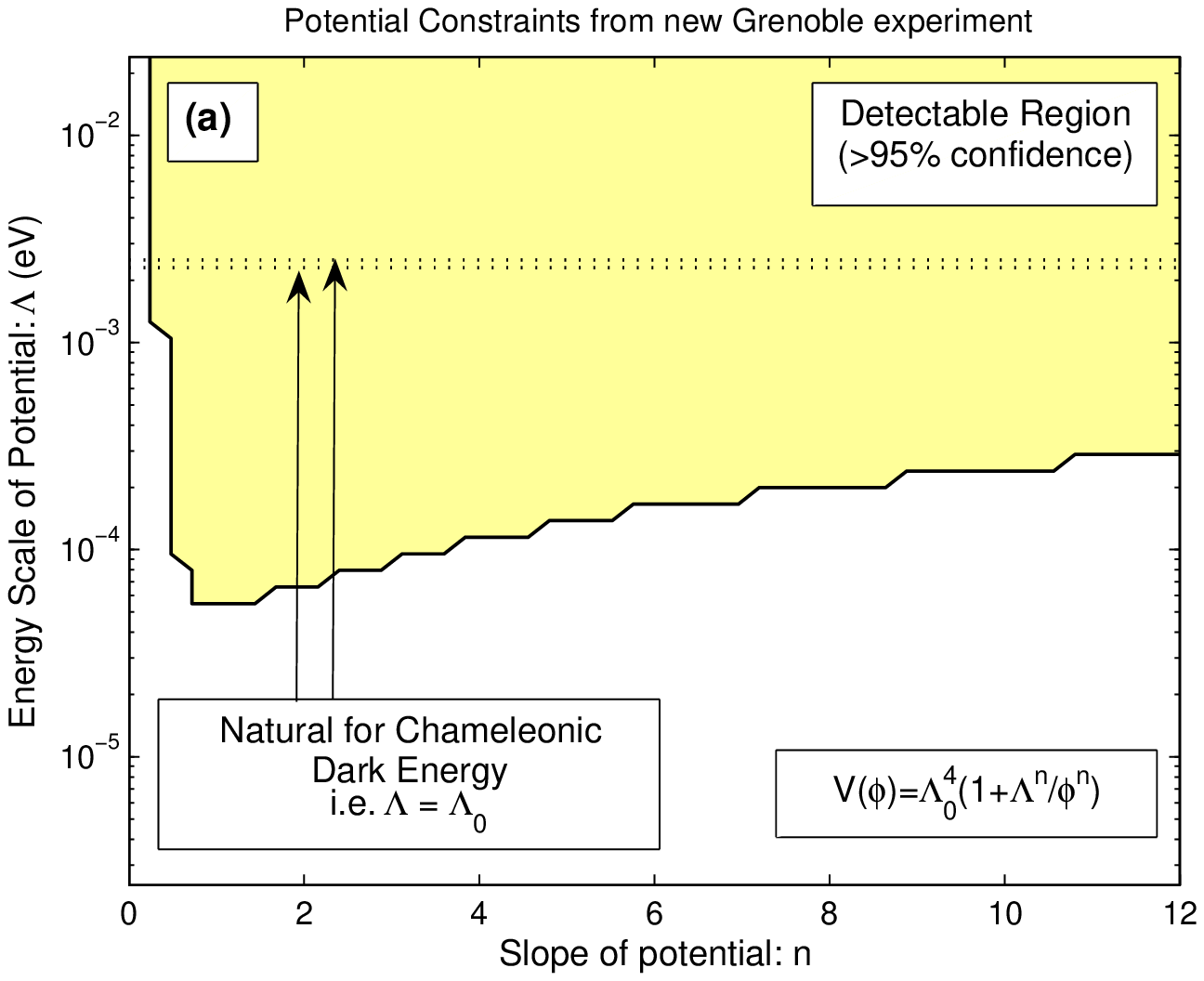}
\includegraphics [width=8.8cm]{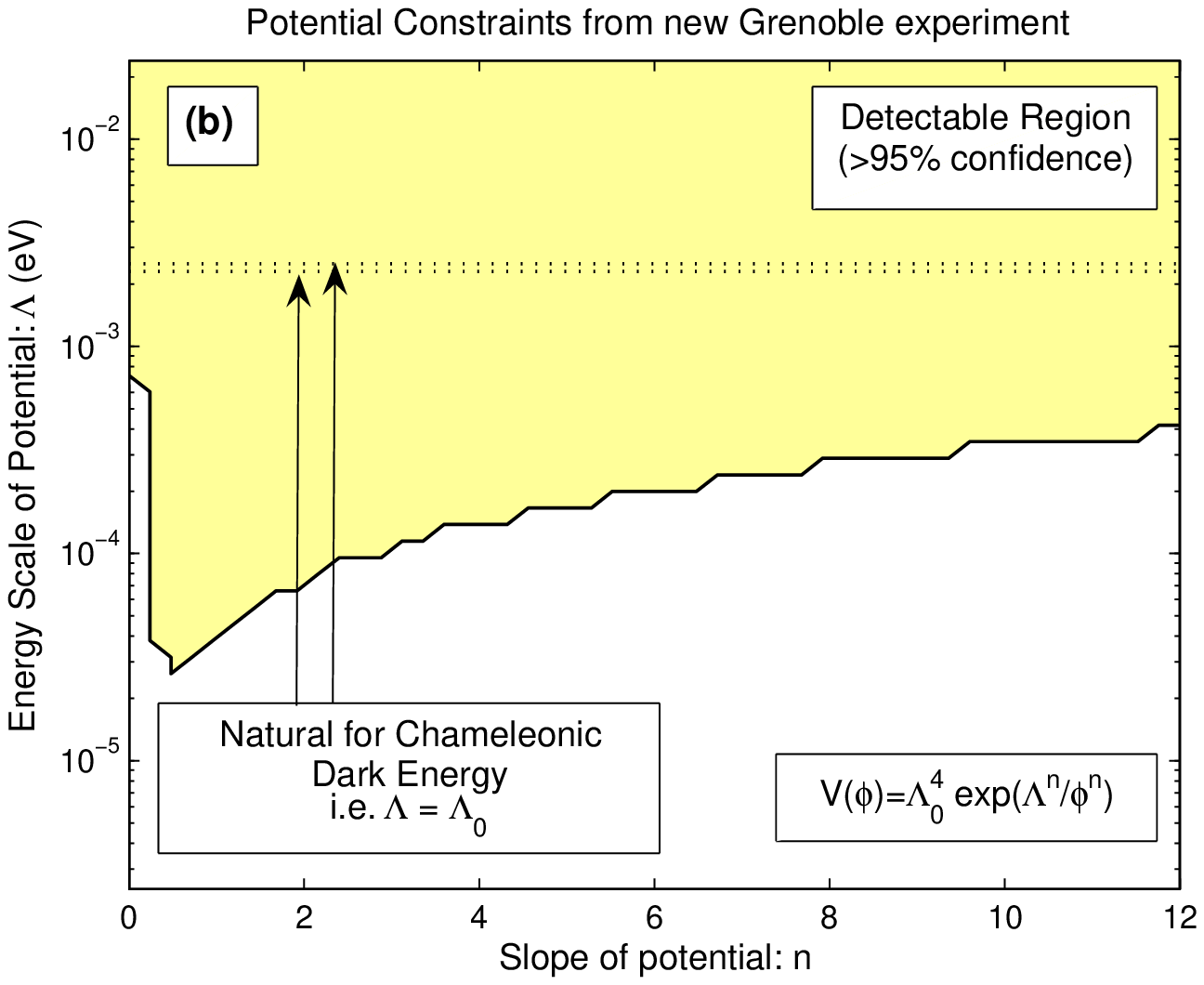}
\caption{Shaded area is the region of $\Lambda$-$n$ parameter
space that could potentially be detected or ruled out with $95\%$
confidence by the new Grenoble experiment proposed by Lambrecht
\emph{et al.} in Ref. \cite{IILnew}.  Figure (a) is for
$V=\Lambda^4_0(1 + \Lambda^{n})/\phi^n$ and figure (b) for $V =
\Lambda^4_0 \exp(\Lambda^n/\phi^n)$ with $\Lambda_0 = (2.4\pm 0.1)
\times 10^{-3}\eV$.  If $\Lambda  = \Lambda_{0}$, which would be
natural is the chameleon field is responsible for the late time
acceleration of the universe, then $\Lambda$ must lie between the
two dotted lines i.e. $\Lambda = (2.4 \pm 0.1) \times 10^{-3}\eV$.
It should be noted that for these constraints to apply the
chameleon to matter coupling $\beta = M_{\rm Pl}/M$ must be large
enough for test masses to have thin-shells and small enough for
the inverse chameleon mass in the background to be large compared
with the separations probed i.e. $m_b^{-1} \gg 10 \mu{\rm m}$.
Both of these requirements will depend to varying degrees on the
quality of the laboratory vacuum in which the experiment is
performed.} \label{fig:expnew}
\end{center}
\end{figure}

In order for there to be a detectable chameleonic force one must
ensure that the laboratory vacuum is good enough so that the chameleon
mass in the background, $m_b$, is $\ll \hbar c / 10\,\mu{\rm m}$.
Provided that this is the case and the $1{\rm pN}$ sensitivity can be
reached with background and systematic effects controlled at the same
level, the Grenoble experiment proposed in Ref. \cite{IILnew} would,
by making measurements for distances $5\mu{\rm m} < d < 10\,\mu{\rm m}$ be able
to detect, or rule out, a large number of chameleon
theories. Specifically if $V = \Lambda^4_0 \exp(\Lambda^n/\phi^n)$ and
$\Lambda = \Lambda_0 = 2.4 \times 10^{-3}\eV$, then this new
experiment could detect or rule out all theories with $1/8 \lesssim
\vert n \vert \lesssim 80$ with at least $99.5\%$ confidence.
Additionally, it would be able to rule or detect all theories with
$V(\phi)=\Lambda^4_0(1 + \Lambda^{n})/\phi^n$, $\Lambda = \Lambda_0 =
2.4 \times 10^{-3}\eV$ and $n \leq -4$ or $n > 1/3$ with the same
confidence.  Figure \ref{fig:expnew} shows $95\%$ confidence limits on
the values of $n$ and $\Lambda$ that this new test should ultimately
be able to place on chameleon theories.

The region of $M$ space that will be accessible to this new experiment
will depend on, amongst other things, $\Lambda$, $n$ and the quality
of the laboratory vacuum and the thickness and composition of the test
masses. Firstly, it is required that the test-masses have thin-shells.
In this experiment the test masses are $15\,{\rm mm}$ thick glass
plates.  The thin-shell conditions generally hold for all $\beta =
M_{\rm Pl}/M \gtrsim \Oo(1)$.

\begin{figure}[tbh]
\begin{center}
\includegraphics[width=8.8cm]{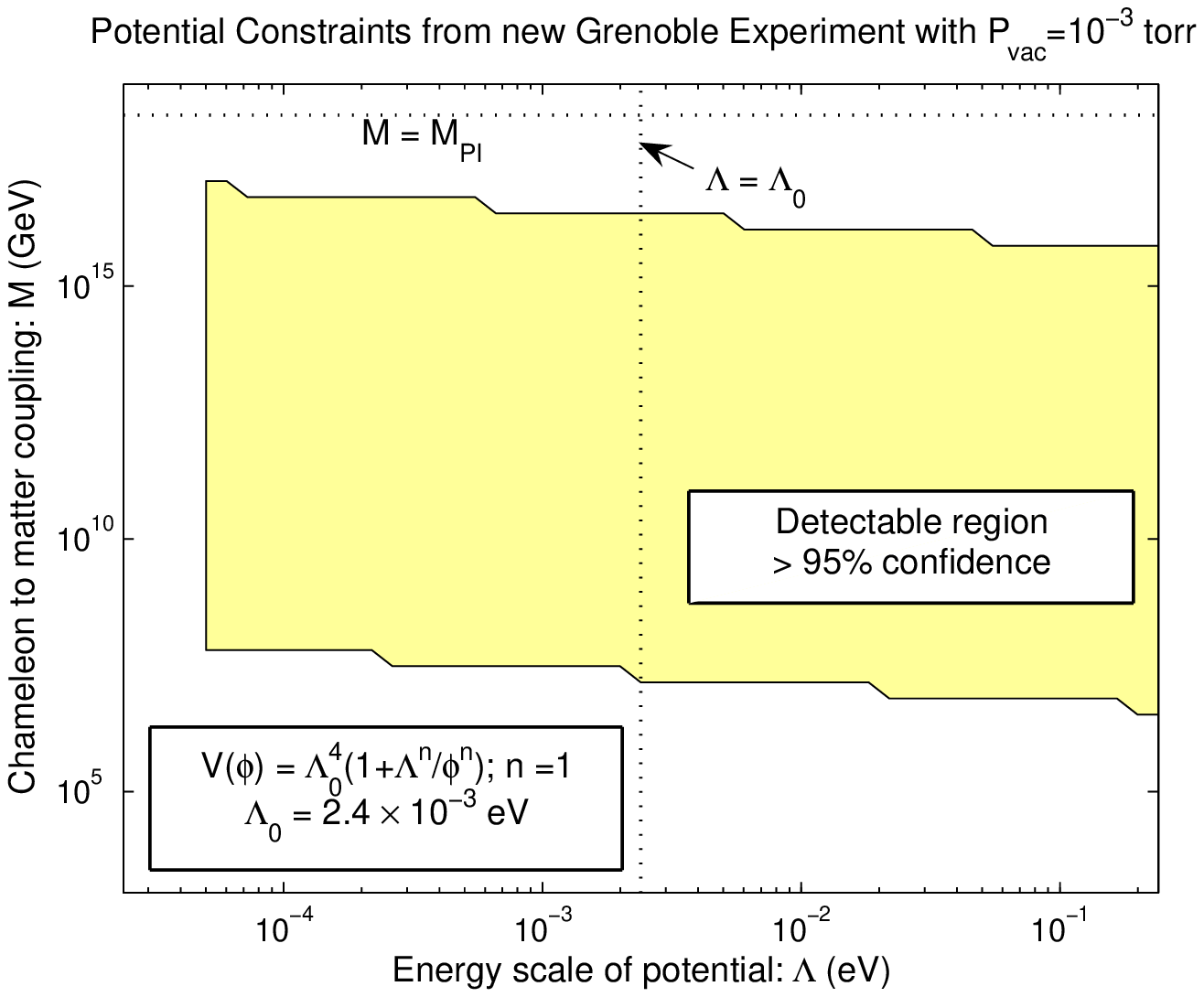}
\includegraphics [width=8.8cm]{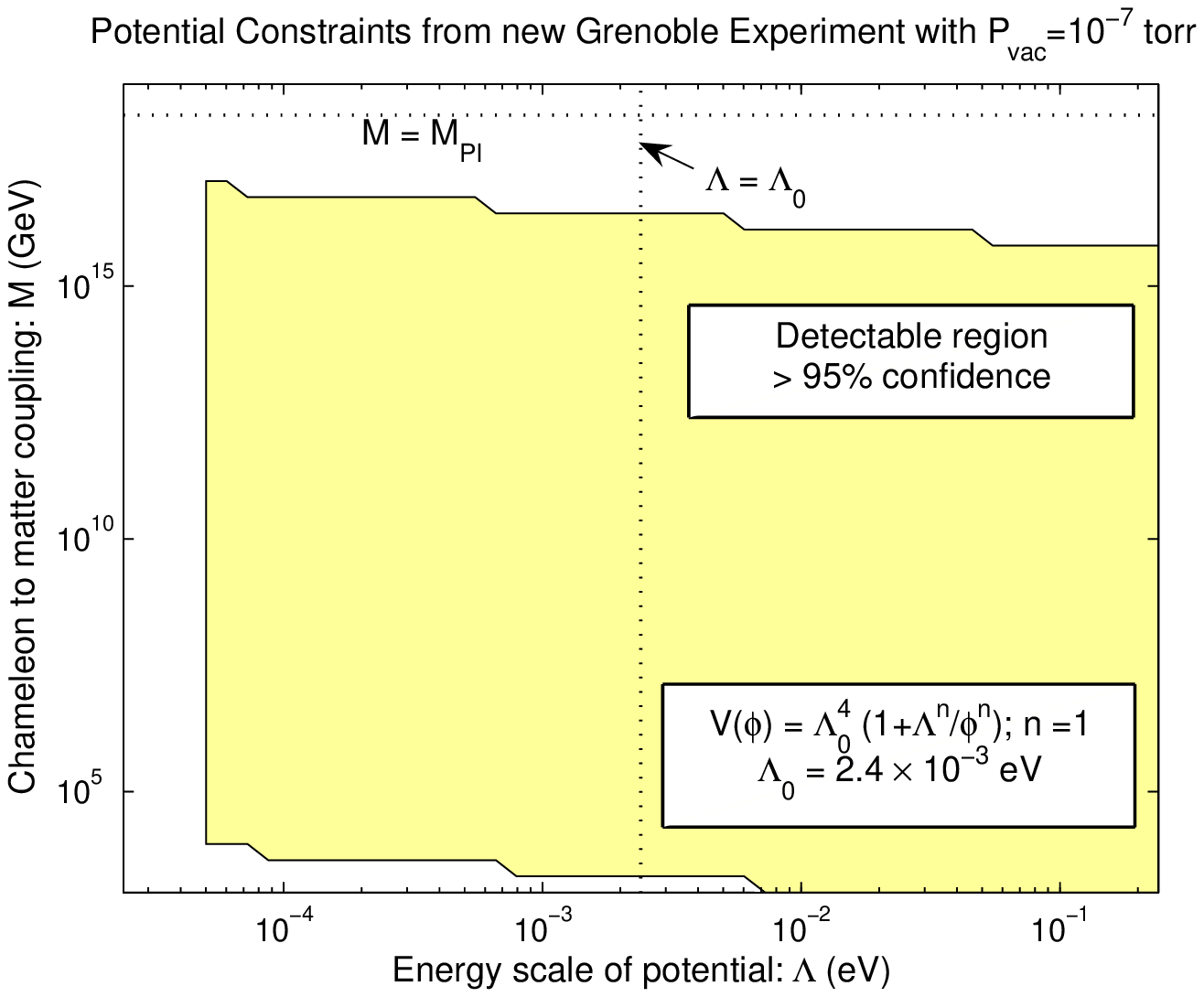}
\includegraphics[width=8.8cm]{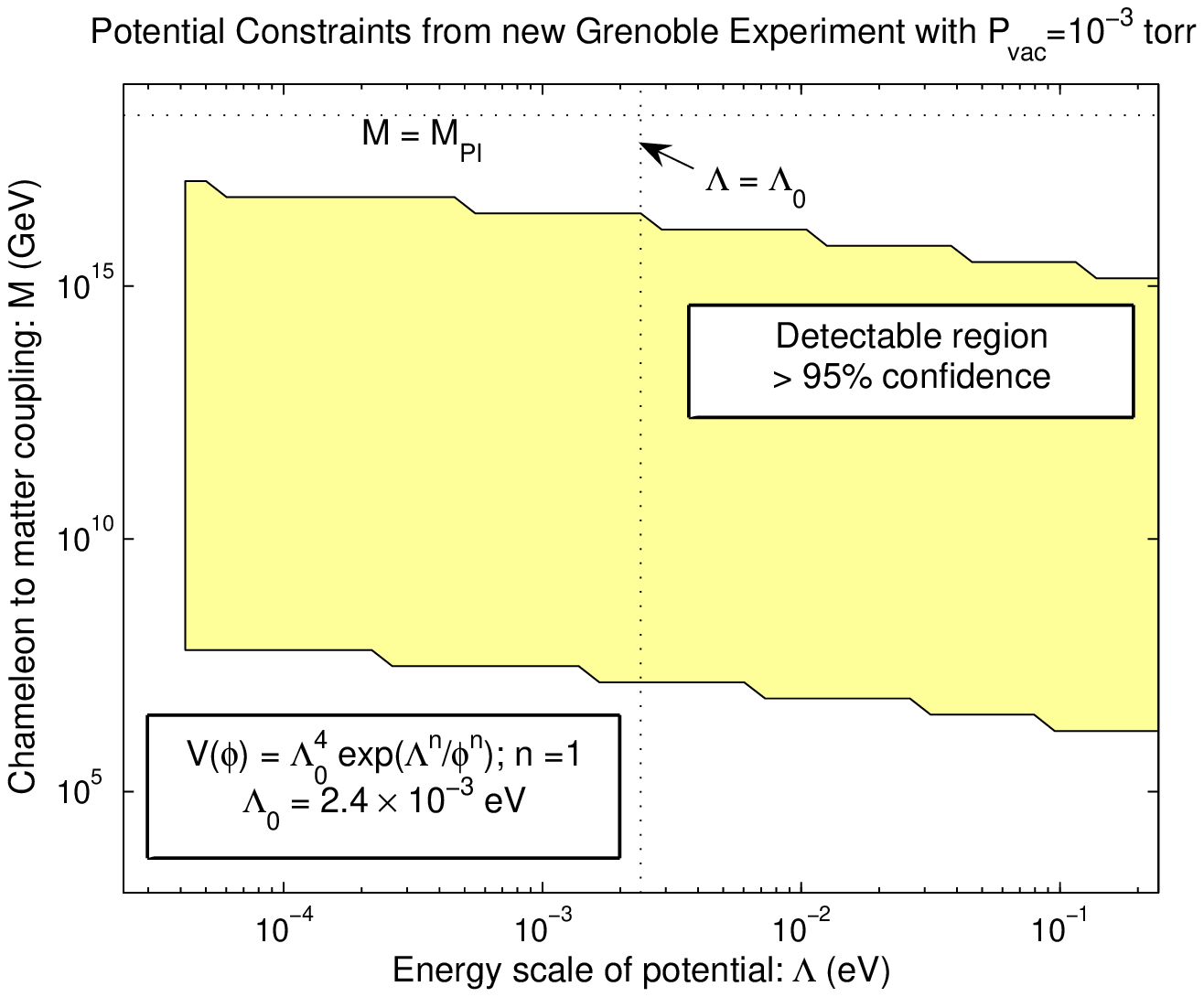}
\includegraphics [width=8.8cm]{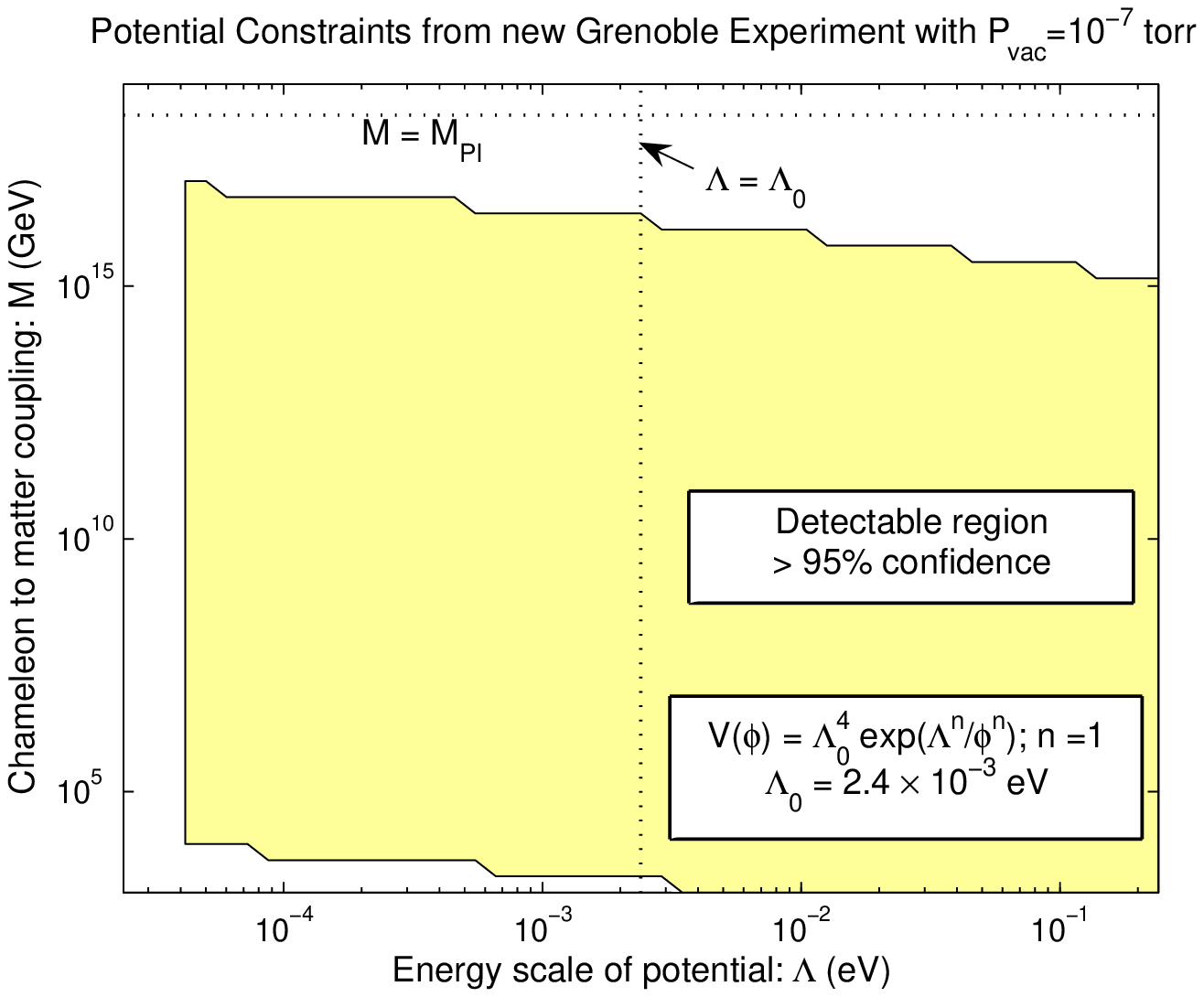}
\caption{Shaded area is the region of $\Lambda$-$M$ parameter space that could potentially be detected or ruled out with $95\%$ confidence by the new Grenoble experiment proposed by Lambrecht \emph{et al.} in Ref. \cite{IILnew} for two choice of the vacuum pressure. The plots shown are for $n=1$ with either $V=\Lambda^4_0(1 + \Lambda^{n})/\phi^n$ or $V=\Lambda^4_0\exp(\Lambda^n/\phi^n)$. The situation for other $\Oo(1)$ values of $n$ is similar. $\Lambda_0 = (2.4\pm 0.1) \times 10^{-3}\eV$. If $P_{vac} = 10^{-3}{\rm torr}$ then for both choices of potential, the Grenoble experiment could detect all $n=1$ theories with $\Lambda = 2.4 \times 10^{8}$ and $10^{7}{\rm GeV} \lesssim M \lesssim 10^{16}{\rm GeV}$. If $P_{vac}$ is lowered to $10^{7}{\rm torr}$, then values of $M$ smaller than $10^{4} {\rm GeV}$ could be detected.     }
\label{fig:genml}
\end{center}
\end{figure}

Constant forces are not detected by this experiment. We define $m_c$
to be the chameleon mass inside the test masses. If $m_c d \ll 1$ for
$5\mu{\rm m} \leq d \leq 10\mu{\rm m}$ then the force between the
plates will be virtually constant, and therefore undetectable.
Additionally if $m_b d \gg 1$, the chameleonic force will be
undetectably small.

Figure \ref{fig:genml} shows the region of
$M$-$\Lambda$ parameter space that the new Grenoble experiment could
potentially detect for $n=1$ and two choices for the vacuum pressure,
$P_{vac}$.  We have taken $V=\Lambda_0(1+\Lambda^n/\phi^n)$ and
$n=1$. The picture is very similar for other $\Oo(1)$ values of $n$
and for $V=\Lambda_0^4\exp(\Lambda^n/\phi^n)$.  It is clear that the
better the vacuum pressure, the smaller the values of $M$ that can be
detected.  If $P_{vac} = 10^{-3}{\rm torr}$ then for both choices of
potential, the Grenoble experiment could detect all $n=1$ theories
with $\Lambda = 2.4 \times 10^{8}$ and $10^{7}{\rm GeV} \lesssim M
\lesssim 10^{16}{\rm GeV}$. If $P_{vac}$ is lowered to $10^{7}{\rm
torr}$, then values of $M$ smaller than $10^{4} {\rm GeV}$ could be
detected.

Since values of $M$ smaller than $10^{4}\,{\rm GeV}$ are not generally
consistent with particle physics \cite{chamPVLAS}, a $10^{7}{\rm
torr}$ vacuum would be sufficient to detect or rule out all chameleon
theories with $V=\Lambda^4_0(1+\Lambda^{n}/\phi)$ or
$V=\Lambda^4_0\exp(\Lambda^n/\phi^n)$ and $\Lambda = \Lambda_0 =
(2.4\pm 0.1) \times 10^{-3} {\rm eV}$ and $M \lesssim 10^{16}\,{\rm
GeV}$.

Provided the laboratory vacuum is of sufficient quality, the Grenoble
experiment proposed by Lambrecht \emph{et al.} \cite{IILnew} has the
potential to detect or rule out almost all of the strongly coupled
chameleon theories with $\Lambda = \Lambda_0 = (2.4\pm 0.1)\times
10^{-3}{\rm eV}$ consider here.  Before this could be done however,
the controversy that surrounds the correct method for calculating the
thermal Casimir force would have to be settled, and the total Casimir
force modeled theoretically to an accuracy of better than $1\%$ at $d
\approx 5 - 10\mum$.

\subsection{New Lamoreaux Experiment}\label{Lamsec}

\indent From Figures \ref{fig:req2}c and \ref{fig:req3}c it is
clear that the chameleonic force predicted by $n>0$ theories with
either $V=\Lambda^4_0(1 +\Lambda^{n}/\phi^n)$ or $V=\Lambda^4_0
\exp(\Lambda^n/\phi^n)$ and $\Lambda \approx \Lambda_0 = 2.4
\times 10^{-3} \eV$ dominates over the expected thermal correction
when $d \approx 70\,\mu{\rm m} - 300\,\mu{\rm m}$.  At such large
separations the chameleonic force between two parallel plates is
more than 100 times the size of the zero-point Casimir force.  If
one wished to detect or rule out chameleonic forces but to avoid
the controversy that surrounds the thermal Casimir force, one
could therefore search for new forces at separations where the
chameleonic force dominates over the total Casimir force.

\begin{figure}[tbh]
\begin{center}
\includegraphics[width=8.8cm]{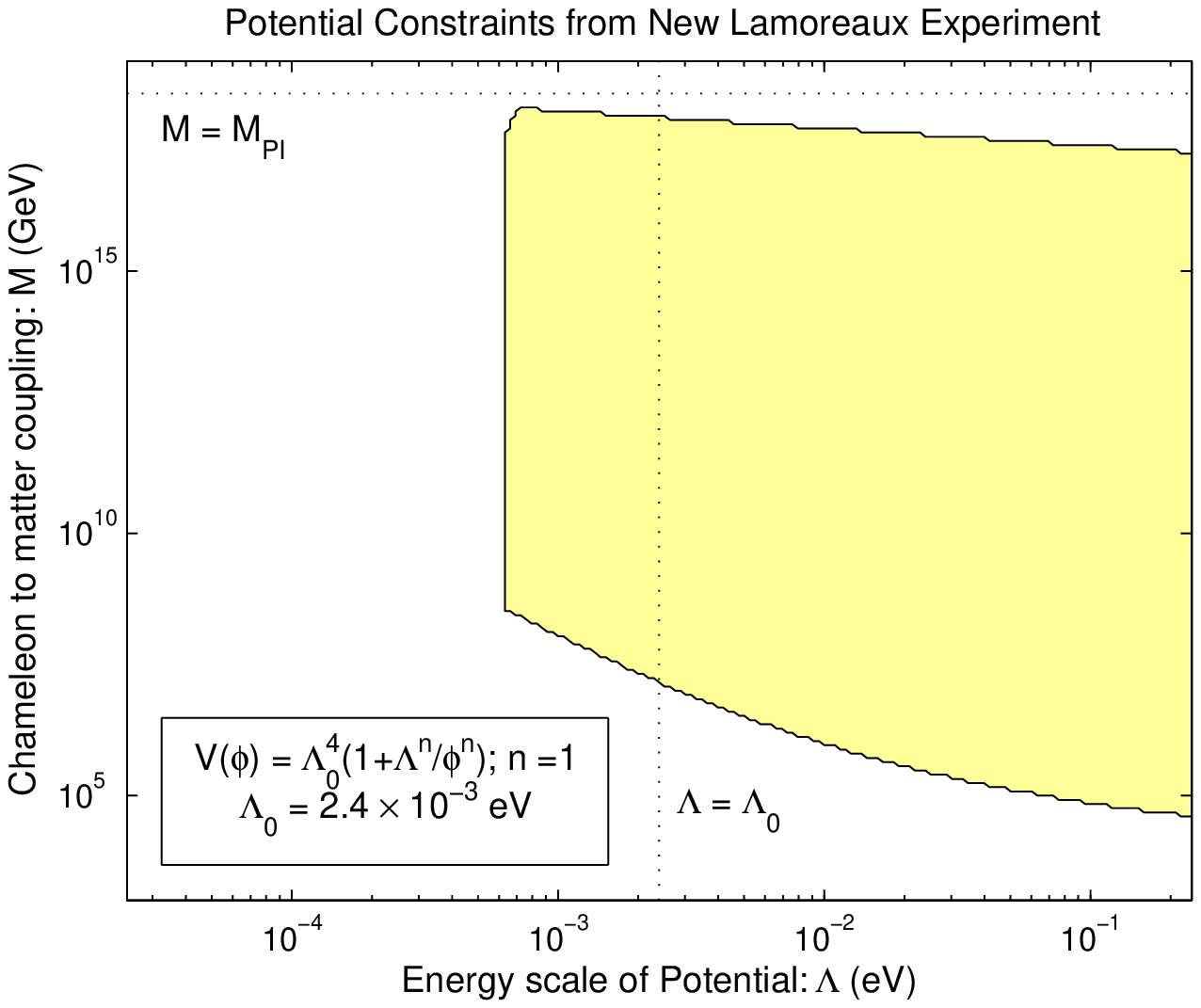}
\includegraphics[width=8.8cm]{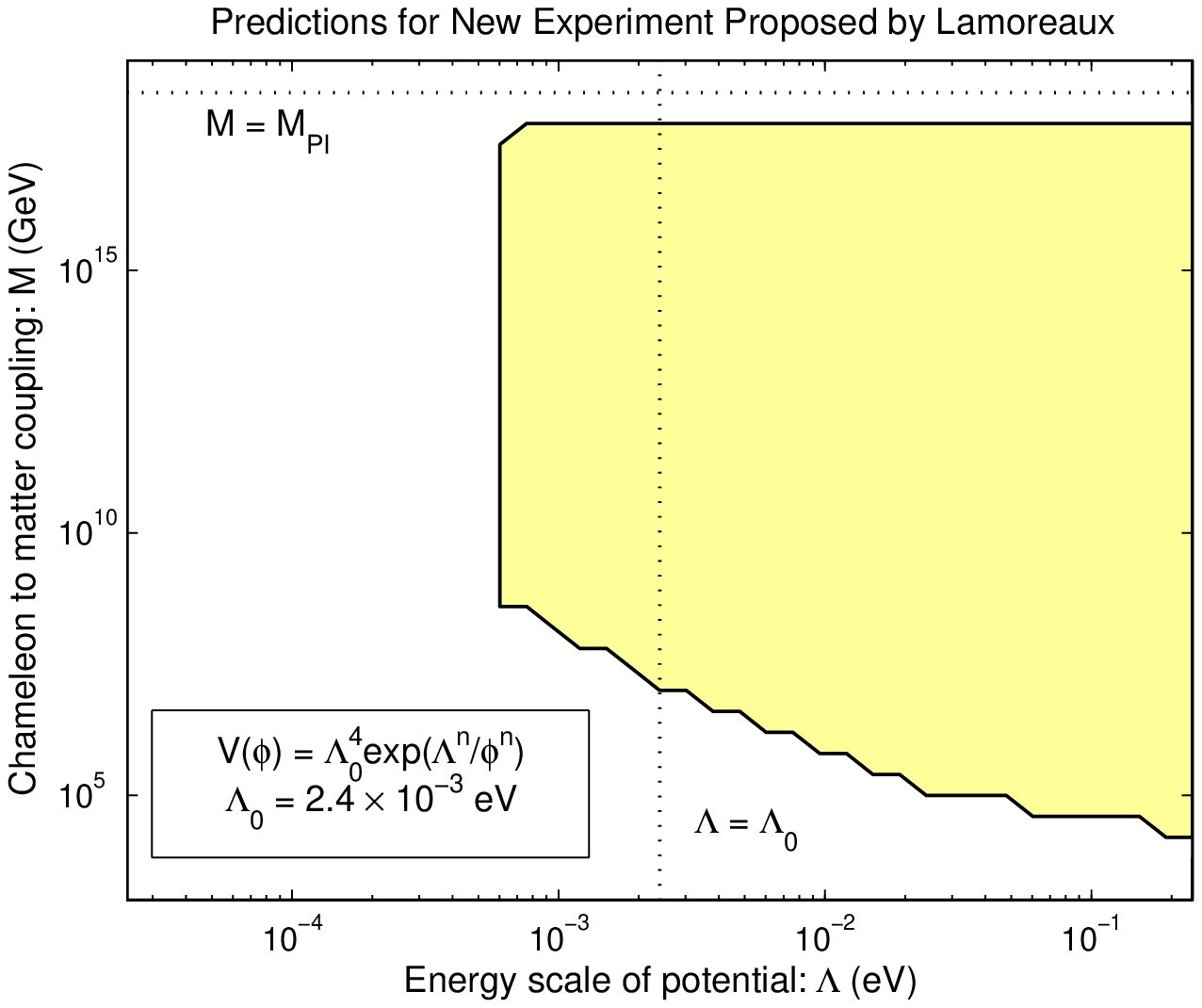}
\caption{Shaded area is the region of $M$-$\Lambda$ parameter space that could potentially be detected or ruled out the new experiment proposed by Lamoreaux \cite{lampriv, lamimprov} i.e. $\Delta F_{\phi}^{\rm tot}(33\mum) = F_{\phi}^{\rm tot}(33\mum)-F_{\phi}^{\rm tot}(d_{\rm cal} = 1\,{\rm cm}) >  0.1$~pN. The plot on the left is for $V=\Lambda^4_0(1 + \Lambda^{n})/\phi^n$, $\Lambda_0 = 2.4 \times 10^{-3}\eV$ and $n=1$, and the one on the right is for $V=\Lambda^4_0\exp(\Lambda^n/\phi^n)$, $\Lambda_0 = 2.4 \times 10^{-3}\eV$ and $n=1$.  The graphs for other values of $n$ close to $1$ are similar.}
\label{fig:lamLM}
\end{center}
\end{figure}

Lamoreaux has proposed a number of improvements to his 1997 torsion
balance experiment \cite{lampriv, lamimprov}. The new experiment would
be stable for separations as large as $1\,{\rm cm}$ \cite{lampriv}.
The improved experiment would make use of one flat plate and a curved
one with radius of curvature $3-4$\, {\rm cm}\cite{lampriv}.  The
pressure of the vacuum is intended to be $5 \times 10^{-7}\,{\rm
torr}$ \cite{lampriv}.  Ultimately, this new experiment should be able
to detect changes in forces that are as small as about $0.1{\rm pN}$
\cite{lampriv}.  We assume that the experiment is electrostatically
calibrated at $d_{\rm cal} \approx 1\,{\rm cm}$, and that the
chameleonic force can be distinguished from any residual electrostatic
forces (which would behave as $1/d$ and $1/d^2$).

We define $\Delta F_{\phi}^{\rm tot}(d) = F_{\phi}^{\rm tot}(d) -
F_{\phi}^{\rm tot}(1\,{\rm cm})$. The total Casimir force is less than
$0.05\,{\rm pN}$ for $d \gtrsim 33\,\mu{\rm m}$.  Therefore, even
without a precise knowledge of the thermal corrections to the Casimir
force, the new test proposed by Lamoreaux should be able to detect the
chameleonic force provided $\Delta F_{\phi}^{\rm tot}(d = 33\,\mu{\rm
m}) > 0.1\,{\rm pN}$.

Figure \ref{fig:lamLM} shows the region of the
$M-\Lambda$ parameter space of $n=1$ chameleon theories that the new
Lamoreaux experiment will be able to detect or rule out
i.e. $\Delta F_{\phi}$ i.e. the values of $M$ and $\Lambda$
for which $\Delta F_{\phi}(d =33\,\mu{\rm m})>0.1\,{\rm pN}$. We have plotted
for both $V = \Lambda^4_0(1+\Lambda^n/\phi^n)$ and
$V=\Lambda_0^4 \exp(\Lambda^n/\phi^n)$. We note that for $n=1$, this new test
would be able to detect $10^{7}\,{\rm GeV} < M < 10^{18}\,{\rm GeV}$ for
$\Lambda = \Lambda_0$.  Furthermore, it would be sensitive to values of
$\Lambda$ as small as $7 \times 10^{-4}\,{\rm eV}$.   With either potential,
for $n> 0$, the chameleonic force between a sphere and a
plate drops more slowly than $1/d$.

\begin{figure}[tbh]
\begin{center}
\includegraphics[width=8.8cm]{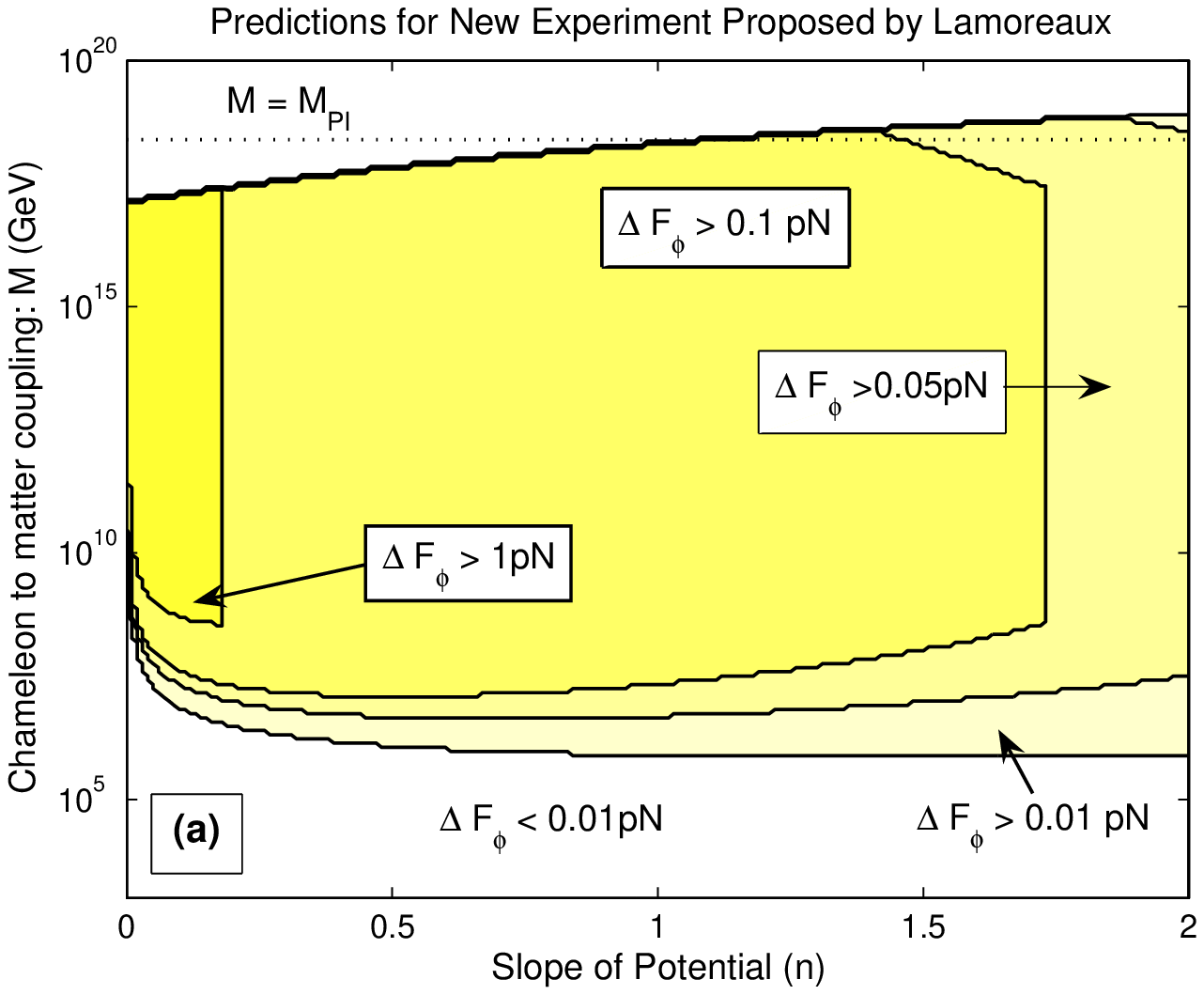}
\includegraphics [width=8.8cm]{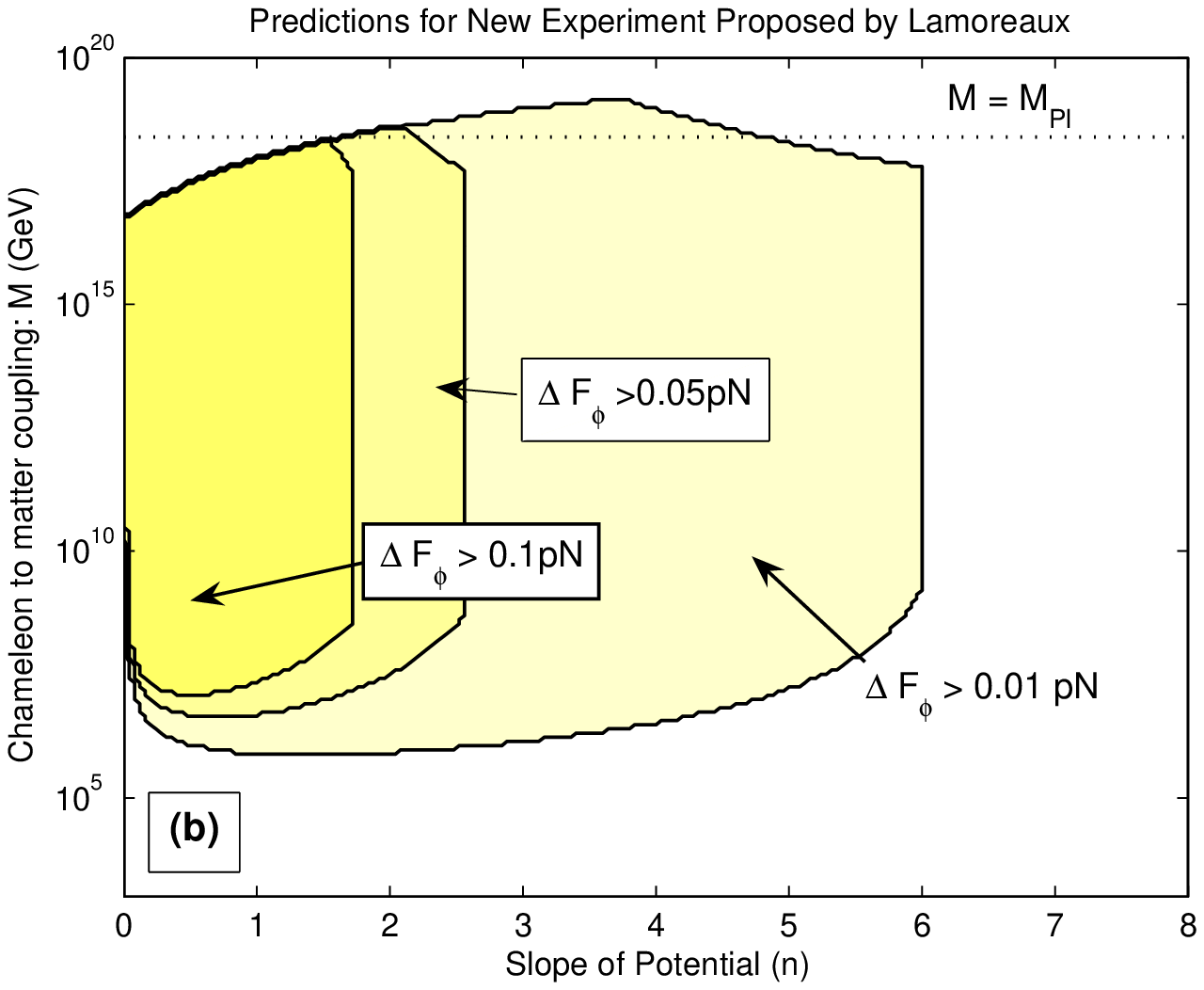}
\caption{Shaded areas show the contours of the predicted values
$\Delta F_{\phi}^{\rm tot}(82\mum) = F_{\phi}^{\rm
tot}(82\mum)-F_{\phi}^{\rm tot}(d_{\rm cal} = 1\,{\rm cm})$ in the
new experiment proposed by Lamoreaux \cite{lampriv,lamimprov}.  We
have taken $V(\phi) = \Lambda_0^4(1+\Lambda^n/\phi^n)$ and
$\Lambda = \Lambda_0 = 2.4 \times 10^{-3}\eV$.  The experiment
should be sensitive to $\Delta F_{\phi}^{\rm tot} > 0.1\,{\rm
pN}$.  At this separation the total Casimir force is $< 7.6 \times
10^{-3}$~pN, and so, with improved sensitivity, it should
ultimately be able to distinguish chameleonic forces as $0.01$~pN
from the Casimir background without an incontravertible model for
the thermal contribution to the Casimir force.} \label{fig:lamnew}
\end{center}
\end{figure}

For $n=1$ and $\Lambda\approx \Lambda_0 = 2.4 \times 10^{-3}\eV$,
larger values of $M$ (i.e. smaller chameleon to matter couplings)
should be detectable by the new Lamoreaux experiment but not the new
Grenoble test \cite{IILnew}. However the new Grenoble test is
sensitive to smaller values of $\Lambda$.  For the Grenoble test to
detect $n=1$ chameleon fields, we found that one would first have to
be able to model the thermal Casimir force to an accuracy of better
than $1\%$. In the new Lamoreaux experiment, however, the chameleonic
force dominates over the total Casimir force when $d=33\,\mu{\rm m}$
and $n=1$, and so useful constraints could be derived \emph{without} a
detailed knowledge of the thermal contributions to the Casimir force.

In the Grenoble experiment, a relatively large range of $n$ could
ultimately be detected. If one wishes to avoid having to deal with
thermal corrections, however, this is not the case for the Lamoreaux
experiment.  In Figure \ref{fig:lamnew} we show the predicted
contours, in $n-M$ parameter space, of $\Delta
F_{\phi}(\Lambda_0^{-1}\approx 82\mum)$ with $\Lambda = \Lambda_0$. We
have only plotted $\Delta F_{\phi}$ contours for values of $n-M$ for
which the test masses have thin-shells. In the absence of a
thin-shell, the chameleon field behaves simply as a Yukawa field with
mass $m_b$. The $0.1$~pN contour represents the limit of detectability
for the new experiment proposed by Lamoreaux.  At a separation $d =
\Lambda_{0}^{-1} \approx 82\mum$, the total Casimir force (including
thermal correction) between the sphere and the plate is less than $7.6
\times 10^{-3}$~pN.  With the sensitivity of $0.1$~pN, the Lamoreaux
test would be unable, without an undisputed model for the thermal
Casimir force, to detect $n \lesssim 1.6$.  If the sensitivity of the
torsion balance could be improved by an order of magnitude, the
Lamoreaux would be capable of detecting or ruling out chameleon
theories with $n \lesssim 6$.  Note that, at this separation $\Delta
F_{\phi}^{\rm tot}$, for $n \leq -8$ theories would still be too small
to detect.  The Lamoreaux experiment is most sensitive to chameleon
theories with $n \sim \Oo(1)$; this is because $\dd F_{\phi}^{\rm
tot}/\dd\, d$ has a slow (but not too slow) drop-off as $d \rightarrow
\infty$ in these theories.

\section{Conclusions}\label{sec:con}

We have investigated the possibility of using experiments which measure the Casimir force to constrain
theories where a scalar field is coupled to matter and where there is a chameleon mechanism. Our
primary aim was to extract the bounds that Casimir force tests currently place on chameleon theories and to make predictions for what near future Casimir experiments will be able to detect.

Chameleon theories are particular interesting because, for certain choices of potential, they can be agent responsible for the late-time acceleration.  For chameleonic dark energy, one must generally require a potential of the form $V = \Lambda_0^4 F(\phi/\Lambda)$ where $\Lambda_0 = (2.4\pm 0.1) \times 10^{-3}\eV$, $F = 1$ cosmologically today and $F^{\prime}(1), F^{\prime \prime}(1) \sim O(1)$ sets the scale of $\Lambda$.  To the best of our knowledge,  chameleon models of dark energy almost always feature at least one scale energy scale, $\Lambda_0$, and so do not in themselves alleviate the fine-tuning problems associated with dark energy.  If these models are to be seen as in some sense `natural' then one would probably have to postulate the existence of new physics at energy scales of the order of $\Lambda_0$.  It is clear from this and previous works  that chameleon theories with $\Lambda \approx \Lambda_0 = (2.4\pm 0.1) \times 10^{-3}\eV$ could have remained undetected thus far are \emph{not} generally ruled out.  This, combined with the lack of a generally accepted solution to the naturalness problem of the dark energy scale, means that we should not discount the possibility that there really is some new physics associated with the $\meV$ scale.  Albeit without any obvious connection to chameleon fields, this possibility has also be raised in the context of the super-symmetric large energy dimensions (SLED) proposal \cite{SLED}.  What makes the SLED proposal and chameleonic dark energy model so interesting is that, unlike so many other explanations of dark energy, they make predictions that are eminently testable and falsifiable by near future laboratory experiments.   If all we ever learn about dark energy comes from astronomical observations then cosmic variance alone means it may be difficult to ever fully understand its behaviour and its origins.  If dark energy can be detected under the controlled conditions of laboratory experiments, however, then the prospects for understanding it could potentially be much better.

In this work, we have aimed to remain as general as possible in our treatment of chameleon theories although when specific predictions have been required we have taken the potential to be of the form of either $V(\phi) = V_1 = \Lambda^4_0\left(1 + \frac{\Lambda^{n}}{\phi^n}\right)$ or $V(\phi) = V_2 = \Lambda_0^4 \exp(\Lambda^n/\phi^n)$.  For different values of $n$ and $\Lambda$ these potentials were shown to result in a wide range of different predictions for Casimir force experiments. Additionally, one can generally think of a more general potential  as behaving locally as one of these two potentials for some $n$, $\Lambda$ and $\Lambda_0$. These choices for the form of $V$ are therefore very handy for understanding the extent to which Casimir force experiments constrain general chameleon theories.

We found that the magnitude of the chameleonic force that would be detectable by Casimir force measurement, generally constrains $\Lambda \lesssim \Oo(100)\Lambda_0 = (2.4\pm 0.1) \times 10^{-3}\eV$. Although for very shallow potential larger values of $\Lambda$ are still allowed, we still found that generally both $\Lambda$ and $\Lambda_0$ must be small in experimentally viable chameleonic dark energy models.  If one small energy scale in the Universe seems undesirable then two unrelated small energy scales is at least doubly so. It therefore seems natural that $\Lambda$ and $\Lambda_0$ be related i.e. $\Lambda \approx \Lambda_0$. We found that the chameleonic force predicted by theories with $\Lambda = \Lambda_0 = 2.4 \times 10^{-3}\eV$  and $V=V_1$ and $n > 0$, or $n < -10$ or $V=V_2$ with any $n$, lies well below the present days experimental limits set by measurements of the Casimir force.  However, this is not the case for theories with $V=V_1$ and $n = -4$ or $n = -6$ (and $\Lambda = \Lambda_0$) which are strongly ruled out by the latest
$95\%$ confidence limits found by Decca et al. \cite{Decca3}. Additionally, the prediction for $V=V_1$, $n= -8$ theory  lies close to the edge of what is currently allowed.  In all allowed theories, the chameleonic force between two parallel plates with separation $d$ increases more slowly than $1/d^3$ as $d \rightarrow 0$.

Casimir force experiments are generally more sensitive to large
gradients in forces than they are to large but slowly varying
forces.  As a result, there are only able to place constraints on
Yukawa fields with mass, $m_{\phi}$, if they probe
separations,$d$, such that $m_{\phi} d \sim \Oo(1)$.  If the test
masses used in these tests do not have thin-shells, then the
chameleon field behaves as a Yukawa field with mass $m_b$, where
$m_b$ is the chameleon mass in the background. If we take $m_c \gg
m_b$ to be the chameleon mass deep inside the thin-shelled test
masses, then for a large range of separations $m_c^{-1} \lesssim d
\lesssim m_b^{-1}$ we found that the chameleon force increased as
$d \rightarrow 0$ like some inverse power of $d$; crucially this
is  a much steeper variation of the force with $d$ than is
predicted for a Yukawa field.  For $m_c d\lesssim 1$, however, the
chameleonic force between thin-shelled bodies depends only very
weakly on $d$ which makes it difficult to detect. If $m_b d \gg 1$
then the chameleonic force is exponentially attenuated. Casimir
force measurements therefore provide the strongest constraints on
chameleon theories for which $m_c^{-1} \lesssim d \lesssim
m_b^{-1}$ which limits the range of chameleon to matter couplings
that they can detect.  Specifically, if $\Lambda = \Lambda_0$ and
the matter coupling has gravitational strength i.e. $M \sim
\Oo(M_{\rm Pl})$ then it is generally the case that $m_c d
\lesssim 1$ and / or the test masses do not have thin-shells.  As
a result, we saw that Casimir force tests are best suited to
searching for and limiting the properties of strongly coupled
chameleon fields i.e. $M_{\rm Pl}/M \gg 1$. This is useful, as it
is precisely these strongly coupled theories that are the most
difficult to constrain \cite{chamstrong}.   If $M \gtrsim
\Oo(M_{\rm Pl})$ then excellent constrains on chameleon fields
come from laboratory tests of the inverse square law (ISL) such as
the E\"{o}t-Wash experiment \cite{EotWash}. Since the bounds on
chameleon theories coming from the E\"{o}t-Wash experiment have
been discussed in great detail elsewhere \cite{chamKA,chamstrong,
EotWash} we have not focussed on them here, although it should be
noted that many of the formulae derived in this work are also
useful for estimating the size of any chameleonic signal that
could be detected by that test.  It should be noted, however, that
the electrostatic shielding used by such tests, however, also acts
as a chameleon force shield if the matter is coupled to the
chameleon much more strongly than it is to gravity.  Casimir force
measurements and ISL tests therefore probe different values of the
chameleon to matter coupling and taken together place the
strongest constrains on chameleon fields with a gravitational or
super-gravitational strength matter coupling.

For $m_c^{-1} \lesssim d \lesssim m_b^{-1}$ and $V=V_1$ or $V=V_2$
and $\Lambda \approx \Lambda_0$ we found the chameleon force
between two parallel plates was generally about $\Oo(1)\%$ of the
size of the total Casimir force at separations of $d\approx
10\mum$.  At larger separations, the dominant contribution to the
Casimir force is expected to come from thermal effects the precise
form of which is a matter of some controversy.  Even still, in the
parallel plate geometry, the chameleonic and total Casimir forces
were found to be generally equal in magnitude for $\Lambda_d d
\sim \Oo(1)$ where $\Lambda_d = \Lambda_0^2/\Lambda \approx
82\mum$ for $\Lambda \approx \Lambda_0$.  In the sphere plate
geometry the situation is slightly more complicated but generally
the chameleonic force dominates for smaller values of $d$.

For $\Lambda \approx \Lambda_0$, the predicted chameleonic force per unit surface area is very small.  It is therefore desirable to use larger rather than smaller test masses so as to increase the surface area and amplify any chameleonic force.
It is no surprise then that the some of the best current constraints on it are provided by the 1997 Lamoreaux experiment \cite{lam97}, in which test masses with relatively large
dimensions  were used.  The relativity large dimensions of the test masses also means that they have thin-shells for a larger
range of couplings and so this experiment was sensitive to a larger range of coupling than were the tests reported in Refs. \cite{MohRoy} and \cite{Decca1} where smaller test masses were used.

We identified two future experiments with experiment with excellent prospects for detecting or ruling out chameleon fields associated with the energy scale of dark energy. Both experiments share some common features: they both make use of relatively large, $1-10{\rm cm}$, test masses, and both probe larger separations than previous tests. Additionally both experiments make us of torsion balances to measure the forces.  The experiments in question the new tests proposed by Lambrecht \emph{et al.} \cite{IILnew} and under construction in Grenoble and the new Lamoreaux experiment \cite{lampriv, lamimprov}. We found that these tests are complementary to each other. Provided the laboratory vacuum is of sufficient quality, the Grenoble experiment \cite{IILnew} has the sensitivity to detect or rule out almost all of the \emph{strongly} coupled chameleon theories which are at the same time dark energy candidates ($\Lambda = \Lambda_0 = (2.4\pm 0.1)\times
10^{-3}{\rm eV}$.  However, before this could be done, the controversy that surrounds the correct method for calculating the
thermal Casimir force would have to be settled, and the total Casimir force modelled theoretically to an accuracy of better than $1\%$ at distances of $\approx 5 - 10\mum$.  The Grenoble experiment has, in fact, be designed to help settle this controversy and so it is highly feasible that in the longer term they will be able to detect or rule out chameleon fields.

The new experiment proposed by Lamoreaux \cite{lampriv, lamimprov} could be used to probe even larger separations than the Grenoble test.  For $n \sim O(1)$, $\Lambda \approx \Lambda_0$ and $V=V_1$ or $V=V_2$, we predicted that the chameleonic force would dominate over the total Casimir force when $d \gtrsim 33\,\mu{\rm m}$.  By probing separations slightly larger than this, the new test is predicted to have the sensitivity to detect chameleon fields with $0 \lesssim n \lesssim 1.6$ for a larger range of $M$ than could the Grenoble experiment; although the Grenoble experiment sees a much larger range of $M$.  Crucially though, since the chameleonic force would dominate over the Casimir force at these separations, the new test proposed by Lamoreaux could make such a detection \emph{without} a detailed knowledge of the thermal contributions to the Casimir force.

For $m_c^{-1} \lesssim d \lesssim m_b^{-1}$ the chameleonic force that is detectable by Casimir force experiments depends is virtually independent of the strength with which the chameleon couples to matter, $M$, as well as the density and composition of the test masses.  This, of course, means that whilst Casimir force experiments are excellent probes of the properties of the chameleon potential e.g. $n$, $\Lambda$ and $\Lambda_0$, they are unable to provide much information about $M$.  By a happy coincidence, however, the converse is true of experiments that search for the conversion of photons into chameleons, e.g. BRFT, PVLAS, BMV \cite{axion}.  In the region of peak sensitivity the effects detectable by those tests depend only indirectly and as a result fairly weakly on $V(\phi)$ but have a very strong dependence on $M$ \cite{chamPVLAS,chamPVLASlong}.  Casimir force measurements and light propagation experiments therefore provide complementary constraints on the properties of chameleons.

In summary:  Tests of gravity generally assume that any new forces
behave a lot like gravity: exhibiting a similar dependence on the
density of test masses and being difficult to shield.  The force
mediated by chameleon fields obeys neither of these assumptions,
and it is partly for this reason that chameleon fields are so
difficult to detect or rule out using traditional tests of
gravity. We have found that Casimir force measurements already
provide the strongest constraints on chameleons fields with strong
matter couplings.  However, most chameleons theories where the
potential is associated with the energy scale of dark energy, $2.4
\times 10^{-3}\eV$, remain hidden from these tests.  This
underlines just how little we know about potential new physics
associated with this small energy scale. The next generation of
Casimir force tests however will offer greatly increased sensitive
and will probe larger separations where the effect of chameleonic
force relative to the Casimir force is more pronounced. Provided
the controversy surrounding the thermal Casimir force can be
settle, these experiments will have the sensitivity to detect or
rule out almost all strongly coupled chameleon fields which are
dark energy candidates.  The prospect of detecting new physics,
such as chameleon fields associated with the $\meV$ scale of dark
energy, in the laboratory is an exciting one.  If no such
detection is made then the potentials of viable chameleon theories
would have to feature an energy scale $\ll \Lambda_0$ which would
seem highly unnatural.   Whatever the outcome, by probing
separations between $10\mum$ and $100\mum$ the new generation of
Casimir force measurements have the potential to increase what is
currently known about the nature and origins of dark energy.

\acknowledgments

We are grateful to P. Antonini, G. Hammond, D. Krause, S.
Lamoreaux, R. Onofrio and for sharing information about the
experiments, and we also thank H. Gies for helpful comments.  CvdB
and ACD are supported partly by PPARC. DFM is supported by the
Alexander von Humboldt Foundation. DJS is supported by PPARC. PhB
acknowledges support from RTN European programme
MRN-CT-2004-503369.

\appendix

\section{Generalized Thin-Shell Conditions}\label{appA}

In this appendix, we derive the condition for a body to have a thin-shell in a chameleon theory with general $V(\phi)$. Consider an isolated spherical body with density $\rho_c$,
radius $R$ in a background with density $\rho_b$. We take $r$ be the
distance from the centre of the body. The chameleon field evolves in the effective potential:
$$
V_{\rm eff}(\phi;\rho)  = V(\phi) + \frac{\phi}{M}\rho.
$$
We have taken the body to be isolated. By this we mean that the length scale, $L_b$, of the region where $\rho = \rho_b$ is large enough for $\phi \approx \phi_b$ far from the body.  $\phi_b$ is defined to be the minimum of $V_{\rm eff}$ when $\rho = \rho_b$:
\begin{equation}
V^{\prime}(\phi_b) = -\frac{\rho_b}{M}.\label{phibApp}
\end{equation}
We also define $m_b = m_{\phi}(\phi_b) = \sqrt{V^{\prime \prime}(\phi_b)}$ to be the mass of the chameleon near $\phi = \phi_b$. In the region where $\rho=\rho_b$, perturbations in the chameleon field about $\phi_b$ decay exponentially over a length scale $\sim m_{b}^{-1}$.  The isolated condition therefore translates to requiring $m_b L_{b}  \gg 1$.

The statement that a body of density $\rho_c$ has a thin-shell is equivalent to requiring that $\phi \approx \phi_c$ deep inside the body, where $V^{\prime}_{\rm eff}(\phi_c;\rho_c) =0$ defines $\rho_c$ i.e.:
\begin{equation}
V^{\prime}(\phi_c) = -\frac{\rho_c}{M}.\label{phicApp}
\end{equation}
We define $m_c = m_{\phi}(\phi_c)$.

In this set-up, the chameleon field equation, Eq. (\ref{chameqn}), reduces to:
\begin{eqnarray}
\frac{\dd^2 \phi}{\dd r^2} + \frac{2}{r}\frac{\dd
\phi}{\dd r} = V^{\prime}(\phi) - V^{\prime}(\phi_{b}) \qquad r > R, \label{chamout}\\
\frac{\dd^2 \phi}{\dd r^2} + \frac{2}{r}\frac{\dd
\phi}{\dd r} = V^{\prime}(\phi) - V^{\prime}(\phi_{c}) \qquad r < R, \label{chamin}
\end{eqnarray}
We define $\phi_0$ in $r > R$ by the equation:
\begin{equation}
\frac{\dd^2 \phi_0}{\dd r^2} + \frac{2}{r}\frac{\dd
\phi_0}{\dd r} = m_b^2(\phi_0 - \phi_b), \label{phi0out}
\end{equation}
If $\phi \approx \phi_0$ in $r >R$ then the $\phi$ field produced by body behaves as if it were a linear perturbation about $\phi = \phi_b$ in $r >R$.

We now prove that if $\phi \sim \phi_0$ as $r \rightarrow \infty$ and $\phi_0 < \phi_b$ then $\phi < \phi_0$ in $r > R$.
We take $\phi = \phi_0 + \phi_1$, and $\vert \phi_1/(\phi_0 - \phi_b)\vert \rightarrow 0$ as $r \rightarrow \infty$. We
wish to prove that $\phi_1 < 0$.  We write $\phi_1 = \chi e^{-m_b r}/r$ and then:
\begin{equation}
\left(e^{-m_b r}\chi^{\prime}\right)^{\prime} = r\left\lbrace(V^{\prime}(\phi_0  + \phi_1) - V^{\prime}(\phi_b) - m_b^2(\phi_0 + \phi_1 -\phi_b)\right\rbrace.
\end{equation}
As $r \rightarrow \infty$ we certainly have $\phi < \phi_b$. We have required $V^{\prime \prime \prime}(\phi) < 0$ which implies that if $\phi < \phi_b$ then:
$$
V^{\prime}(\phi) - V^{\prime}(\phi_b) - m_b^2(\phi -\phi_b) < 0.
$$
It follows that as $r \rightarrow \infty$ we must have:
$$
\left(e^{-m_b r}\chi^{\prime}\right)^{\prime} < 0,
$$
and so since $e^{-m_b r} \chi^{\prime} \rightarrow 0$ by as $r \rightarrow \infty$ by the requirement that $\phi \sim \phi_0$ as $r \rightarrow \infty$, we must have for large $r$ that:
$$
\chi^{\prime}(r) > 0 \Rightarrow \chi(r) < 0 \Rightarrow \phi_1(r) < 0
$$
Thus $\phi_1(r)< 0$ provided that:
$$
V^{\prime}(\phi_0  + \phi_1(r)) - V^{\prime}(\phi_b) - m_b^2(\phi_0 + \phi_1(r) -\phi_b) < 0
$$
but since required $\phi_0 < \phi_b$ this certainly holds as $r \rightarrow \infty$ and continues to do so provided that$\phi_1(r) < 0$.  It follows $\phi_1(r) < 0$ and thus $\phi < \phi_0$ for all $r > R$. We will need this shortly.

A full solution of the field equations requires $\phi(r=R) \geq \phi_c$, since $V'(\phi)<0$. For a body to have a thin-shell, it must fulfil
$\phi \approx \phi_c$ inside the body which requires that all perturbations in $\phi$ about $\phi_c$ decay over a length scale
that is smaller than $R$. A necessary condition for this is $m_c R \gtrsim 1$. It is enough, however, to have $m(\phi(r=R))R \gg 1$ on the
surface of the body. Since $m(\phi(R))> m(\phi_0(R))$  (as $\phi(R)<\phi_0$ by continuity) then its enough to predict that $m(\phi_0(R))R \gg 1$.
If we assume that there is no thin-shell, then we would have to have $m(\phi_0(R))R \lesssim 1$, but if the assumption of no thin-shell leads us to
predicting $\phi_0(R) < \phi_c$ then because of $m_cR \gg 1$ this cannot be the case.

We now find a sufficient condition for the existence of a thin-shell, by assuming that the body does \emph{not} have a thin-shell and seeing when
this leads to a contradiction.  If a body does not have a thin-shell then in $r > R$:
\begin{equation}
\phi \approx \phi_{0}(r) \equiv \phi_b - \frac{{\cal C} R e^{-m_b (r-R)}}{r},
\end{equation}
for some constant ${\cal C}$. And in $r < R$ Eq. (\ref{chamin}) can be linearised to,
$$
\frac{\dd^2 \phi}{\dd r^2} + \frac{2}{r}\frac{\dd
\phi}{\dd r} = m_b^2 (\phi - \phi_b) + \frac{\rho_c - \rho_b}{M}:
$$
and so in $r <R$ we would have:
\begin{equation}
\phi \approx \phi_b - \frac{\delta\rho_c}{Mm_b^2} + \left(\frac{\delta
\rho_c}{Mm_b^2} - {\cal C}\right)\frac{R \sinh(m_b r)}{r\sinh(m_b R)}, \label{phisolvein}
\end{equation}
where
\begin{eqnarray} {\cal C} &=& \frac{\delta\rho_c e^{-m_b
R}\cosh(m_bR)}{M m_b^2}\left[1-\frac{\tanh(m_b R)}{m_b R}\right] \label{Ceqn} \\
&=& \left(V^{\prime}(\phi_b)-V^{\prime}(\phi_c)\right)
\frac{e^{-m_bR}\cosh(m_bR)}{m_b^2}\left[1-\frac{\tanh(m_b R)}{m_b
R}\right], \nonumber
\end{eqnarray} and $\delta \rho_c =
\rho_c - \rho_b$.  However, if $\phi_0(r=R) = \phi_b - {\cal C} \leq \phi_c$ then we found that there certainly is a thin-shell, which leads to a contradiction.  A sufficient condition for the existence of a thin-shell is therefore:
\begin{equation}
{\cal C} \geq \phi_b - \phi_c.
\end{equation}

\subsection{Linear Thin-Shells}
Consider Eq. (\ref{phisolvein}) which gives the form of $\phi$ in $r <R$ in those cases where it is acceptable to linearize the field equation about $\phi = \phi_b$. If $m_b R \gg 1$ then all variation of $\phi$ inside dies off in a
thin region near the surface of the body (over thickness $\sim
1/m_b$).  Furthermore if this linearisation is correct then $\phi_c \approx \phi_b - \delta \rho_c / M m_c^2$, and so $\phi \approx \phi_c$ deep inside the body.  It follows from the definition of a thin-shell that such a body would have one. However, this type of thin-shell behaviour also exists in Yukawa scalar field theories where $m_{\phi} = {\rm const}$; it has nothing to do with the non-linear nature of the field equation which give the theory its chameleonic properties.  We therefore deem this to be a \emph{linear thin-shell}.

If $m_b R\gg 1$ then a body will have a linear-thin shell provided $\phi_c \approx \phi_c$ which implies $m_c \approx m_b$. Outside of a body with a linear thin-shell, $\phi$ has the form:
\begin{equation*}
\phi \approx \phi_b - \frac{{\cal C}R e^{-m_b (r-R)}}{r},
\end{equation*}
where ${\cal C}$ is given by Eq. (\ref{Ceqn}).  Since ${\cal C}$ depends on $\delta \rho_c/M$ it is clear that the chameleonic field far from the body depends on both the density of the body and the strength with which the chameleon couples to it.  This is precisely what one na\"{i}vely expects to see in scalar field theories with a coupling to matter. This behaviour is, however, very much associated with linear field equations and it is \emph{not}, as we shall see, what one finds when the non-linear nature of the chameleon field equation dominates the behaviour of the field.

\subsection{Non-Linear Thin-Shells}
In addition to linear thin-shells,
which exist even in theories with a simple Yukawa scalar field theories,
chameleon theories also exhibit non-linear thin-shell behaviour. If a
body has a thin-shell of any description then all variation in $\phi$
is exponentially small outside a thin region near the surface of the
body.  As was shown in ref. \cite{chamstrong} for $V
\propto \phi^{-n}$ theories, when a body has a non-linear thin-shell, the chameleon field far from the body is independent of $\delta\rho_c/M$.  This is one of the key
features of chameleon theories that allows models with $M \ll M_{Pl}$
to evade the constraints coming from experimental tests of gravity and
searches for WEP violation \cite{chamstrong}.

Outside the body, $\phi$ obeys Eq. (\ref{chamout}) and by integrating this once we find:
\begin{eqnarray}
\frac{1}{2}\left(\frac{\dd \phi}{\dd r}\right)^{2} &=& V(\phi) - V(\phi_b) - V^{\prime}(\phi_b)(\phi-\phi_b) + \int_{r}^{\infty} \frac{2}{x} \left(\frac{\dd \phi}{\dd x}\right)^{2} \dd x,
\end{eqnarray}
In $r > R$, we have $\dd \phi / \dd r > 0$ and so defining $d = r-R$,
$$
y^2=\frac{1}{2}\left(\frac{\dd \phi}{\dd r}\right)^2$$
 and $y_s = y(r=R)$ we have:
\begin{equation}
x = \int_{y(x)}^{y_s} W(z)\dd z, \label{W1thin}
\end{equation}
where
$$
W(y) = (V_{b}^{\prime}-V^{\prime} + 2\sqrt{2}y/r)^{-1}.
$$
Defining
$$
\chi^2 = \int_{r}^{\infty} \frac{2}{x} \left(\frac{\dd \phi}{\dd x}\right)^{2} \dd x,
$$
it is useful to write:
\begin{equation}
W(y)^{-1} = \left[V_{b}^{\prime}-V^{\prime} - m_b\sqrt{2(y^2-\chi^2)}\right] + \left[m_b\sqrt{2(y^2-\chi^2)} + \frac{2\sqrt{2}y}{r}\right].
\end{equation}
If $V = V_b + V_b^{\prime}(\phi-\phi_b) + \frac{1}{2}m_b^2
(\phi-\phi_b)^2$ then the terms in the first set of square
brackets would vanish (note that $\phi_b>\phi$); these terms are
therefore associated with the non-linear nature of the field
equations.  Whatever the potential is, the terms in the second set
of square brackets are manifestly increasing more slowly than $y$
as $y \rightarrow \infty$.  From the definition of $y$, it is
clear that if near the surface of the body:
$$
V_{b}^{\prime}-V^{\prime} \gg \frac{2\sqrt{2}y}{r} = \frac{2}{r}\frac{\dd \phi}{\dd r},
$$
then $\phi$ is varying over scales that are small compared with $r$, and so we have thin-shell behaviour of some description.  If additionally:
\begin{equation}
\left[V_{b}^{\prime}-V^{\prime} - m_b\sqrt{2(y^2-\chi^2)}\right] \gg \left[m_b\sqrt{2(y^2-\chi^2)} + \frac{2\sqrt{2}y}{r}\right], \label{thincc1}
\end{equation}
near the surface of the body then the non-linear terms in the
field equations  dominate the behaviour of $\phi$. This is
therefore the condition for a non-linear thin-shell.

If the non-linear thin-shell condition holds then $\phi$ is a quickly varying function near $r=R$.
Let us define $\delta R \ll R$ to be the scale over which $\dd \phi/\dd r$ varies near $r=R$.
We then have $\chi^2 \sim \Oo(\Delta r/R) y^2$ and so $\chi^2 \ll y^2$ near the surface of a body
for which Eq. (\ref{thincc1}) holds. For $r \ll R$ we then have:
\begin{eqnarray}
V_{b}^{\prime}-V^{\prime} - m_b\sqrt{2(y^2-\chi^2)} \approx Q(y) = V_b^{\prime} -V^{\prime} - m_b \sqrt{2}y, \\
y^2 \approx \bar{y}^2 = V(\phi) - V(\phi_b) - V^{\prime}(\phi_b)(\phi-\phi_b),
\end{eqnarray}
which equality as $y \rightarrow \infty$.  It can then be checked that $V^{\prime \prime \prime}< 0$ implies that:
$$
\frac{\dd \ln Q}{\dd \ln y} > 1,
$$
and so $Q$ increases always increases faster than $y$.  Thus when
Eq. (\ref{thincc1}) holds, $W(y) \rightarrow 0$ faster than $1/y$
as $y$ increases. Thus not only does the integral in Eq.
(\ref{W1thin}) converges as $y_s \rightarrow \infty$, but this
should additionally provide a very good approximation to $y(x)$
whenever $y(x) \ll y_s$ and Eq. (\ref{thincc1}) holds for $y=y_s$.
The far chameleon field perturbation produces by bodies for which
Eq. (\ref{thincc1}) holds near their surface is therefore almost
independent of $y_s$.  It is for this reason that the perturbation
in the chameleon field far from a body with a non-linear
thin-shell is virtually independent of the value of $\phi$ on the
surface of the body and, as a result, of $\rho_c/M$.

We now return to the non-linear thin shell condition, Eq. (\ref{thincc1}), and rephrase it in a more useful form.  We begin by assuming that a body does not have a thin-shell and hence that $\phi \sim \phi_0(r)$ in $r>R$. We then look to see when this assumption leads to a contradiction.  The assumption that there is no thin-shell and $\phi \sim \phi_0$ implies, by Eq. (\ref{thincc1}), that at $r=R$:
\begin{equation}
V^{\prime}(\phi_b)-V^{\prime}(\phi_b-{\cal C}) \lesssim \frac{2{\cal C}}{R^2}\left(1+m_b R +m_b^2 R^2\right).\label{nothinreq2}
\end{equation}
Using Eq. (\ref{Ceqn}) to give ${\cal C}$ for a non-thin shelled body, we find that Eq. (\ref{nothinreq2}) is violated,
implying that there must be a non-linear thin-shell, when:
$$
\frac{V^{\prime}(\phi_b - {\cal
C})-V^{\prime}(\phi_b)}{V^{\prime}(\phi_c)-V^{\prime}(\phi_b)} \gtrsim
f(m_b R),
$$
where
$$
f(m_b R) = 2e^{-m_b R} \cosh(m_b R)\left[1+ \frac{1}{m_b R} +
\frac{1}{m_b^2 R^2}\right]\left(1-\frac{\tanh m_b R}{m_b R}\right).
$$
If $m_b R \ll 1$, $f(m_b R) \sim \frac{2}{3}$ and if $m_b R \gg 1$
then $f(m_b R) \sim 1$; $f(m_b)$ is monotonic in $m_b R$.  The
non-linear thin-shell condition (or the thin-shell condition for
short) is therefore approximately equivalent to:
\begin{equation} {\cal C} = \frac{(\rho_c-\rho_b) f(m_b R) R^2}{2
M[m_b^2R^2 + m_bR+1]}\gtrsim \phi_b - \phi_c -
\frac{(\rho_{c}-\rho_b)(1-f(m_b R))}{M m_c^2}.
\end{equation} or
\begin{equation} \frac{m_c^2 R^2}{m_b^2R^2 + m_bR+1} \gtrsim \frac{2M
m_c^2\left(\phi_b -\phi_c\right)}{(\rho_c-\rho_b) f(m_b R)} -
\frac{2(1-f(m_b R))}{f(m_b R)} \geq 2.
\end{equation}

These conditions could, alternatively, be written as ${\cal C} > {\cal C}_{\rm thin}$ where
$$
\frac{{\cal C}_{\rm thin}(1 + m_b R + m_b^2 R^2)}{V^{\prime}(\phi_b) -
V^{\prime}(\phi_b-{\cal C}_{\rm thin})} = \frac{R^2}{2}.
$$
It is clear from its definition that the ${\cal C}_{\rm thin}$ depends only on $R$, $m_b$ and
the form of $V(\phi)$. It is independent of $\rho_c/M$.  Since, in a
thin-shelled body, almost all variation in $\phi$ takes places in a
thin region near the surface of the body, the non-linear terms in the
field equation should only be important over scales $\ll R$.  Far from a thin-shelled body, $\phi$ should therefore behave if the field equations were linear i.e.:
$$
\phi \sim \phi_b - \frac{{\cal C}^{\ast}Re^{m_b(R-r)}}{r},
$$
for some ${\cal C}^{\ast}$.  If ${\cal C}^{\ast} \ll {\cal C}_{\rm thin}$, then the thin-shell conditions would not be satisfied near the surface of the body, and if ${\cal C}^{\ast} \gg {\cal C}_{\rm thin}$ then by Eq. ({\ref{thincc1}) the non-linear terms in the field equation would begin to dominate far at $r \gg R$. It must therefore be the case that ${\cal C}^{\ast} \approx {\cal C}_{\rm thin}$ for thin-shelled bodies. As $r/R \rightarrow \infty$ then we have, for a thin-shelled body, that:
\begin{equation}
\phi \approx \phi_b - \frac{{\cal C}_{\rm thin}Re^{m_b(R-r)}}{r}.
\end{equation}
As should be the case, the large $r$ behaviour of $R$ is, to leading order, independent of $\rho_c/M$ and depends only on $m_b$, $R$ and, through ${\cal C}_{\rm thin}$ on the form of $V(\phi)$.

\section{Force between two parallel plates}\label{appB}
It was shown in section \ref{sec:force:plates} that the chameleonic force between two parallel plates with thin-shells and separation $d$ is given by:
\begin{equation}
\frac{F_{\phi}}{A} = V(\phi_0) - V(\phi_b) + V^{\prime}(\phi_b)(\phi_b - \phi_0) \leq V(\phi_0)-V(\phi_b),
\end{equation}
where (provided $d \gg 1/m_c$, where $m_c$ is the chameleon mass inside the plates) $\phi_0(d)$ is given by:
\begin{equation}
\sqrt{2} \int_{0}^{\infty} W(y;\phi_0)\dd y = \frac{d}{2}, \label{inteqnapp}
\end{equation}
where we have defined $y = \sqrt{V-V_{0} - V^{\prime}_b(\phi-\phi_{0})}$ and $1/W(y;\phi_0) = (V^{\prime}_b-V^{\prime}(\phi)) \geq 0$. We define $P(y)$ by:
$$
\frac{\dd \ln W}{\dd \ln y} = -\frac{2V^{\prime \prime}(V-V(\phi_{0})-V^{\prime}_{b}(\phi-\phi_0))}{(V^{\prime}(\phi)-V^{\prime}_b)^2} \equiv -P(y),
$$
As $y \rightarrow \infty$, the condition that $V^{\prime \prime
\prime} < 0$, implies that $P(y) > 1$ as $y \rightarrow \infty$.  We
define $y_{-1}$ by $P(y_{-1}) = 1$; the dominant contribution to the
integral in Eq. (\ref{inteqnapp}) then comes from $O(y_{-1})$ values
of $y$.  Defining $\phi_{-1}$ by $y(\phi_{-1}) = y_{-1}$, we find a
good estimate of $\phi_{-1}$ by expanding $P(y)$ about $y$ to order
$(\phi-\phi_0)^2$:
\begin{eqnarray} V^{\prime \prime \prime}_0(\phi_{-1}-\phi_{0})^2
\approx V^{\prime}_{0}-V^{\prime}_{b}.
\end{eqnarray} Now, linearization of the field equations about
$\phi_{0}$ is a good approximation provided that:
$$
\frac{V^{\prime \prime \prime}_{0}(\phi -\phi_{0})^2}{2
(V_{0}^{\prime}-V^{\prime}_{b})} \ll 1.
$$
The point $\phi = \phi_{-1}$ therefore lies close to the edge of the
region where the linear approximation is valid. Since this is the
case, we approximate the integral in Eq. (\ref{inteqnapp}) by
expanding $W(y)$ out about $y=0$ ($\phi=\phi_{0}$):
\begin{equation} (V^{\prime}_b - V^{\prime})^2 = 1/W^2(y) \sim
(V_b^{\prime}-V_{0}^{\prime})^2(1 + 2 a^2 y^2 + k^2 a^4 y^4
+O(y^6)). \label{Wapprox1}
\end{equation} where we have defined
$$
a^2 = \frac{m_{0}^2}{(V_{b}^{\prime}-V_{0}^{\prime})^2}, \qquad k^2 =
\frac{V_0^{\prime \prime \prime}(V_0^{\prime}-V_b^{\prime})}{m_0^4}.
$$
If $k^2 \leq 2$ then to $O(y^4)$ we can rewrite $1/W^2(y)$ as:
\begin{equation} \frac{1}{W(y)^2} \approx
(V_b^{\prime}-V_0^{\prime})^2\left[(1+c^2 y^2)^{2p} + O(y^6)\right],
\label{Wapprox2}
\end{equation} where $c^2 = a^2/p$ and $(2p-1)/p = k^2$, and $0 \leq
k^2 \leq 2$ implies $p\geq1/2$.

We define $n_{eff} = (2-k^2)/(k^2-1)$ so that $p=1+1/n_{\rm eff}$.
$W(y)$ drops off faster than $1/y$ for $y^4 > 1/k^2$.

The approximation of $W(y)$ given by Eq. (\ref{Wapprox2}) is therefore
approximately the same as that given by Eq. (\ref{Wapprox1}) for $y^4
\lesssim 1/k^2$, as long as the new $O(y^6)$ terms introduced in
Eq.(\ref{Wapprox2}) are smaller than the $O(y^4)$ terms at $y^4 =
1/k^2$.  This requires: $p > 4/7$ i.e. $k^2 > 1/4$.  To ensure that we
are likely to be justified in ignoring the $O(y^6)$ terms, we
therefore require $k^2 \geq 1/3$. Therefore for $1/3 \leq k^2\leq 2$
we find that:
$$
\frac{d}{2} \approx \frac{\sqrt{2}}{V_b^{\prime}-V_{0}^{\prime}}\int_0^{\infty} \frac{\dd y}{(1+c^2 y^2)^{1+1/n_{\rm eff}}}.
$$
Performing this integral, we arrive at:
\begin{equation}
m_0 d \approx \sqrt{\frac{2(n_{\rm eff}+1)}{n_{\rm eff}}} B\left(\frac{1}{2},\frac{1}{2}+\frac{1}{n_{eff}}\right), \label{m0midk}
\end{equation}
where $B(\cdot,\cdot)$ is the Beta function, $n_{\rm eff} =
(2-k^2)/(k^2-1)$ and $1/4 \leq k^2<2$ i.e. $n_{\rm eff} > 0$ or
$n_{eff} \leq -7/3$.  When $V = \Lambda^4 + \Lambda^4(\Lambda/\phi)^n$
and $m_0 \gg m_b$ this approximation for $m_0 d$ is actually exact and
$n_{\rm eff}=n$; note that for such a potential we require $n \leq -4$ or
$n > 0$ for a valid chameleon theory to emerge. Note that the
condition $m_0 \gg m_b$ implies:
$$
m_b d \ll \sqrt{\frac{2(n_{\rm eff}+1)}{n_{\rm eff}}}
B\left(\frac{1}{2},\frac{1}{2}+\frac{1}{n_{\rm eff}}\right).
$$

We now consider the small $k^2$ case.  If $k^2$ is small it is either because $k_{0}^2 \equiv V_{0}^{\prime \prime \prime}V_{0}^{\prime}/m_{0}^4$ is small, or because $1-V_{b}^{\prime}/V_{0}^{\prime}$ is small.  In the former case, the approximation:
$$
\frac{1}{W(y)^2} \approx (V_0^{\prime}-V_{b}^{\prime})^2(1 + 2a^2 y^2 + k^2 a^4 y^4 +O(y^6)),
$$
is generally no longer valid to $O(y^4)$ at the point $y=y_{-1}$, where $W(y) \sim C/y$, for some $C$.  When this approximation does hold at $y=y_{-1}$, we have $a^2 y_{-1}^2 \approx 1/k^2$.    The $O(y^6)$ terms are of the order $-g k^2 a^6 y^6/3$ where,
$$
g = \left(1-\frac{V^{(4)}_0(V_0^{\prime}-V_{b}^{\prime})}{m^2_0 V_{0}^{(3)}}\right).
$$
For the $O(y^6)$ term to be smaller than the $O(y^4)$ at $y=y_{-1}$ we
need $\vert g k^2 a^2 y^2_{-1} \vert = \vert g \vert < 3k^2$.  If
$V^{(4)}_0V_0^{\prime}/ V_{0}^{(3)} V_{0}^{(2)} \sim O(1)$, then this
condition generally holds provided that $3k^2 \gtrsim 1$.  It is for
this reason that we took $k^2 \geq 1/3$ in the previous case.  If
$k^2$ is smaller than a $1/3$, however, then it is generally the case
that $\vert g \vert < 3k^2$ does \emph{not} hold.  If $k^2$ is small
because $k_0^2$ is, then the behaviour of $W(y)$ near the point where
it drops off like $1/y$ is determined by higher derivatives in $V$
than $V^{(3)}_0$.  However, chameleon potentials with $k_0^2 \ll 1$ are not
generally very natural, and so we do not consider them here.
What is important to understand is behaviour of $\phi_0$ as $\phi_0
\rightarrow \phi_b$.

 As $d \rightarrow \infty$, $\phi_0$ converges to $\phi_b$ and so $k^2 \rightarrow 0$.  It is important to understand how $\phi_0$ and hence $F_{\phi}/A$ behave when $d \gtrsim O(1/m_b)$.  For large separations, $\phi_0 \rightarrow \phi_b$.  For $\phi$ close to $\phi_0$ then we have:
$$
y^2 + y_0^2 \approx \frac{m_b^2}{2}(\phi-\phi_b)^2 + \frac{V^{\prime \prime \prime}_b}{6}(\phi-\phi_b)^3,
$$
where
$$
y_0^2 = V_0 - V_b - V_b^{\prime}(\phi_0-\phi_b) = \frac{F_{\phi}}{A}.
$$
Expanding $W^2(y)$ we find:
$$
\frac{1}{W^2(y)} = (V_b-V)^2 \approx m_b^4(\phi-\phi_b)^2 + m_b^2V_b^{\prime \prime \prime}(\phi-\phi_b)^3 +O((\phi-\phi_b)^4).
$$
We therefore have:
\begin{equation}
\frac{1}{W^2(y)} \sim 2m_b^2(y^2 + y_0^2) - \frac{2V_b^{\prime \prime \prime}}{3m_b} (2(y^2+y_0^2))^{3/2}.
\end{equation}
We then have:
$$
\frac{m_b d}{2} \approx \int_0^{\infty} \frac{ \dd t}{\sqrt{t^2 + 1}\sqrt{1+ l^2 (t^2+1)^{1/2}}},
$$
where
$$
l^2 = -\frac{2\sqrt{2}V^{\prime \prime \prime}_b y_0}{3 m_b^3} \approx \frac{2k^2}{3}.
$$
For $m_b d \gg 1$ we then have:
\begin{eqnarray*}
\frac{m_b d}{2} &\approx& \int_0^{\infty} \frac{ \dd t}{\sqrt{1+l^2 \cosh t}} \\
&\approx& 2\sinh^{-1} \frac{\sqrt{2}}{l} - \frac{l^2}{4} + O(l^4) \approx \ln(8) - 2\ln(l) + O(l^4).
\end{eqnarray*}
Thus as $d \rightarrow \infty$ we have:
\begin{equation}
\frac{m_b d}{2} \approx \ln(12) - \ln(k^2).
\end{equation}
This approximation therefore certainly requires $l^2 \lesssim 1$ to be valid i.e. $k^2 \lesssim 3/2$. Furthermore we must also have $m_b \approx m_0$. If this holds then: $m_0^2 \approx m_b^2 (1+k^2)$ and so $m_0^2/m_b^2 \approx 1$ requires: $k^2 \ll 1$, which in turn requires $m_b d \gg 2\ln(12) \approx 5$.  Thus for $m_b d \gg 5$ we have
\begin{equation}
\frac{F_{\phi}}{A} \sim \frac{72 m_b^6 e^{-m_b d}}{V^{\prime \prime \prime\,2}_b}.
\end{equation}

Finally, we consider the behaviour of $m_0 d$ and hence $F_{\phi}/A$ for $k^2 > 2$.  In this case we write:
$$
\frac{1}{W^2(y)} \approx (V_0^{\prime}-V_b^{\prime})^2 (1 + 2 a^2 y^2 +  2 a^4 y^4 + (k^2-2)a^4 y^4 + O(y^6)),
$$
and so to $O(y^4)$ we have:
$$
\frac{1}{W^2(y)} \approx (V_0^{\prime}-V_b^{\prime})^2 e^{2 a^2 y^2} (1+(k^2-2)a^4 y^4).
$$
We therefore find that:
\begin{equation}
m_{0} d \approx \frac{\pi^{3/2}}{2\sqrt{2}(k^2-2)^(1/2)}\left[J_{-1/4}^2\left(\frac{1}{2\sqrt{k^2-2}}\right) + Y_{-1/4}^2\left(\frac{1}{2\sqrt{k^2-2}}\right)\right],
\end{equation}
where $J_{-1/4}$ and $Y_{-1/4}$ are Bessel functions.  For small $4(k^2-2)$ this gives:
$$
m_{0} d \approx \sqrt{2\pi}\left(1-3(k^2-2)/8\right),
$$
and if $k^2 \gg 2$ we have:
$$
m_{0}d \approx  \frac{B\left(\frac{1}{4},\frac{1}{4}\right)}{\sqrt{2 k}} \approx \frac{5.24}{\sqrt{k}},
$$
where, as above, $B(\cdot,\cdot)$ is the Beta function. Note that the $1/3 < k^2 < 2$ and $k^2 > 2$ approximations for $m_0 d (k^2)$ are continuous at $k^2 = 2$.

\section{Force between a Sphere and a Plate for Power-Law Potentials with $0 < n \leq 2$}\label{appC}
In this appendix we derive the force between a sphere and a plate for a power-law potential with $0 < n \leq 2$.  The potential is taken to have the form: $V= \Lambda^4 + \Lambda^4(\Lambda/\phi)^n$. When $m_0 \gg m_b$, Eq. (\ref{kmideqn}) is exact and we have:
$$
m_0 d = \sqrt{\frac{2n+1}{n}} B\left(\frac{1}{2},\frac{1}{2}+\frac{1}{n}\right) \equiv \sqrt{n(n+1)}K_{n}^{\frac{n+2}{2n}}.
$$
This result is valid so long as:
$$
m_b d \ll \sqrt{\frac{2n+2}{n}} B\left(\frac{1}{2},\frac{1}{2}+\frac{1}{n}\right).
$$
For $0 < n \leq 2$ the RHS of the above expression takes values
between about $2.5$ and $3.2$.  Provided $m_b d \ll 3$, the force $\dd
F_{\phi}$ between two parallel surfaces with separation $s$ and area
$\dd A$ is:
\begin{equation} \frac{\dd F_{\phi}}{\dd A} \approx K_{n} \Lambda^4
\left(\Lambda s\right)^{-\frac{2n}{n+2}}. \label{Fsmall}
\end{equation}
Let us now take one of the surfaces to be a sphere of radius $R$. The minimum separation of the sphere and plate is taken to to be $d$. Provided $m_b R \ll 1$, $d \ll R$, we showed in section \ref{sec:force:sphere} that the total force, $F^{\rm tot}_{\phi}$, is, to a good approximation, given by:
\begin{equation}
F_{\phi}^{\rm tot} \approx 2\pi R \int_d^{\infty} \frac{\dd F_{\phi}(s)}{\dd A} \dd s. \label{FFF}
\end{equation}
When $n > 2$ and $m_0 \gg m_b$, $\dd F_{\phi}(s)/\dd A$ drops off faster than $1/s$. No matter what $n$ is, we found in appendix \ref{appB} that when $m_0 \approx m_b$, $\dd F_{\phi}(s)/\dd A \propto e^{-m_b d}$. Therefore when $n > 2$, the dominant contribution to the $F_{\phi}^{\rm tot}$ comes from values of $s \approx d$. If, as is often the case, $m_b d \ll 1$, we can therefore calculate $F_{\phi}^{tot}$ to leading order knowing only the form of $\dd F_{\phi}(s)/\dd A$ for $m_b s \ll 1$. This is given by Eq. (\ref{Fsmall}).  This calculation is performed in section \ref{sec:force:sphere} and so is not repeated here.

In this appendix we consider the behaviour of $F_{\phi}$ when $0 < n \leq 2$. In these theories, when $m_b s \ll 1$, $\dd F_{\phi}(s)/\dd A$ drops off more \emph{slowly} than $1/s$. When $m_b s \gtrsim 5$, $\dd F_{\phi}(s)/\dd A \propto e^{-m_b d}$.  We therefore expect that $F_{\phi}(s)/A$ first drops off faster than $1/s$  when $s = s_{-1} \sim \Oo(1/m_b) < 5/m_b$.  The dominant contribution to the integral in Eq. (\ref{FFF}) then comes from values of $s$ that are $\Oo(s_{-1})$ i.e. $\Oo(1/m_b)$.

We found in Appendix \ref{appB} that the assumption, $m_0 \approx m_b$ required $m_b d \gtrsim 5$. It is therefore safe to assume that $m_0(s_{-1})$ is large compared to $m_b$.   The $1/3 \leq k^2 \leq 2$ approximation for $m_0 s$, Eq. (\ref{m0midk}) can therefore be applied.  We define $x = V^{\prime}_b/V^{\prime}_{0}$; $x \rightarrow 1$ as $s \rightarrow \infty$ and $x \rightarrow 0$ as $s \rightarrow 0$. We then have:
$$
n_{\rm eff} = \frac{n + (n+2)x}{1-(n+2)x},
$$
and so by Eq. (\ref{m0midk}) we have:
\begin{equation}
m_b s = x^{\frac{n+2}{2(n+1)}} m_0 s \approx L(x) \equiv x^{\frac{n+2}{2(n+1)}}\sqrt{\frac{2(n+1)}{n+(n+2)x}} B\left(\frac{1}{2}, -\frac{1}{2} + \frac{n+1}{n+(n+2)x}\right). \label{mbseqn1}
\end{equation}
Now
\begin{equation}
\frac{\dd F_{\phi}}{\dd A} = V_0 - V_b - V_b^{\prime}(\phi_0-\phi_b) = V_b\left(x^{-\frac{n}{n+1}}-1+n(x^{\frac{1}{n+1}}-1)\right). \label{Fofx}
\end{equation}
By Eq. (\ref{FFF})
\begin{eqnarray*}
F_{\phi}^{\rm tot} = 2\pi R \int_{0}^{\infty} \frac{\dd F_{\phi}(s)}{\dd A} \dd s  - 2\pi R \int^d_{0} \frac{\dd F_{\phi}(s)}{\dd A} \dd s
\end{eqnarray*}
For $0 < n < 2$, and provided $m_b d \ll 1$, the second term on the right hand side of the above expression evaluates to:
$$
2\pi R \int^d_{0} \frac{\dd F_{\phi}(s)}{\dd A} \dd s \approx \frac{2\pi K_n \Lambda^3 R(n+2)}{2-n}  \left(\Lambda d\right)^{\left(\frac{2-n}{n+2}\right)}.
$$
We define:
$$
F_0 = 2\pi R \int_{0}^{\infty} \frac{\dd F_{\phi}(s)}{\dd A} \dd s.
$$
Using Eq. (\ref{Fofx}) we have:
\begin{eqnarray}
F_{0} &=& 2\pi R \frac{V_b}{m_b} \int_0^{1} \left(x^{-\frac{n}{n+1}}-1+n(x^{\frac{1}{n+1}}-1)\right) \dd (m_b s), \nonumber\\
&=& 2\pi R \frac{V_b}{m_b} \int_0^{1} \left(x^{-\frac{n}{n+1}}-1+n(x^{\frac{1}{n+1}}-1)\right) L^{\prime}(x)\dd x,\label{F0L}
\end{eqnarray}
where the last equality follows from Eq. (\ref{mbseqn1}) and $L^{\prime}(x) = \dd L / \dd x$; $L(x)$ is given by Eq.  (\ref{mbseqn1}).

We now wish to evaluate $F_0$ approximately. The dominant contribution to $F_{0}$ comes from values of $x$ near the point where:
\begin{equation}
\left(x^{-\frac{n}{n+1}}-1+n(x^{\frac{1}{n+1}}-1)\right)L(x) \label{exprrr1}
\end{equation}
takes its maximum value.  For all $0 < n < 2$, this maximum occurs for values of $x < 0.5$.  Our assumption that $m_b/m_s = x^{(n+2)/(2(n+1))} \ll 1$ is therefore justified.

We define $x_{\rm max}$ to be the value of $x$ at the maximum of Eq. (\ref{exprrr1}); the smaller $n$ is, the larger $x_{\rm max}$ becomes. For $0 < x \lesssim O(x_{\rm max})$ we find that:
\begin{equation}
\sqrt{\frac{2(n+1)}{n+(n+2)x}} B\left(\frac{1}{2}, -\frac{1}{2} + \frac{n+1}{n+(n+2)x}\right) \approx \sqrt{\frac{2(n+1)}{n}}B\left(\frac{1}{2},\frac{1}{2}+\frac{1}{n}\right)\frac{(1+(\beta_n-1) x)}{1-x}. \label{betaapprox}
\end{equation}
where
$$
\beta_n = \frac{n+2}{2n^2}\left[2(n+1)\left(\Psi\left(\frac{1}{n}\right) - \Psi\left(\frac{1}{2}+\frac{1}{n}\right)+n\right)-n\right]
$$
and where $\Psi$ is the Digamma function i.e. $\Psi(x) = \Gamma^{\prime}(x)/\Gamma(x)$.  We use Eq. (\ref{betaapprox}) to approximate the right hand side of Eq. (\ref{mbseqn1}), and thus to approximate $L(x)$ as:
$$
L(x) \approx x^{\frac{n+2}{2(n+1)}} \sqrt{\frac{2(n+1)}{n}}B\left(\frac{1}{2},\frac{1}{2}+\frac{1}{n}\right)\frac{(1+(\beta_n-1)x)}{(1-x)}.
$$
Putting this approximation for $L(x)$ into Eq. (\ref{F0L}) we find
\begin{eqnarray}
F_{0} &\approx& 2\pi R D_{n} \left(\frac{n+2}{2-n}\right) \sqrt{\frac{2}{n^2}}B\left(\frac{1}{2},\frac{1}{2}+\frac{1}{n}\right) \left(\frac{\sqrt{n(n+1)}V_b}{m_b}\right), \nonumber
\end{eqnarray}
where:
$$
D_{n} = \frac{4n(n+1)}{(n+4)(n+2)}\left(1+\frac{2-n}{3(n+2)}\beta_n\right).
$$
We note that $D_2 = 1$.  $F_0$ may therefore be written as:
\begin{equation}
F_{0} \approx 2\pi \left(\frac{n+2}{2-n}\right) \Lambda^3 R K_n D_n \left(\frac{a_n \Lambda}{m_b}\right)^{\frac{2-n}{n+2}}
\end{equation}
where we have defined
$$
a_n = \sqrt{\frac{2(n+1)}{n}} B\left(\frac{1}{2},\frac{1}{2}+\frac{1}{n}\right).
$$
Thus for $0 < n < 2$ the total force between a sphere and a plate with separation $d \ll m_b^{-1} \ll R$ is:
\begin{equation}
F_{\phi}^{\rm tot}(0<n<2) = 2\pi \Lambda^3 R K_{n} \left(\frac{n+2}{2-n}\right) \left[ D_{n} \left(\frac{a_n \Lambda}{m_b}\right)^{\frac{2-n}{n+2}} - \left(\Lambda d\right)^{\frac{2-n}{n+2}}\right].
\end{equation}
By taking the limit of this equation  as $n \rightarrow 2$ we find if $n = 2$:
$$
F_{\phi}^{\rm tot}(n=2) = 2\pi \Lambda^3 R K_{2} \left[\log\left(\frac{a_2}{m_b d}\right) -4 D_2^{\prime}\right],
$$
where $D_{2}^{\prime} = \dd D_n / \dd n \vert_{n=2}$. We find that $4D_{2}^{\prime} = 2 + 2\log 2$, $a_2 = 2\sqrt{3}$ and $K_2 = \sqrt{2}$. Thus:
\begin{equation}
F_{\phi}^{\rm tot}(n=2) = 2\sqrt{2}\pi \Lambda^3 R \log\left(\frac{\sqrt{3}}{2 e^{2} m_b d}\right).
\end{equation}

\end{document}